\documentclass[apj,twocolumn, floatfix,table]{aastex631} \usepackage{amsmath}
\usepackage{longtable} 
\usepackage{aas_macros}
\usepackage{placeins}

\usepackage[]{xcolor}

\usepackage{ulem}

\definecolor{darkyellow}{RGB}{150,120,0}

\definecolor{meridithgreen}{RGB}{0, 150, 0}

\def\eff{\mathrm{eff}}

\shorttitle{Shooting a sparrow with a cannon}
\shortauthors{Joyce et al.}

\begin{document}

\title{Stellar Evolution in Real Time II: R Hydrae and an Open-Source Grid of\\$>3000$ Seismic TP-AGB Models Computed with MESA}

\author[0000-0002-8717-127X]{Meridith Joyce}
\affiliation{Konkoly Observatory, HUN-REN CSFK, Budapest, Konkoly Thege Mikl\'os \'ut 15-17, Hungary}
\affiliation{CSFK, MTA Centre of Excellence, Budapest, Konkoly-Thege Mikl\'os \'ut 15-17, H-1121, Budapest, Hungary}

\author[0000-0002-8159-1599]{L\'{a}szl\'{o} Moln\'{a}r}
\affiliation{Konkoly Observatory, HUN-REN CSFK, Budapest, Konkoly Thege Mikl\'os \'ut 15-17, Hungary}
\affiliation{CSFK, MTA Centre of Excellence, Budapest, Konkoly-Thege Mikl\'os \'ut 15-17, H-1121, Budapest, Hungary}
\affiliation{ELTE E\"otv\"os Lor\'and University, Institute of Physics and Astronomy, 1117, P\'azm\'any P\'eter s\'et\'any 1/A, Budapest, Hungary}

\author[0000-0001-7902-8134]{Giulia Cinquegrana}
\affiliation{School of Physics \& Astronomy, Monash University, Clayton VIC 3800, Australia}
\affiliation{ARC Centre of Excellence for All Sky Astrophysics in 3 Dimensions (ASTRO 3D)}

\author[0000-0002-3625-6951]{Amanda Karakas}
\affiliation{School of Physics \& Astronomy, Monash University, Clayton VIC 3800, Australia}
\affiliation{ARC Centre of Excellence for All Sky Astrophysics in 3 Dimensions (ASTRO 3D)}

\author[0000-0002-4818-7885]{Jamie Tayar}
\affiliation{Department of Astronomy, University of Florida,  USA }

\author[0000-0003-3759-7616]{D\'{o}ra Tarczay-Neh\'{e}z}
\affiliation{Konkoly Observatory, HUN-REN CSFK, Budapest, Konkoly Thege Mikl\'os \'ut 15-17, Hungary}
\affiliation{CSFK, MTA Centre of Excellence, Budapest, Konkoly-Thege Mikl\'os \'ut 15-17, H-1121, Budapest, Hungary}

\begin{abstract}
We present a comprehensive characterization of the evolved thermally pulsing asymptotic giant branch (TP-AGB) star R Hydrae, building on the techniques applied in \textit{Stellar Evolution in Real Time I} \citep{TUMi} to T Ursae Minoris. We compute over 3000 theoretical TP-AGB pulse spectra using \texttt{MESA} and the corresponding oscillation spectra with \texttt{GYRE}. We combine these with classical observational constraints and nearly 400 years of measurements of R~Hya's period evolution to fit R Hya's evolutionary and asteroseismic features. Two hypotheses for the mode driving R Hya's period are considered. Solutions that identify this as the fundamental mode (FM) as well as the first overtone (O1) are consistent with observations. Using a variety of statistical tests, we find that R Hya is most likely driven by the FM and currently occupies the ``power down"  phase of an intermediate pulse (TP $\sim$ $9$--$16$). We predict that its pulsation period will continu
e to shorten for millennia. Supported by calculations from the Monash stellar evolution code, we find that R Hya has most likely undergone third dredge-up in its most recent pulse. The \texttt{MESA}$+$\texttt{GYRE} model grid used in this analysis includes exact solutions to the linear, adiabatic equations of stellar oscillation for the first 10 radial-order pressure modes for every time step in every evolutionary track. The grid is fully open-source and packaged with a data visualization application. This is the first publicly available grid of TP-AGB models with seismology produced with \texttt{MESA}. 
\end{abstract}


\correspondingauthor{Meridith Joyce}
\email{meridith.joyce@csfk.org}

\section{Introduction} 
\label{sec:intro}
R~Hydrae is a type of star known as a \textit{Mira variable}, an astronomical class dedicated to the ``miraculous'' stars that would seemingly vanish and reappear to the naked eye.
R~Hydrae was one of the first Mira stars discovered, and it has been controversial since the 17th century. 
Since the star is only visible to the naked eye when near its maximum brightness, R~Hydrae, or R~Hya, was in fact independently discovered multiple times. Although it was first documented in 1662, its variability was not recognized until 1704 \citep{Hoffleit-1997}.  

Most Miras show period ``meandering", i.e., semiregular period changes on decadal timescales, but a handful of these stars display longer, monotonic changes in period on top of this \citep[see, e.g.,][]{Zijlstra-2002b,Templeton-2005,Merchan-2023}. Since its discovery, R~Hya has become one of the few Mira stars that shows significant, long-term period changes, and it stands out further among these thanks to the duration of its observed period decrease. We can trace R~Hya's period decay back to the start of the late 18th century through various data sources. While there is a dearth of recordings spanning most of the rest of the 18th century, the star has been identified in some earlier maps from the 17th century \citep{Zijlstra2002}. These suggest that the period was stable at around 495\,d, thus placing a constraint on the duration of the period decrease of 300 years or less.

Large shifts in pulsation period suggest significant changes in the structure of the stellar envelope. Since Mira stars are asymptotic giant branch (AGB) stars, they undergo thermal pulses (TPs), which can induce rapid structural changes. Indeed, \citet{wood-zarro-1981} showed that the period changes of R~Hya could be fitted by thermally pulsing evolutionary models. In their proposed scenario, the earliest observations---particularly the gap in the 18th century---corresponded to the maximum expansion phase after the He-shell flash, after which the star has been shrinking ever since. Based on this model, the actual He-shell flash would have happened about 550 years ago \citep[see also][]{Uttenhaler2010}. However, the models of \citet{wood-zarro-1981} were rather limited in their assumptions: the models varied only the core mass under a fixed, $2.0\,M_\odot$ envelope, for example, and did not allow for a detailed characterization of the star.

Taking into account the fact that the pulsation period appeared to reach a standstill in the 1950s, alternative hypotheses for R Hya's evolutionary state emerged. \citet{Zijlstra2002} proposed that the star might be undergoing envelope relaxation, which changes not only the structure of the outer layers, but the pulsation mode as well. In this case, we would have been observing a mode switch. This relaxation process could be driven by weak chaos, which may arise from the non-linear nature of the pulsation, as was shown to be the case for RV~Tau-type and semiregular variable stars \citep{buchler1996,Serre-1996}. 

Nevertheless, the thermally pulsing scenario has recently resurfaced. \citet{Fadeyev-2023,Fadeyev-2024} computed TP-AGB models and fitted them to the observed periods, assuming that the recent, apparent cessation in period decline indicated the end of radial contraction. 
The results of those studies agreed with those of \citet{wood-zarro-1981} in finding that the period decrease corresponded to the second phase of radial decrease. However, \citet{Fadeyev-2023} used only the extrema of the periods and the length of period decrease as modeling constraints. Soon after, \citet{Fadeyev-2024} made a more detailed comparison to the pulse morphology, but still assumed that R~Hya's radius had reached its minimum and therefore produced solutions requiring much higher core and stellar masses than previous works did.  

Ambiguity therefore remains regarding the cause of the changing period of R~Hya. Understanding the physical processes behind such changes is important not just for understanding R~Hya or AGB stars in general, but from the perspective of understanding the enrichment history and chemical evolution of the Galaxy. TP-AGB stars produce large amounts of gas and dust that disperse into the interstellar medium, with dust formation being driven by the carbon-to-oxygen (C/O) ratio.
Observations indicate that R~Hya has produced dust within the timescale of a thermal pulse, as detected by multiple instruments \citep{Hashimoto-1998,Decin-2008,Decin-2020,Homan-2021}. These observations show detached shells that indicate changes in the mass-loss rate: namely, a strong decrease in \textbf{dust} production about 230 years ago \citep{Decin-2008,ZhaoGeisler2012}. This coincides, to within a few decades, with the start of R~Hya's period decrease, indicating that this change could be related to the TP. 

Eventually, thermal pulses become vigorous enough to induce the third dredge-up (TDU), a convective mixing process that brings newly formed elements to the surface, including carbon, fluorine and \textit{s}-process elements. The changes in surface chemistry induced by TDU further impact mass loss, dust formation, and Galactic chemical enrichment \citep{Busso1999,Herwig05, KarakasLattanzio2014}.
 
There is, however, some uncertainty regarding whether the TP we are likely observing has undergone TDU. A clear indicator for TDU is the presence of technetium on the surface, which was detected in some spectroscopic measurements \citep{Merrill-1952yes,Orlov-1983,Little-1979,Lebzelter2003,tumi-uttenthaler2011,Uttenthaler-2019-Tc} but not in others \citep{Merrill1952no,Uttenhaler2010}. Unfortunately, these claims are difficult to verify: typically, only the existence of the lines is indicated without quantitative information such as the significance of the detection, though one exception claimed $\log\epsilon(\mathrm{Tc}) = -2.0$ \citep{Orlov-1983,Kipper-1991}. The non-detection published by \citet{Uttenhaler2010} was later revised by \citet{tumi-uttenthaler2011}. In the latter, they suggested that their earlier observations suffered from the ``line weakening'' effect, where absorption lines in a Mira star (not limited to Tc) may become considerably weaker than similar lines i
n non-Mira M-type stars \citep{Merrill-1962}. While \citet{tumi-uttenthaler2011} reclassified R~Hya as Tc-rich, R~Hya shows the least amount of Tc in the Tc-rich group.

This inconclusiveness calls for more detailed modeling of the short-term evolution of the star. With the availability of state-of-the-art stellar evolution and asteroseismic software and dedicated computing resources, grid-based modeling and ``big data'' approaches to characterizing AGB stars have become tractable. Foundations for these approaches were laid in the study of T~Ursae Minoris, another Mira star experiencing strong period changes. In \citet{TUMi}---the conceptual predecessor to this study---a hybrid evolutionary and asteroseismic modeling approach revealed that T~UMi began a thermal pulse only a few decades ago. T~UMi is now in the first phase of radial decrease, when the burning H-shell is extinguished, but the effects of the He-shell flash have not yet reached the surface. Detailed fits to the observed periodicities, the period change rate, and the transition to double-mode pulsation made it possible not only to determine many physical parameters of the star, su
ch as its birth mass, to unprecedented precision, but to make testable predictions for T UMi's behavior over the next several decades. 

The literature includes a number of TP-AGB calculations using a variety of stellar evolution codes and covering a range of masses and metallicities. Early examples of stellar calculations in this regime include models presented by \citet{Fagotto1994evolutionary}. \citet{bono1997evolutionary} and \citet{bono00} used the FRANEC code to study metal-rich stellar evolution up to $Z <0.04$ and for intermediate masses. \citet{Mowlavi98} and \citet{mowlavi1998grids} also computed models up to $Z < 0.1$ using the Geneva code. \citet{salasnich00} published metal-rich, $\alpha$-enhanced evolutionary tracks with the Padova code. Massive metal-rich stellar evolution was investigated by \citet{Meynet06,meynet_2008} and \citet{Siess07}. The effects of the $\Delta Y/\Delta Z$ relation were studied by \citet{valcarce13}, while \citet{Ventura20} looked at dust and gas production in high-$Z$ AGB stars. Other evolutionary codes and grids of calculations started to include the TP-AGB phase in the
m as well, such as the Granada code \citep{claret07}, GARSTEC \citep{Weiss09}, COLIBRI \citep{Marigo13,Marigo17}, and MIST \citep{Choi2016}, as well as the models of \citet{Bertelli08}. \citet{Karakas14He} and \citet{Karakas16} investigated nucleosynthesis and dredge-up in TP-AGB stars. Recently, the Monash \citep{Faulkner68evolution, Lattanzio86, Frost96,campbellThesis} and \texttt{MESA} \citep{MESAI, MESAII, MESAIII, MESAIV, MESAV, MESAVI} codes were also used to study high-$Z$, TP-AGB stellar evolution in \citet{Karakas2022} and  \citet{Cinquegrana22paper2}.

However, such calculations have varying, and often limited, degrees of open-source accessibility. Further, with the exception of stellar evolution databases that provide approximate global asteroseismic parameters at every time step (e.g., $\nu_\text{max}, \Delta \nu$ in MIST; \citealt{Choi2016}), none of these existing AGB grids provide exact frequency solutions to the oscillation equations. 

The Hungarian euphemism ``\'agy\'uval l\H{o} ver\'ebre'' translates in English to ``shooting a sparrow with a cannon.'' In this paper, we expand upon methods developed by \citet{TUMi} in order to determine the physical characteristics and precise evolutionary state of R~Hya by analyzing over 3000 asteroseismic stellar evolution models run from the zero-age main sequence to the end of the TP-AGB. We thus present the most comprehensive evolutionary, asteroseismic, and chemical/nucleosynthetic analysis of R~Hydrae to date, accompanied by the first fully open-source grid of asteroseismic AGB calculations. 

In Section~\ref{sec:observations}, we collate and discuss all available observational constraints for R~Hya, including new period measurements introduced in this analysis. In Section~\ref{sec:modelgrid}, we describe the construction of the model grid. In Sections \ref{sec:twomodes} and \ref{sec:shape_of_pulse}, we explore different possibilities for R~Hya's pulsation mode and evolutionary stage, respectively. In Section \ref{sec:fittingprocedure}, we describe the model fitting procedure and several statistical methods for determining preferred solutions. In Section \ref{sec:results}, we present heat maps showing the relative likelihood of models of R~Hya as a function of initial mass and initial metallicity, and in Section \ref{sec:discussion}, we discuss these results in the context of previous parameter determinations in the literature. 
In Section \ref{sec:advanced_tactics}, we briefly explore the effects of non-adiabaticity and non-linearity on our solutions, which are otherwise computed using the adiabatic and linear approximations.
In Section \ref{sec:nucleo}, we discuss the chemical and nucleosynthesis-related features of our preferred models. In Section \ref{sec:Monash}, we provide and discuss a suite of ``clone'' models computed with the Monash stellar evolution code designed to supplement our preferred models by exploring their proclivity to undergo TDU. 
In Section \ref{sec:moneysec}, we discuss the implications of our preferred models for the past and future behavior of R~Hya and compare this study to our previous work on T Ursae Minoris. We conclude and summarize this analysis in Section \ref{sec:conclusions}.
We discuss future work in Section \ref{sec:future}.
All new observational measurements processed for this analysis can be found in Appendix \ref{appendix:data}. The effects of parameter variations in our fitting and statistical procedures are detailed in Appendix \ref{appendix:param_var}. Instructions on how to use the data visualization application packaged with this grid are provided in Appendix \ref{appendix:visualizer}, along with demonstration of some of its features.

\begin{figure*}
\centering
\includegraphics[width=\textwidth]{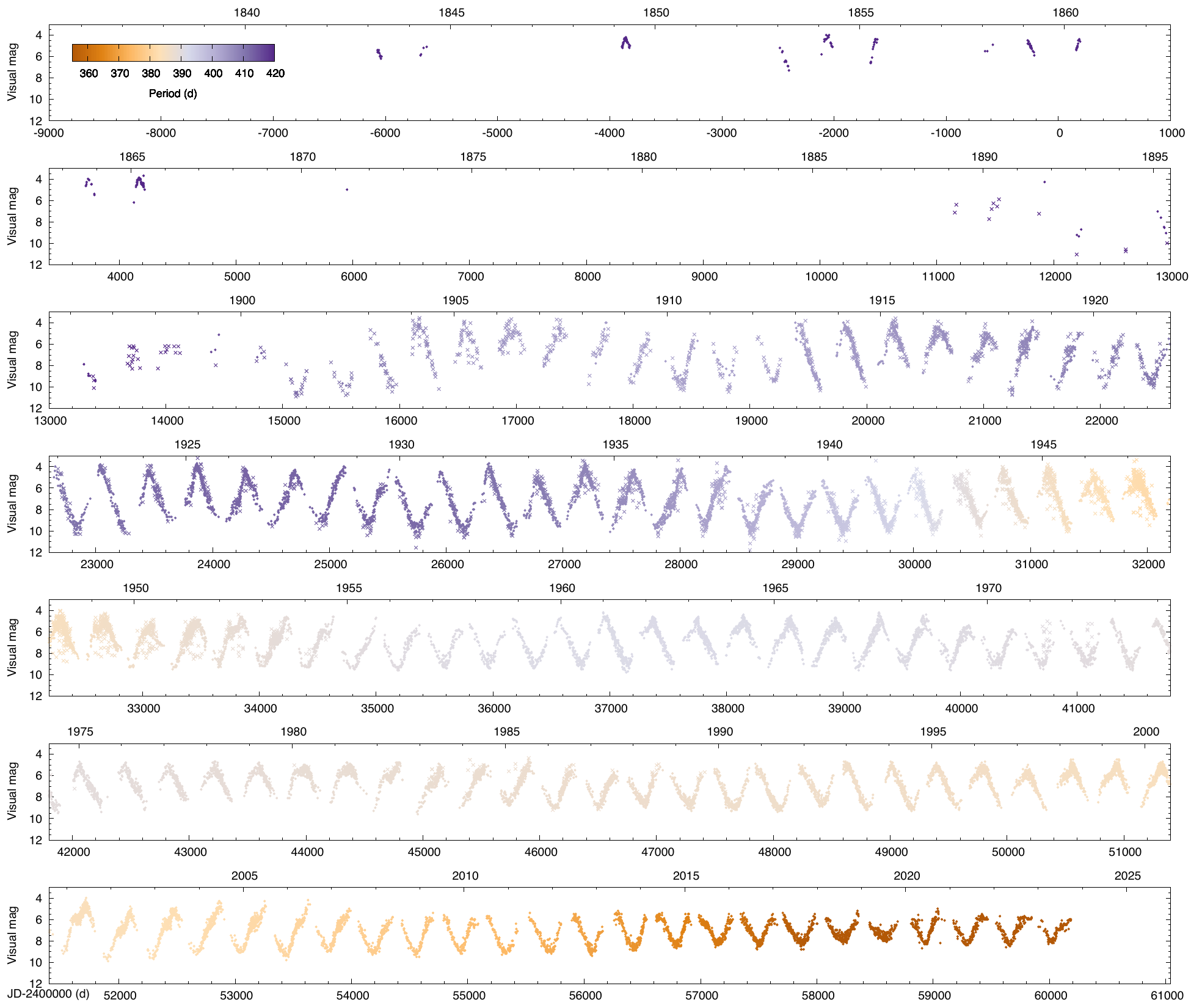}
\caption{Visual light curve of R Hya from the AAVSO (points) and scaled photographic data from DASCH (crosses). The coloring refers to the pulsation period.} 
\label{fig:lightcurve}
\end{figure*}

\section{Observational Constraints}
\label{sec:observations}
We collected all relevant observations on R~Hya to constrain the parameter space for the models. We first summarize the various constraints on the fundamental physical parameters of the star, then we discuss constraints we can obtain from the available time-domain photometry. Further indirect constraints (e.g., age, mass) that we compare to the model grids are discussed in Section~\ref{sec:discussion}. 

We note that if we assume that the period change is caused by a TP, then changes in the observations correspond to real changes in the radius, luminosity and $T_{\rm eff}$ of the star. As such, measurements taken 2--3 decades ago may include small shifts in those parameters compared to contemporary values.   

\subsection{Physical parameters}
\label{subsec:phys_par}
The presence of TiO and VO molecular bands in the spectra of R~Hya indicate that it is an O-rich Mira \citep{Lockwood-1969}. Many other oxygen-bearing molecules have been observed in the circumstellar envelope of the star as well, as described more recently in \citet{Wallström-2024}. 

The distance to R~Hya was determined to be $125\pm10$\,pc by \citet{Haniff-1995} based on period--luminosity (PL) relations. This agrees well with the result of \citet{dusty-vlbi-2022}, who found $d = 126_{-2}^{+3}$\,pc based on a reanalysis of \textit{Gaia} DR3 data. \citet{dusty-vlbi-2022} also derived a luminosity of $L = 10300 \pm 300\,L_\odot$ and effective temperature of $T_{\rm eff} = 3100$\,K using DUSTY models fitted to the spectral energy distribution (SED) of the star. Other authors found lower luminosity and temperature values based largely on broad-band near-infrared color data: \citet{deBeck-2010} derived $L=7375\,L_\odot$ and $T_{\rm eff} = 2128\pm140$\,K; \citet{Haniff-1995} found $T_{\rm eff} = 2680\pm70$\,K; \citet{Feast-1996} found $T_{\rm eff} = 2830\pm170$\,K.

Given the proximity and large physical size of the star, its radius can be derived via interferometry. \citet{Feast-1996} published $R = 401 \pm 46\,R_\odot$, whereas \citet{Haniff-1995} published two different model solutions corresponding to $R = 442 \pm 65\,R_\odot$ or $R = 384 \pm 54\,R_\odot$. We note that these measurements refer to the contemporary radius value and not to the maximum extent of the star when it was pulsating with longer periods.

\begin{table*} 
\centering 
\caption{Summary of Observational Information on R Hydrae} 
\begin{tabular}{cccl}  
\hline \hline
Property & Value & Reference & Notes  \\
\hline
\multicolumn{4}{c}{Directly derived physical constraints}\\
\hline
Luminosity ($L_\odot$) & $11600\pm1000$ & \citet{Zijlstra2002} & 
using $d=165$\,pc \citep{Eggen-1985}\\
Luminosity ($L_\odot$) & $10300\pm300$ & \citet{dusty-vlbi-2022} & 
VLBI distance and SED modeling\\
Luminosity ($L_\odot$) & 7375 & \citet{deBeck-2010} & PL relations \\
Temperature &  $2830\pm170$\,K & \citet{Feast-1996} & (\textit{J--K}) color\\
Temperature &  $2128\pm140$\,K & \citet{deBeck-2010} & (\textit{V--K}) color\\
Temperature &  $2680\pm70$\,K & \citet{Haniff-1995} &  bolometric flux and angular diameter\\
Temperature &  2600\,K & \citet{teyssier-2006} & IR color-temperature \\
Radius ($R_\odot$) & $442\pm65$ & \citet{Haniff-1995} &  
D model (FM, fainter, hotter base model)\\
Radius ($R_\odot$) & $384\pm54$ & \citet{Haniff-1995} &  E model (O1, brighter, cooler base model) \\
Radius ($R_\odot$) & $401\pm46$ & \citet{Feast-1996} & based on \citet{Haniff-1995} \\
Composition & rich in $^{99}$Tc & \citet{Uttenthaler-2019-Tc} & indicates it has undergone third dredge-up \\
$\dot{P}$ (d/yr) & $ -0.493\pm0.011$ & this work & from photometry \\
\hline
\multicolumn{4}{c}{Indirect physical constraints}\\
\hline
Age & 6.9 Gyr (3.6–8.5 Gyr) & \citet{Eggen-1998} & P--age relation \\
Age & 6.5 Gyr (4.1–7.4 Gyr) & \citet{Zhang-2023} & P--age relation \\
Age & 1.01 Gyr (0.17–1.83 Gyr) & \citet{Grady-2019} & P--age relation \\
Age & $0.5^{+2.6}_{-0.3}$ Gyr & \citet{Trabucchi-2022} & P--age relation \\
Mass ($M_\odot$) & 1.35 (1.0--1.5) & \citet{Decin-2020} & $^{17}$O/$^{18}$O isotope ratio \\
Mass ($M_\odot$) & 1.5--2.5 & \citet{Homan-2021} & close binary model, total mass of $\sim$ 2.5 $M_\odot$  \\
Mass ($M_\odot$) & 1.26--3.28 & \citet{Trabucchi-2022} & P-M relation (300 d FM period) \\
Mass ($M_\odot$) & 2.45--6.28 & \citet{Trabucchi-2022} & P-M relation (500 d FM period) \\
$[\rm{Fe/H}]$ & 0.1--0.65 & \citet{Feast-2000} & P--[Fe/H] relation (extrapolated) \\
$[\rm{Fe/H}]$ & 0.0 (-0.6 -- +0.5) & \citet{APOGEE-2015} & range based on Galactic kinematics \\
\hline
\multicolumn{4}{c}{Astrometric constraints}\\
\hline
$\pi_{\rm DR3}$ &   $6.74 \pm 0.46$ mas & \citet{GaiaEDR3} & Gaia DR3 parallax; not converted to distance \\
Distance   &  165 pc & \citet{Eggen-1985} & Hyades supercluster\\
Distance   &  130 pc & \citet{whitelock-2008} & PL relation\\
Distance  &  $126^{+3}_{-2}$ pc & \citet{dusty-vlbi-2022} & VLBI data\\
Distance & $150 \pm 10$ pc & \citet{BJ-EDR3-2021} & Bailer-Jones EDR3 distance; no Apsis data \\
$R_{\rm gal}$ & $8041\pm159$\,pc & this work & galactocentric distance \\
z & $114\pm2$\,pc & this work & vertical distance from the plane \\
$v_\gamma$ & $-15.9 \pm 1$ km/s & \citet{tumi-uttenthaler2011} & literature RV data \\
U velocity & $-41.1\pm 1.0$\,km/s & this work & based on Gaia DR3 and $v_\gamma$ \\
V velocity &  $-9.24\pm0.04$\,km/s & this work & based on Gaia DR3 and $v_\gamma$ \\
W velocity &  $ +2.05\pm0.04$\,km/s & this work & based on Gaia DR3 and $v_\gamma$ \\
\hline
\end{tabular}
\label{table:obs_constraints}
\end{table*}

\begin{figure*}
\centering
\includegraphics[width=\textwidth]{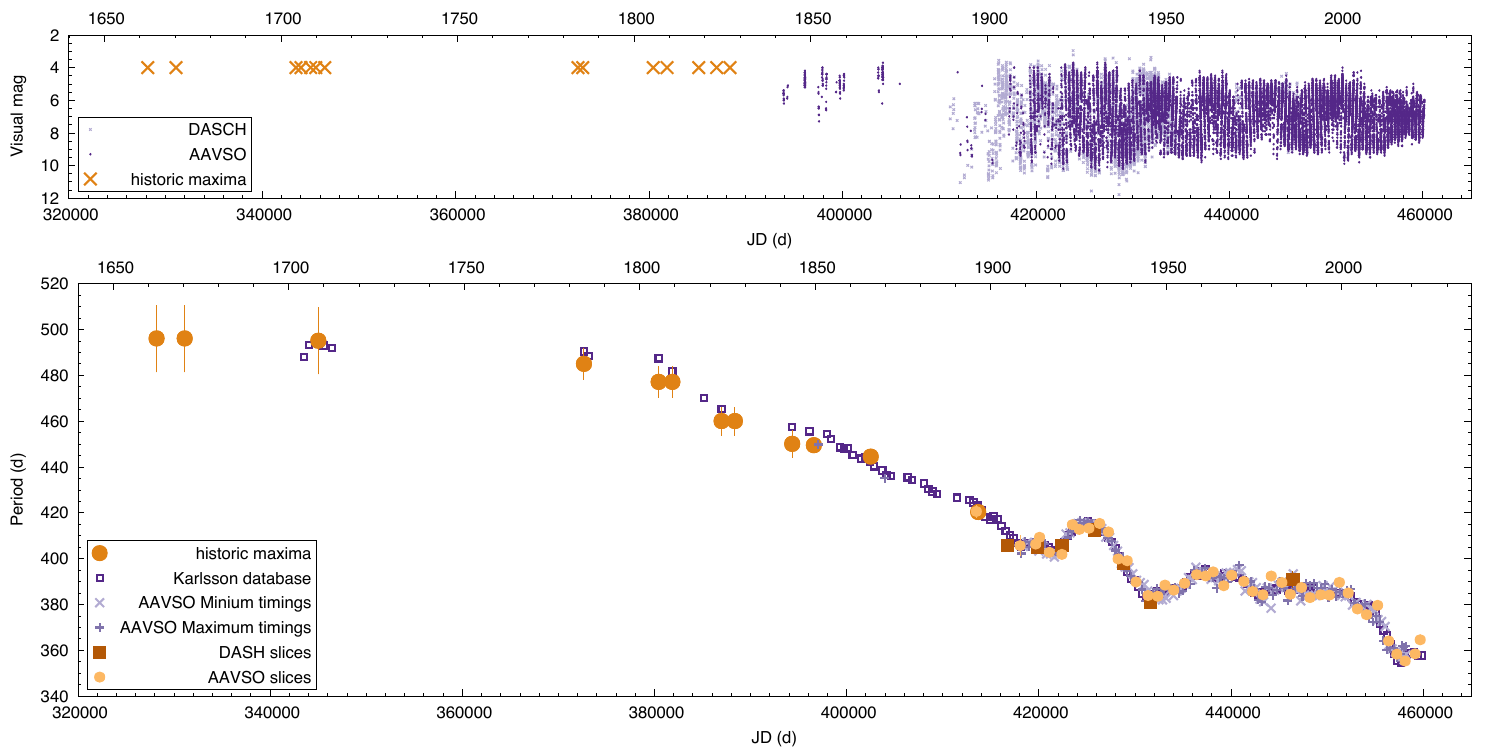}
\caption{Light curve and period evolution of R Hya. Upper panel: the same light curve as in Fig.~\ref{fig:lightcurve}, extended with the time stamps of positive detections based on stellar maps by \citet{Zijlstra2002}. Lower panel: the calculated pulsation period values. Large brown circles are from \citet{Zijlstra2002}, with error ranges estimated by the authors here. Blue squares are from the Mira O--C database. The rest of the calculations are new, produced in the present analysis, as described in Section~\ref{sect:per-evo}.}
\label{fig:period_ev}
\end{figure*}

\subsection{Period evolution}
\label{sect:per-evo}
The steady decrease of the pulsation period has been known for over a century, with regular observations going back to the 1840s. \citet{Zijlstra2002} successfully recovered the pulsation period for the second half of the 17th century and late 18th century from historical maps, and concluded that R~Hya appeared to have a constant period of 495 d before the start of the period decline. Based on their TP models, \citet{wood-zarro-1981} proposed that R~Hya is evolving from the maximum expansion and luminosity phase of the TP, and the change in $\dot{P}$ matched the change of slope in their model.

After the extensive decline, the period seemed to be stabilizing at 385 d in the 1990s. This raised the possibility of a mode switch due to envelope relaxation \citep{Zijlstra2002}, or an end to the radial decline \citep{Fadeyev-2023}. 

Since then, however, more than 20 years of additional photometry has become available. Figure \ref{fig:lightcurve} shows the recorded light variations of R~Hya over 380 years, extending into 2023. In this plot, we combine the visual brightness estimates collected by the American Association of Variable Star Observers (AAVSO) with the measurements of the Digital Access to a Sky Century at Harvard project (DASCH\footnote{\url{https://dasch.cfa.harvard.edu/}}) based on archival photographic plates \citep{dasch2012}. We scaled the average brightness and average pulsation amplitude of the DASCH data to match that of the visual estimates. 

The AAVSO light curve immediately suggests that the pulsation amplitude has continued to decrease over the last 15--20 years. Our analysis of the new data set indicates that the period change of R~Hya did not stop, but rather began to decline steeply again since 2000. Figure~\ref{fig:period_ev} shows the full span of observations, including the historic detections and period estimates compiled by \citet{Zijlstra2002}. 

We estimated the more recent periods in two ways for Figure~\ref{fig:period_ev}. First, we cut the AAVSO and DASCH data into 2000 d long segments and calculated the pulsation frequencies of each segment with Period04 \citep{Lenz-2005}. We fitted the pulsation frequencies along with two harmonics. For the AAVSO data, we used a sliding window with a step size of 1000 days. We also used the pulsation cycle lengths calculated from maximum and minimum timings published by \citet{AAVSO-2021} and by \citet{Karlsson-2013} in his extensive Mira O--C database\footnote{\url{http://var.saaf.se/mirainfooc.php}}. The latter helpfully includes the 19th century observations collected in the seminal paper of pioneering female astronomer Annie Jump Cannon \citep{Cannon-1909}\footnote{We encourage the reader to check this paper for bright long-period variables manually, because it is not indexed by SIMBAD.} in digitized form. Unfortunately, the database comes without error values. The authors t
hemselves note that in many cases they had no information on the uncertainties in the maxima timings, which could be as high as multiple days, depending on observers and observing conditions. 

Based on the extensive period data, we find that, instead of stagnating in its period decline, the star is experiencing meandering, a well-known behavior of Mira stars in which the pulsation period changes quasi-periodically on a decadal to centennial scale. The combined effect of the meandering and a continuing evolutionary decrease led to an apparent but temporary stabilization of the pulsation period. The new observations extend the length of the period change to nearly 250 years and can be fitted with a linear rate of $\dot{P} = -0.493 \pm 0.011$\,d/yr. 

\subsection{Separation of period change and meandering}
\label{subsec:meadering}
The physical cause of meandering has not yet been established. One hypothesis suggests that meandering is caused by thermal relaxation oscillations, potentially connected to the TP \citep{Templeton-2005}. However, current TP-AGB evolutionary models are unable to reproduce meandering; they are designed to calculate smooth changes in the stellar structure during oscillations caused by He-shell flashes that take place over few-hundred year timescales. Since the physics of meandering is not included in \texttt{MESA}, we can only fit the global period decline trend underlying the meandering with our models. This means that our theoretical curves may appear to be poor fits to some individual period measurements, and the goodness-of-fit calculations will be affected by meandering-based offsets that are not reflective of the true evolutionary trend.

Because of this, we tested whether separating the meandering from the TP-induced period changes resulted in better model fits. We did this by fitting a quadratic polynomial to the post-1845 period values. We fitted both the AAVSO light curve segment periods and the \citet{Karlsson-2013} data set: the differences were only noticeable for the very sparse late-19th century segment of the data. We elected to work with the AAVSO segments since we could determine period uncertainties for them, whereas the maximum timings did not include any uncertainty data. The best-fitting curve is the following: $0.00086819\,t^2 -0.5839\,t + 441.279$, where $t$ is defined in years relative to the epoch of JD2400000. With the presumed TP-induced period change removed, we can compare the meandering signal to that of other Miras. A frequency analysis reveals a dominant periodicity of 36.8 yr. 

However, applying this fit involves making an assumption about the shape of the period decline. Instead, we adjust our fitting method to handle the elevated fit uncertainties coming from the presence of extra variations caused by physics that is not included in the models.

\begin{figure}
\centering
\includegraphics[width=\columnwidth]{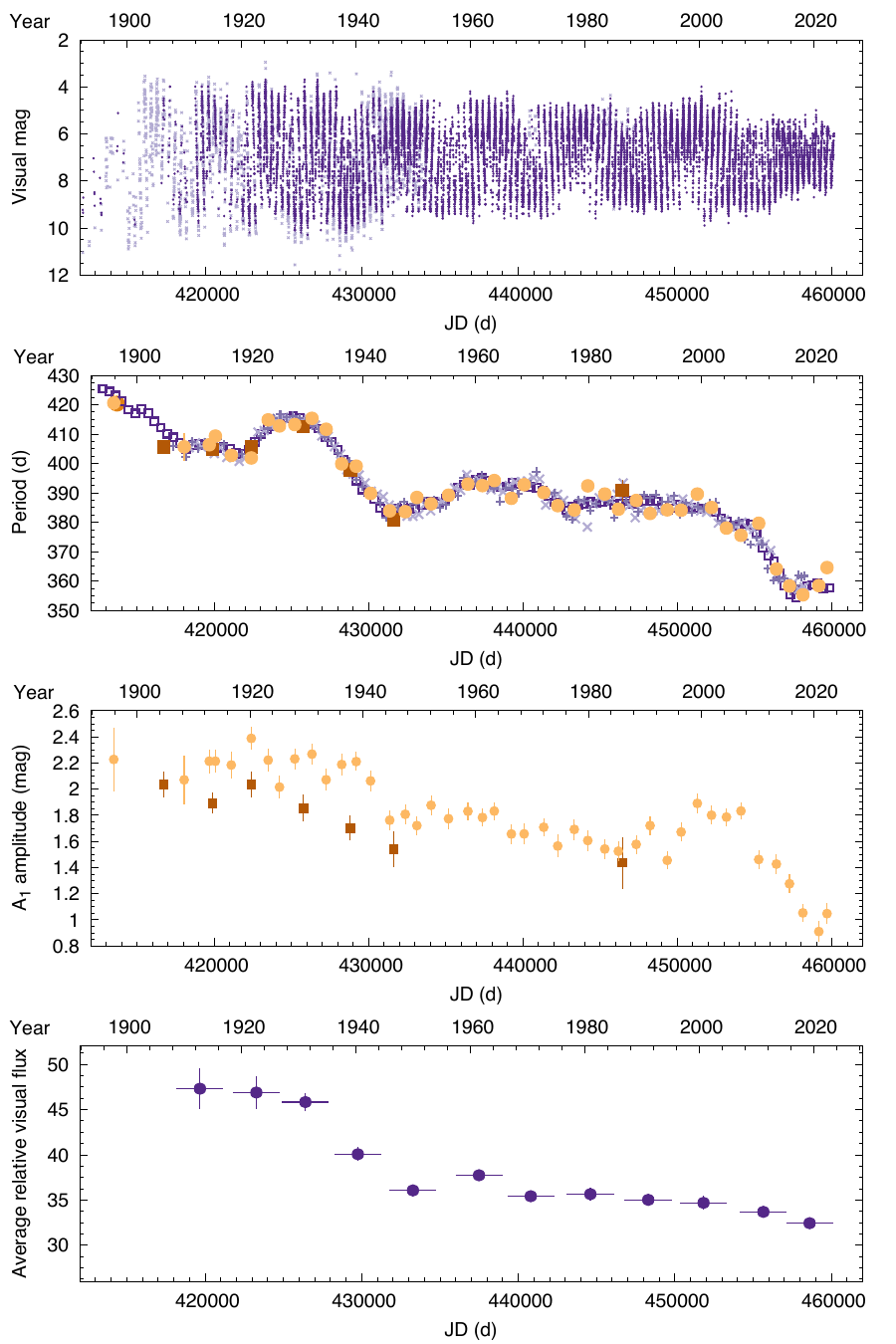}
\caption{Changes in physical characteristics for R~Hya between 1890--2023. From top to bottom: the visual light curve; the pulsation period; the Foruier amplitude of the pulsation frequency; the estimated pulsation-averaged visual flux relative to the solar value. Values produced in the present analysis, as described in Sections~\ref{sect:per-evo} and \ref{sec:amp-evo}.  }
\label{fig:period_ev_dense}
\end{figure}

\subsection{Changes to the amplitude and average brightness}
\label{sec:amp-evo}
The pulsation amplitude of the star appears to be shrinking in parallel with its period. Over the last 130 years, enough photometric data have been collected to analyze the temporal evolution of the pulsation amplitude and average brightness of the star. We use the Fourier coefficients calculated in Section~\ref{sect:per-evo} and find that the Fourier amplitude of the pulsation frequency and the frequency itself are strongly correlated. As the frequency increases (period shrinks), the amplitude decreases as well. 

We also examined the average brightness of the star. If R~Hya is in a TP, its luminosity should be decreasing along with its radius. At first glance, the average brightness appears to be increasing in Figure~\ref{fig:period_ev}, but that is only an artefact of using magnitude units. The visual brightness of the star changes by multiple orders of magnitude, and the bright phases get compressed by the logarithmic scaling. If we convert it into flux units, the downward change in pulsation-averaged brightness becomes apparent. We note that we did not apply bolometric corrections to the visual flux. The variation of Mira stars is much larger in visual than in bolometric units due to flux conversion into infrared near minimum light, therefore the average flux relative to solar we plot here is also lower than the luminosity of the star. Nevertheless, we observe a strong correlation with the average flux and the pulsation period. 

\subsection{Metallicity and kinematics}
\label{sect:kinematics}
Mira stars have cool atmospheres with a rich array of absorption features, including molecular lines. However, this makes determining their [Fe/H] exceedingly difficult. Here we can turn to clusters that contain Mira variables, and use the cluster metallicities as proxies. The period--[Fe/H] relation proposed by \citet{Feast-2000} suggests an [Fe/H] range of 0.1 to 0.65~dex, indicating super-solar metallicity. However, this is based on  extrapolation, since the clusters used by \citet{Feast-2000} all have sub-solar metallicities.

The location and kinematic information of R~Hya can further constrain its physical parameters. We calculated the galactocentric coordinates and velocities of R~Hya using the Gaia DR3 data \citep{GaiaDR3} and the python notebooks for the Gala python package \citep{gala,gala-zenodo}. Since there is no radial velocity data for R~Hya in DR3, we substitute the value of --15.9\,km/s measured by \citet{tumi-uttenthaler2011}. The Galactic velocity components of the star, relative to the Local Standard of Rest \citep{LSR-2010}, are $U_{\mathrm{LSR}} = -41.1\pm 1.0$\,km/s, $V_{\mathrm{LSR}} = -9.24\pm0.04$\,km/s, and $W_{\mathrm{LSR}} = 2.05\pm0.04$\,km/s, respectively. These low velocities place the star in the thin disk population on a Toomre diagram \citep[e.g.,][]{Feltzing-2003, Yan-2021}. 

We find R~Hya's galactocentric distance and the vertical distance from the galactic plane to be $R=8041\pm159$\,pc and $z=114\pm2$\,pc, respectively. If we compare this to the observed metallicity distribution functions of the APOGEE survey, we find that the star falls into the group centered on solar metallicity, with the distribution extending from [Fe/H] = - 0.6 all the way up to +0.5 (lower panel of Fig.~5 in \citealt{APOGEE-2015}, $|z|<0.5$~kpc, $7<R<9$~kpc slice). This is consistent with the range we obtained from the period--[Fe/H] relation, and suggests that R~Hya may indeed be super-solar in terms of metallicity. 

\subsection{Possible companion stars}
R~Hya is more accurately the R~Hya system. It has one resolved companion with an estimated mass of 0.8~$M_\odot$ at a separation of 3500 au, which is too far away and too small to alter R~Hya's evolution \citep{Mason-2001,Anders-2019}. However, there are some indications of a much closer companion as well. High-resolution observations revealed bi-conical outflows around the star that is potentially shaped by a close-by companion \citep{Decin-2020}. \citet{Homan-2021} observed a disk around the star with an inner edge at a distance of 6 AU (or $\sim3$ stellar radii). Based on the geometry and velocities of the disk, they estimate a current central mass of $\sim2.5\,M_\odot$, which includes a proposed inner companion with a tentative lower mass limit  0.65~$M_\odot$. 

The presence of the distant companion offers the possibility of measuring the metallicity of the system via that companion instead of R~Hya itself. This star, Gaia DR3 6195030801634430336, is a K-type dwarf \citep{Smak-1964}, and metallicities of (late) K- and M-type dwarfs could be estimated either from \textit{J--K} photometry or from moderate-resolution spectroscopy \citep{Johnson-2012,Mann-2013}. Unfortunately, we did not find the appropriate observations.
Nevertheless, Gaia DR3 6195030801634430336 offers an opportunity to verify the metallicities predicted by the AGB models for R~Hya. We therefore urge observers to target this dwarf companion. 

While the possibility put forth in \citet{Homan-2021} is worthy of future consideration, theirs is currently the only claim of a close companion. We proceed treating R~Hya using a single-star evolutionary assumption, which is also appropriate for well-separated, non-interacting binaries (e.g., \citealt{Joyce2018balphaCen}). When we refer to R~Hya henceforth, we are referring to the primary component of the wide binary.

\section{Model Grid}
\label{sec:modelgrid}
We use the Modules for Experiments in Stellar Astrophysics (\texttt{MESA}; \citealt{MESAI, MESAII, MESAIII, MESAIV, MESAV, MESAVI}) stellar evolution program to compute evolutionary tracks. The precise version used in this analysis is commit \texttt{1d059d5}\footnote{The commit is associated with pull request 541 and documented here: \url{https://github.com/MESAHub/mesa/pull/541}}, which closely follows \texttt{MESA} stable release version 23.05.1. This slightly more recent \texttt{MESA} iteration includes modifications necessary to make grids of AGB calculations tractable. In particular, this commit involves the reintroduction of velocity drag, which helps to prevent the development of instabilities caused by excess motion in the outer layers of the star during the TP-AGB phase (see, e.g., \citealt{Farag2023}). We use the \texttt{GYRE} stellar oscillation program \citep{GYRE} version 7.0 in adiabatic mode to compute exact solutions for the lowest ten radial order ($\ell = 0$
) $p$-modes accompanying each time step.  

We launch two variations of the following grid:
\bgroup\let\qquad\quad\def\quad{,}\def\newskip{\qquad\qquad}
\begin{align*}
M_\text{init} &= \{0.8 \ldots 5.0\}, \qquad \delta M = 0.1 M_\odot ,\\
Z_\text{init} &\in \bigl\{ 0.0001 \quad 0.0005 \quad 0.0010 \quad 0.0013 \quad 0.0018, \\
&\newskip 0.0020 \quad 0.0025 \quad 0.0030 \quad 0.0036 \quad 0.0040, \\
&\newskip 0.0050 \quad 0.0060 \quad 0.0070 \quad 0.0080 \quad   0.0095 ,\\
&\newskip 0.0100 \quad 0.0125 \quad 0.0135 \quad 0.0140 \quad 0.0200 ,\\
&\newskip 0.0216 \quad 0.0247 \quad 0.0300 \quad 0.0344 \quad 0.0400 ,\\
&\newskip 0.0450 \quad 0.0500 \quad 0.0600 \quad 0.0700 \quad 0.0800 ,\\
&\newskip 0.0900 \quad 0.1000 \quad 0.1100 \quad 0.1200 \quad 0.1300 ,\\
&\newskip\hphantom{0.0000,0.0000,0.0000,\},} 0.1400 \quad 0.1500 \bigr\}
\end{align*}
\egroup
for a total of $37 \times 43 = 1554$ models each. In the first variation, the initial helium abundance, $Y_\text{init}$, is fixed to $Y_i = 0.3$ uniformly. In the second, $Y_i$ is scaled according to $Z_\text{init}$ via the canonical relation 
\begin{equation}
    \rm Y_i = Y_0 + \frac{\Delta Y}{\Delta Z} \times Z_i, 
\label{eq:deltaY_deltaZ} 
\end{equation}
where $Y_0$ is the primordial He abundance and $\frac{\Delta Y}{\Delta Z}$ the He-to-metal enrichment ratio. We take $Y_0 = 0.2485$ \citep{Aver13} and $\frac{\Delta Y}{\Delta Z} = 2.1$ \citep{casagrande2007helium}.

The percentage of models that entered the TP-AGB phase in the fixed-helium grid is roughly 97\%, whereas $\sim90\%$ of the helium-varied models made it to this point. However, more than half ($\sim60$\%) of helium-varied models failed at some point before reaching the end of the TP-AGB due to convergence difficulties (but nearly always after intersecting R~Hya's parameters---see Section \ref{sec:visualizer}). In contrast, the convergence failure rate of the static helium grid was under 10\%, but the majority of those models were instead truncated by a maximum run time allowance of 72 hours (again, far later in their evolution than needed to fit R~Hya).  
When using the $Y_i \sim Z_i$ treatment, the highest metallicities ([Fe/H]${}\gtrsim+0.5$) can require stars comprising 50\% or more helium. Such exotic (and improbable) compositions are better studied with precision modeling---small numbers of carefully built models with tailored assumptions---rather than grids involving large numbers of models that use generalized assumptions, as presented here.

For these reasons, we focus on the fixed-$Y_i$ grid throughout the rest of the paper, but make both sets of models available. While changes in the treatment of helium do impact age results and some best-fitting pulse indices, overall trends in the solution spaces remain largely unchanged (see Section \ref{sec:discussion}). Results from both grids are presented when their differences are significant. 

Though they ran successfully, models adopting the two lowest metallicity values, $Z_\text{init} = 0.0001$ and $Z_\text{init} = 0.0005$, are excluded from figures and further analysis for two reasons:
(1) there are no cases of plausible fits to R~Hydrae for either of these; and
(2) they correspond to [Fe/H] $= -2.2$ and $-1.53$, respectively, whereas the remaining metallicities are roughly uniformly spaced in [Fe/H] from $-1.0$ to $+1.0$.  

In almost every case, models with masses below $1.0 M_\odot$ failed to converge for both helium configurations. Hence, $M = 0.8 M_\odot$ and $M=0.9 M_\odot$ are also excluded from figures and further consideration. The choice of an upper bound of $5 M_\odot$ was informed by a previous estimate placing R~Hya's mass at $4.8 M_\odot$ \citep{Fadeyev-2023}. The metallicity minimum was set based on the absence of reasonable models below [Fe/H]${}\sim-1.2$ in exploratory calculations, and the metallicity maximum was set high enough to accommodate the theoretical metallicity maximum stars can reach, per \citet{Cheng-Loeb-2023}.

\subsection{Description of physical assumptions}
\begin{figure}
\centering
\includegraphics[width=\columnwidth]{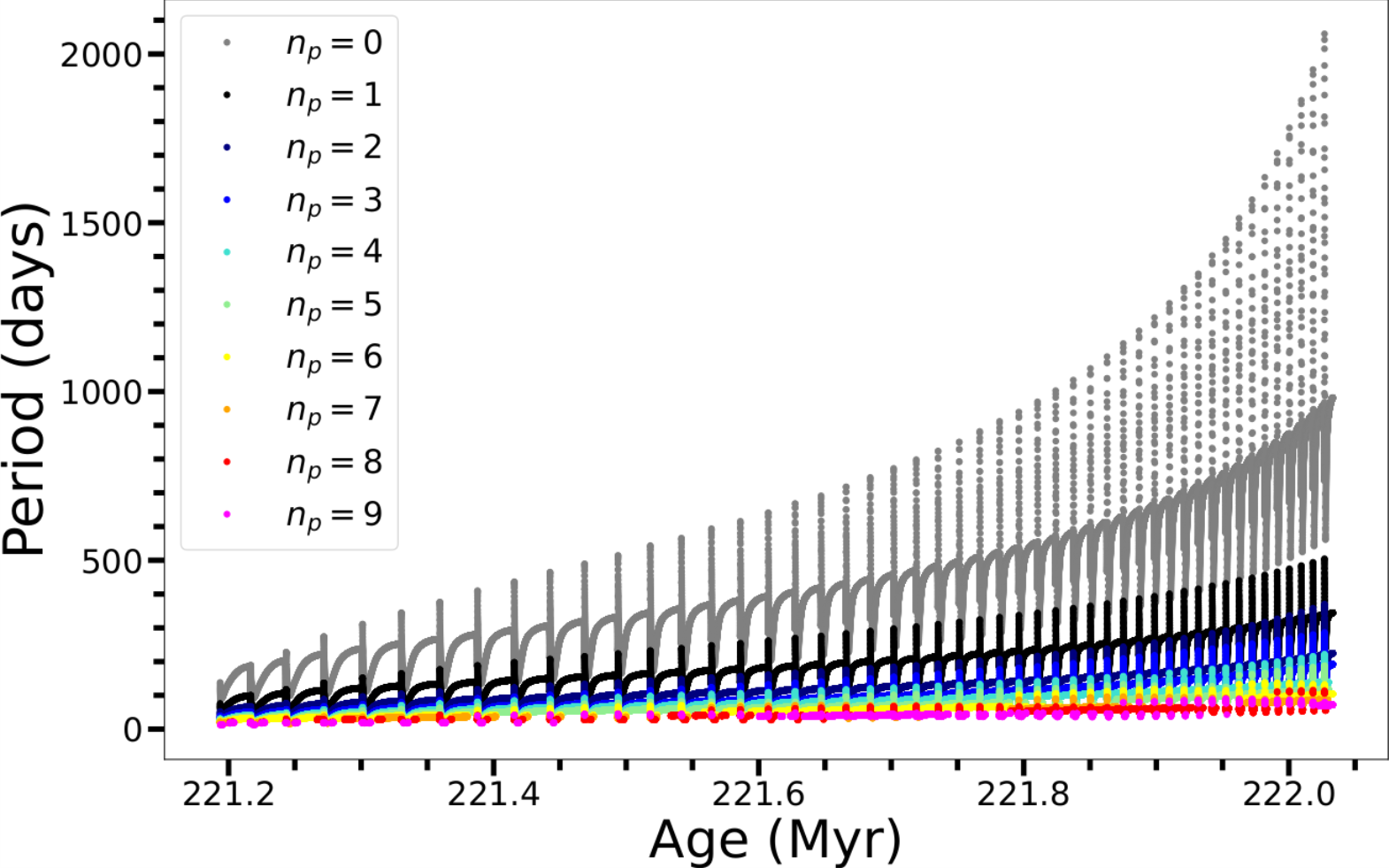}
\caption{
Demonstration of seismic information available in grid: Period in days for each of the first 10 radial ($\ell = 0$) pressure modes, $n_p=0,\ldots,9$, is shown as function of age in Myr for an AGB track of mass $3.7 M_{\odot}$ and $Z=0.0216$.} 
\label{fig:seismic_curves}
\end{figure}
The physical assumptions adopted in the evolutionary 
calculations include the \citet{GS98} opacities and solar mixture, an Eddington grey $T$--$\tau$ relation for the atmospheric boundary conditions \citep{Eddingtonttau}, the \texttt{`Henyey'} mixing length scheme \citep{Henyey1964} with a fixed value of 
$\alpha_\text{MLT}=1.931$ per the calibration to AGB stars performed in \citet{CinquegranaJoyce2022}, the \texttt{`pp\_extras.net'}, \texttt{`co\_burn.net'} and \texttt{`approx21.net'} nuclear reaction network options, the Schwarzschild criterion for convective stability, and the \texttt{`Reimers'} and \texttt{`Blocker'} schemes for cool RGB winds and cool AGB winds, respectively \citep{Reimers1975, Bloecker1995}. Because we use the Schwarzchild criterion, we do not take thermohaline mixing or semi-convection into account. We do not consider rotation. We do not study the evolution of binaries or multiples but reiterate that single-star evolutionary calculations are acceptable for well-separated, non-interacting binaries like R~Hya and its distant companion, Gaia DR3 6195030801634430336.  

Though it is well known that the mass loss treatment in AGB models can significantly impact the number of pulses a star undergoes before exhaustion (e.g., \citealt{Ventura2005b, StancliffeJeffery2007, Marigo2020}), we studied only one value for the mass loss coefficient in the Bl\"ocker wind scheme, $\eta_\text{Bl\"ocker} = 0.01$. \textbf{This value was chosen in part for consistency with the mass loss scheme and efficiency assumed in the Monash stellar evolution code (see Section \ref{sec:Monash}) and to follow the ``default'' conventions assumed in, e.g., the \texttt{MIST} isochrones and stellar tracks \citep{Choi2016}. It is expected that lowering the mass loss efficiency would increase the total number of thermal pulses, and raising it would do the opposite. The anticipated effects of changing the mass loss prescription are less obvious.
To the best of our knowledge, no systematic study of the impact of varying mass loss prescriptions in \texttt{MESA} has been conducted, though the literature would clearly benefit from such a study.}

Likewise, while it was shown in \citet{Cinquegrana2022} that the AESOPUS opacities \citep{MarigoAringer2009} are most appropriate for AGB stars, convergence difficulties and a desire for grid uniformity necessitated a reversion to the (better tested) \citet{Ferguson2005} low-temperature opacities, which were adopted throughout. Likewise to mitigate convergence difficulties, the models do not include convective overshoot. While this omission will produce slightly incorrect core masses, it does not affect our determinations of the best fit to R~Hya, which are based on surface and near-surface parameters. However, we direct the reader to \citet{CinquegranaJoyce2022}, \citet{Cinquegrana2022} and \citet{Cinquegrana2023} for best practices regarding the treatment of opacities, convective boundary mixing, and other physics for AGB models computed with \texttt{MESA}. 

To keep computing time reasonable, the structural and time resolution coefficients, \texttt{mesh\_delta\_coeff} and \texttt{time\_delta\_coeff}, respectively, are not changed from their defaults.\footnote{ Except in the case of a handful of non-adiabatic calculations with GYRE, for which \texttt{mesh\_delta\_coeff} was halved, corresponding to a doubling of \texttt{MESA}'s structural resolution.} The time resolution is not a concern, as the models are mathematically ``well behaved'' during the relevant regions and cubic splines are used for interpolation (see Section \ref{sec:fittingprocedure}). 
However, improving the default structural resolution by a factor of 10, i.e., setting \texttt{mesh\_delta\_coeff} = 0.1, has been shown to produce changes in predicted $p$-modes of up to $0.4\,\mu{\rm Hz}$ on the main sequence (M. Joyce \& Y. Li 2024, in prep). While the extent of this effect in supergiants is still being investigated, we introduce a 60 d buffer in period when we set the requirements for agreement between the observations and models (Table \ref{table:obs_domain}) to take this convergence-related modeling uncertainty into account.

The assumptions adopted for the adiabatic asteroseismic calculations did not deviate significantly from \texttt{GYRE} defaults. One notable difference is use of the \texttt{`MAGNUS\_GL6`} shooting scheme, or sixth-order Gauss--Legendre Magnus method, in the numerical options. This is more reliable for finding modes in evolved stars as compared to the default \texttt{diff\_scheme} option, which is second-order. The complete lists of parameter values for all \texttt{MESA} and \texttt{GYRE} controls are available in those files (i.e., \texttt{inlists}) on Github\footnote{ \url{https://github.com/mjoyceGR/AGB_grid_visualizer}} and Zenodo\footnote{Zenodo link available upon publication}.

An example of a seismic pulse spectrum produced with this configuration is shown in Figure \ref{fig:seismic_curves}. The periods, in units of days, are shown as a function of time during the TP-AGB phase. Each curve represents one of the 10 lowest-frequency (longest period) radial $p$-modes, $n_p = \{0,..,9\}$; $\ell,m=0$. An equivalent data set is available for every model in the grid. 

\subsection{Modeling Procedure}
\label{sec:modelingprocedure}

Because the pulsation period measurements of R Hya constitute our primary observable, it is important to have an exact solution for the star's radial oscillations at every time step in the evolutionary calculations. However, this is not an out-of-the-box capability of \texttt{MESA} itself; rather, \texttt{MESA} provides structural profiles that can be used as initial conditions for seismic calculations. The \texttt{GYRE} program generates a frequency spectrum characterizing a structural model's response to perturbation, which can then be compared directly to asteroseismic observations.  

It is computationally expensive to run an evolutionary model that associates a theoretical frequency spectrum to every time step for the entire duration of the TP-AGB, especially with the temporal resolution required to study thermal pulses in detail (5--10 years in the case of T UMi; \citealt{TUMi}).
This issue was dealt with in \citet{TUMi}, hereafter Paper I, by first manually isolating the regions of interest on evolutionary tracks, then re-running seismic calculations for the selected regions only. Among the many shortcomings of this technique are the lack of automation---making the computation of large grids intractable---and the generation 
intermediate structural data (\texttt{profile}) files that served as initial conditions for separate, corresponding \texttt{GYRE} runs. 

We have improved upon the techniques of Paper I in three ways:
\begin{enumerate}
\item We ran \texttt{GYRE} as part of the evolutionary calculations themselves rather than as a separate, post-processing step, thus decreasing the number of data output and writing operations (time\footnote{Due its greater parallelization, \texttt{GYRE} benefits more from multiple threads than \texttt{MESA} does. While low-order $p$-mode calculations are not particularly expensive, it is useful to allocate more threads to simulations calling \texttt{GYRE} on-the-fly. A model running on two threads that calls \texttt{GYRE} at every time step will run more slowly than the same model, running on the same number of threads, that does not.} and data volume reduction);   
\item We relied on more sophisticated fitting and statistical techniques in lieu of greater time resolution (significant run time reduction); and
\item We developed several interactive data visualization tools that led to improvement in the accuracy and scope of the results.\footnote{in Python, publicly available at the first author's Github: \url{https://github.com/mjoyceGR/AGB_grid_visualizer} }
\end{enumerate}
Whereas Paper I treated \texttt{MESA} and \texttt{GYRE} calculations as independent, we used a technique called \textit{ \texttt{GYRE} on-the-fly} for this analysis. This method was adapted from a 2022 \texttt{MESA} Summer School group laboratory exercise authored by Earl Bellinger, available freely on Zenodo \citep{earl_patrick_bellinger_2022_7118662}\footnote{Exercise instructions available here: \url{https://earlbellinger.com/mesa-summer-school-2022/}} Some of this functionality is also present in select \texttt{MESA test\_suite} cases, available with the code, though documentation is more limited in those cases.
 
Calculations are broken into three evolutionary stages: 
ZAMS (zero-age main sequence) to TACHeB (terminal-age core helium burning), or phase 1, 
TACHeB to the onset of the TP-AGB phase (phase 2), and 
TP-AGB through termination (phase 3). 
Termination is determined by one of the following:
\begin{itemize}
    \item[-] reaching the end of core helium exhaustion and ascent onto the WDCS (White Dwarf Cooling Sequence), required by the termination string \texttt{`phase\_WDCS'};  
\item[-] the condition  \texttt{min\_timestep\_limit}, which is an indication of convergence failure; 
\item[-] the condition \texttt{max\_number\_retries} $= 500$, which is an indication of convergence difficulty; or
\item[-] run time exceeding 72 hours on 16 cores ($Y_i$-fixed grid) or 120 hours on 16 cores ($Y_i$-varied grid). 
\end{itemize}
The latter two conditions are practical (not physical) choices that balance the goal of fitting R Hya specifically---for which the behavior of the TP-AGB models beyond R Hya's classical constraints does not matter---with the production of a useful grid. The limit of 3 days' run time is the main source of truncation for the fixed helium gird. The limit on \texttt{max\_number\_retries} is the main source of truncation for the helium-varied grid.  
Model summary files are generated at the conclusions of phases 1 and 2 as well as at the conclusion of phase 3 if the model is not terminated early. These are given names of the form \texttt{TACHeB.mod} (end of phase 1), \texttt{AGB\_seed.mod} (end of phase 2), and \texttt{AGB\_terminal.mod} (end of phase 3), respectively. Dividing the evolution into three parts allows us to relegate the asteroseismic calculations and use of velocity drag to phase 3 only. The \texttt{AGB\_seed.mod} models, also made available with this grid, can also be used to start new calculations from the onset of the TP-AGB. \texttt{GYRE} is called once at each time step in phase 3. 

As noted at the top of Section \ref{sec:modelgrid}, we perform a sweep in two model input parameters only: initial mass and initial $Z$, with two different treatments of helium. We set $\alpha_\text{MLT}= 1.931$ for all tracks.
While we explored several values for the mass loss efficiency parameter $\eta$ initially, we did not add this as a dimension to the current grid (though it can be introduced later). The models we provide here cost just over 4.5 million CPU hours.
    
\section{Mode Hypotheses}
\label{sec:twomodes}
\begin{figure}
\centering
\includegraphics[width=\columnwidth]{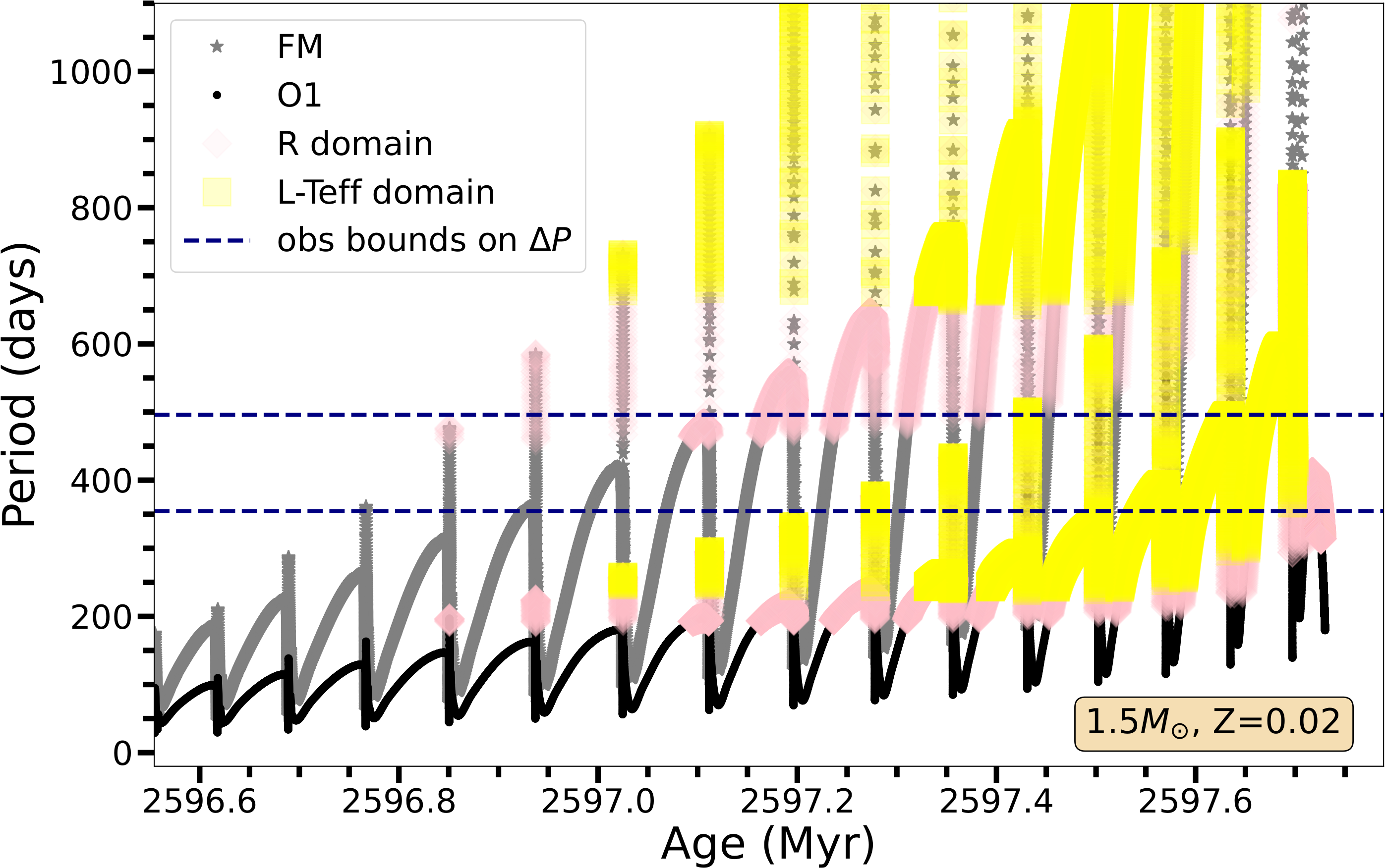}
\caption{The period evolution of the FM (grey) and O1 (black) modes (units of days) from the same model ($M=1.5 M_\odot, Z=0.02$) are shown. The boundaries on the observed change in period are shown with dashed blue horizontal lines.
Regions where the theoretical periods intersect R~Hya's observational radial constraints are highlighted in pink. Regions where the periods intersect with R~Hya's luminosity and effective temperature constraints are highlighted in yellow. This figure demonstrates the degeneracy between the FM and O1: there are regions along both curves that intersect some, or all, of the observational constraints at multiple different times. }
\label{fig:twomodes}
\end{figure}

In principle, we cannot know the pulsation mode(s) of a star simply by observing its brightness variations---a fact keenly demonstrated by the current controversy surrounding Betelgeuse (\citealt{Betelboi, MacLeod2023, BetelNote2023}; in contrast, \citealt{Saio2023}). However, we can narrow the possibilities for an observed mode based on other features of the star or known behaviors of its class.

First, we know that the mode amplitudes in R Hya---or indeed in any Mira---are too large to be dipole or higher degree modes, so these are not computed or considered. As \citet{Yu-2020} has shown, although non-radial modes can exist in some luminous red giants (just below the Mira stars in brightness and period), they will appear in triplet peaks. We do not observe such peaks in R~Hya, and the plateau segment in R~Hya's light curve is long enough to provide the required frequency resolution to detect them if they existed.

Mira stars predominantly pulsate in the lowest radial modes, but mode identification can be difficult. Earlier, \citet{Haniff-1995} and \citet{Feast-1996} both claimed that most Miras pulsate in the first overtone, based on their theoretical period--radius relations. However, they assumed that Mira masses do not significantly exceed 1.0 to $1.5\,M_\odot$. This notion was challenged soon after \citep[see, e.g.,][]{Barthes1998}.

We now know that stars with the longest pulsation periods are limited to pulsating in the first overtone or the fundamental mode, as was shown for long-period variables in the LMC  by \citet{Soszynski2013}. Model calculations by \citet{Trabucchi-2017,Trabucchi2021} suggest that the longest periods observed in Miras correspond to the fundamental mode specifically. While first overtone stars almost never exceed periods longer than $\sim$300~d, the range of the observed periods in R~Hya, when adopting the most generous observational and modeling uncertainties, does approach this regime. Further, since the star is undergoing a TP, the overtone may stay excited during peak expansion. We have shown that TP-AGB stars may switch modes during a TP in \citet{TUMi}, but we still know very little about the evolution of mode excitation over a whole TP. 

Still, we can safely restrict our considerations to two hypotheses regarding R~Hya's observed pulsation mode: the fundamental or first overtone. We remain agnostic between these in our assumptions, which enables the consideration of a much larger parameter space of possible fits to R Hya. 
The viability of either hypothesis, and hence one source of difficulty in determining the best fit, is demonstrated in Figure \ref{fig:twomodes}. The Figure shows the fundamental mode (FM; grey curve with larger amplitudes) and first overtone (O1; black curve with smaller amplitudes) periods as a function of time for an example model. Regions of intersection with the classical observational constraints and maximum/minimum of the period measurements are indicated on the Figure. As there are clearly several regions along both curves where the theoretical predictions intersect a subset, or all, of the observational constraints, there is a degeneracy between FM and O1 solutions.

Figure \ref{fig:twomodes} also indicates that many different regions of the TP(s) intersect the observational parameter space. However, we show in the next section that most parts of the TP can be ruled out by looking at the shape of the observed period decline more closely.

\section{The Shape of a Thermal Pulse}
\label{sec:shape_of_pulse}
\begin{figure}
\centering
\includegraphics[width=\columnwidth]{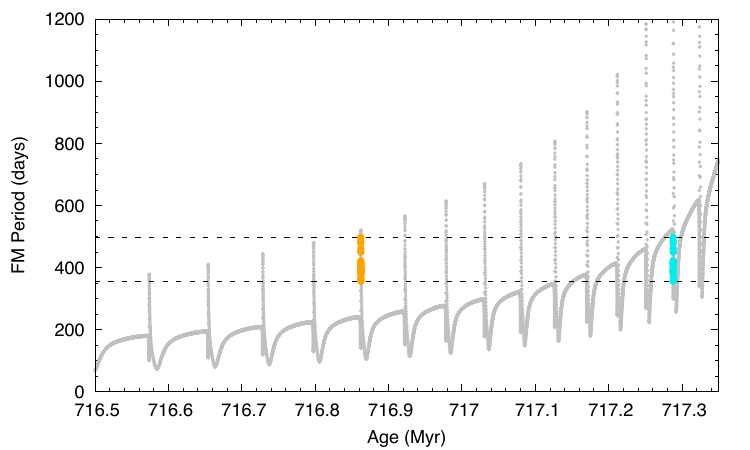}
\includegraphics[width=\columnwidth]{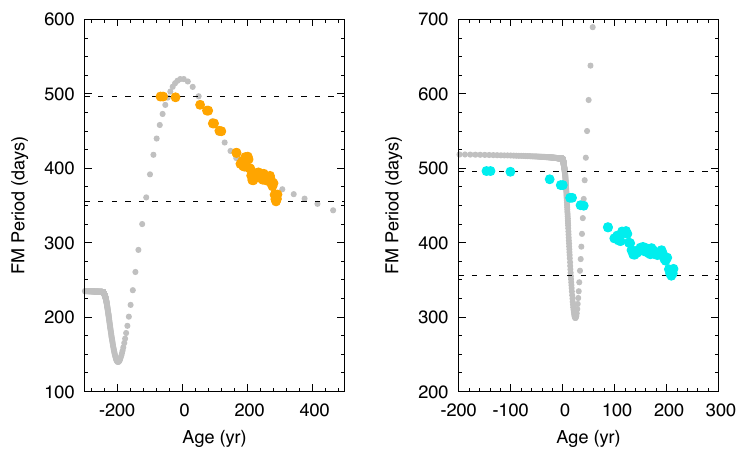}
\caption{\textbf{UPPER:} Two different hypotheses are shown regarding the segment of a thermal pulse that the period observations of R~Hya trace. The orange and blue highlighted regions are the period observations, shifted in age to align with the power-down (left; orange) and onset, or knee (right; blue), of different TPs from the same model ($1.9M_\odot, Z=0.0018$). \textbf{LOWER:} These same regions, zoomed in (ages in relative units). In the left panel, we find the power-down phase to be a plausible fit to R~Hya's period measurements for this mass and metallicity. On the right, we see that the slope of the onset (knee) region is much too steep to be a good fit to the measurements, and this is true for all masses, metallicities, and mode assumptions. }
\label{fig:slope_hypothesis}
\end{figure}
By far, the most constraining observational feature of R Hya is its 400-year period evolution. Comparing R~Hya's observed period vs time relation to preliminary seismic evolutionary models, we find that the magnitude of the period decline over such a short (by stellar evolutionary standards) timescale can only be consistent with the most dynamic phases of a thermal pulse. 
Figure \ref{fig:slope_hypothesis} shows the fundamental mode period as a function of age for a $1.9M_{\odot}$, $Z=0.0018$ model. In the top panel, we see in orange and cyan two candidates for the feature the observed period measurements could be tracing: in orange (left), the data trace a theoretical pulse's descent from maximum radius (maximum period), sometimes referred to as the ``power down'' phase; in cyan (right), the data trace the decline from the onset of the TP (referred to as the ``knee'' feature in \citealt{TUMi}). 
However, when we look more closely at these two TP sub-regions in the lower panels of Figure \ref{fig:slope_hypothesis}, it becomes clear that the knee (right) has far too steep a slope to fit the measured period decline. The power-down region, however, provides a quite good fit to R~Hya's periods for this FM model.

When we compute consistency metrics formally (see Section \ref{sec:fittingprocedure}), we find that TP-onset solutions fail to be good fits to the observations regardless of the mass and metallicity assumed in the model: the slope in this region is always significantly steeper than observed $\dot{P}$. Hence, the data most likely trace a power-down region, with the goodness of fit to the slope varying based on the global properties of the model (mass, $Z$) as well as the pulsation mode assumed.

\section{Fitting and Statistics}
\label{sec:fittingprocedure}
Several issues complicate the determination of the goodness-of-fit of theoretical TP-AGB models to R~Hya: 
(1) the heterogeneous uncertainties on the individual period measurements and non-uniform spacing of the data---the last 160 years of period measurements are significantly more precise and densely sampled than the earlier measurements; 
(2) the large difference in precision between the period measurements and the measurements of temperature, luminosity, and radius; 
(3) the fact that there are, for example, only a handful of brightness (temperature, radius) measurements taken over the duration of the period measurements, meaning there is no one-to-one correspondence between the measurements of all parameters; and
(4) the fact that the meandering component of the period evolution cannot be reliably disentangled from the overall period decline, meaning the inflection of the most densely observed region is not well constrained (as discussed in Section \ref{subsec:meadering}).  

We considered several different methods for determining the best-fitting model(s) and computed multiple agreement statistics. To mitigate the first two issues, we weighted individual period measurements by their uncertainties and made choices that de-emphasize agreement with luminosity, effective temperature, and radius (hereafter $L$, $T_\eff$, $R$) measurements relative to agreement with period measurements when all constraints are used.  
To deal with the third issue, we computed one statistic that treats the adopted values and uncertainties in measurements of $L$, $T_\eff$, $R$ as though they apply to each of the period measurements individually. 

All of the fit statistics are based on point-wise distance metrics, so models that minimize the difference between observed and predicted values for some choice of period alone or $\{P, L, T_\eff, R\}$ are preferred. The fourth issue is pernicious in this framework, as any such distance metric naively computed will end up biased towards the portion of the period curve that is least well constrained and known to reflect physics that our models do not capture. To counteract this, we introduce a constraint that enforces slope steepness, described in more detail in step 2 of the next section.
\begin{table}\centering
  \begin{tabular}{|c|c|}
    \hline
    \textbf{Parameter} &\textbf{Value}  \\ \hline
   $\log L_{\odot}$ &   $4.0 \pm 0.4 $ \\ $T_{\text{eff}}$ &   $2500\pm1000$ K \\ 
   $R_{\odot}$ &  $425^{+290}_{-215} $ \\    
   $P$ &  uniform $\pm 60$ d uncertainty \\ 
   $P_\text{max}$  & within $\pm60$ d of measured $P_\text{max}$ \\    \hline
  \end{tabular}
  \caption{Observational constraints and uncertainties used to define the domain for the modeling and fitting procedure. Values adopted are based on the approximate medians of values quoted in Table \ref{table:obs_constraints}. }
  \label{table:obs_domain}
\end{table}

\subsection{Fitting Procedure}
\label{sec:data_processing}
The calculations are performed as follows:
\begin{enumerate}
    \item We compare the TP-AGB (phase 3) component of each model to observational boundaries and isolate the regions where agreement is found. Agreement is strictly enforced, meaning a track must intersect the values and uncertainties prescribed in order to be considered in the next stages of analysis.  Given the high variance among values quoted for the classical constraints in the literature, we use a relatively loose interpretation of the constraints listed in Table \ref{table:obs_constraints}.
    The observational boundaries enforced in the analysis pipeline are summarized in Table \ref{table:obs_domain}.

    Deciding which models are candidates based on the theoretical periods requires making an assumption about the mode we are actually observing. We must therefore compare each model against the observational boundaries twice: once using the FM periods and once using O1 periods (for reasons discussed in Section \ref{sec:twomodes}, these are the only two possibilities.).  

    \item Once the observationally consistent region of a TP-AGB pulse spectrum is identified, we use a signal processing routine from \texttt{scipy.signal} \citep{SciPy2020} to isolate the thermal pulses in this domain. Because thermal pulses contain sharp local maxima, this can be automated by specifying prominence and time step spacing criteria, but no one combination of these parameters is reliable for $\sim$3000 models, each of which contains anywhere from $\sim5$ to $\ge50$ pulses.

    Peaks that do not correspond to the period (luminosity) maximum within the power-down phase are sometimes identified, but most of these are discarded on the basis of either morphology or inconsistent physics: for example, if the hydrogen-to-helium burning ratio is not consistent with undergoing an active thermal pulse, set by $L_\text{H}/L_\text{He} < 10$. 

    In some cases, a secondary feature known as the helium sub-flash is identified, as demonstrated in Figure \ref{fig:subflash}. As we have no physical reason to exclude sub-flash detections, they are are permitted. However, in the majority of cases the sub-flashes are excluded \textit{de facto} due to their shallower slopes, enforced as follows.
\begin{figure}
    \centering
    \includegraphics[width=\columnwidth]{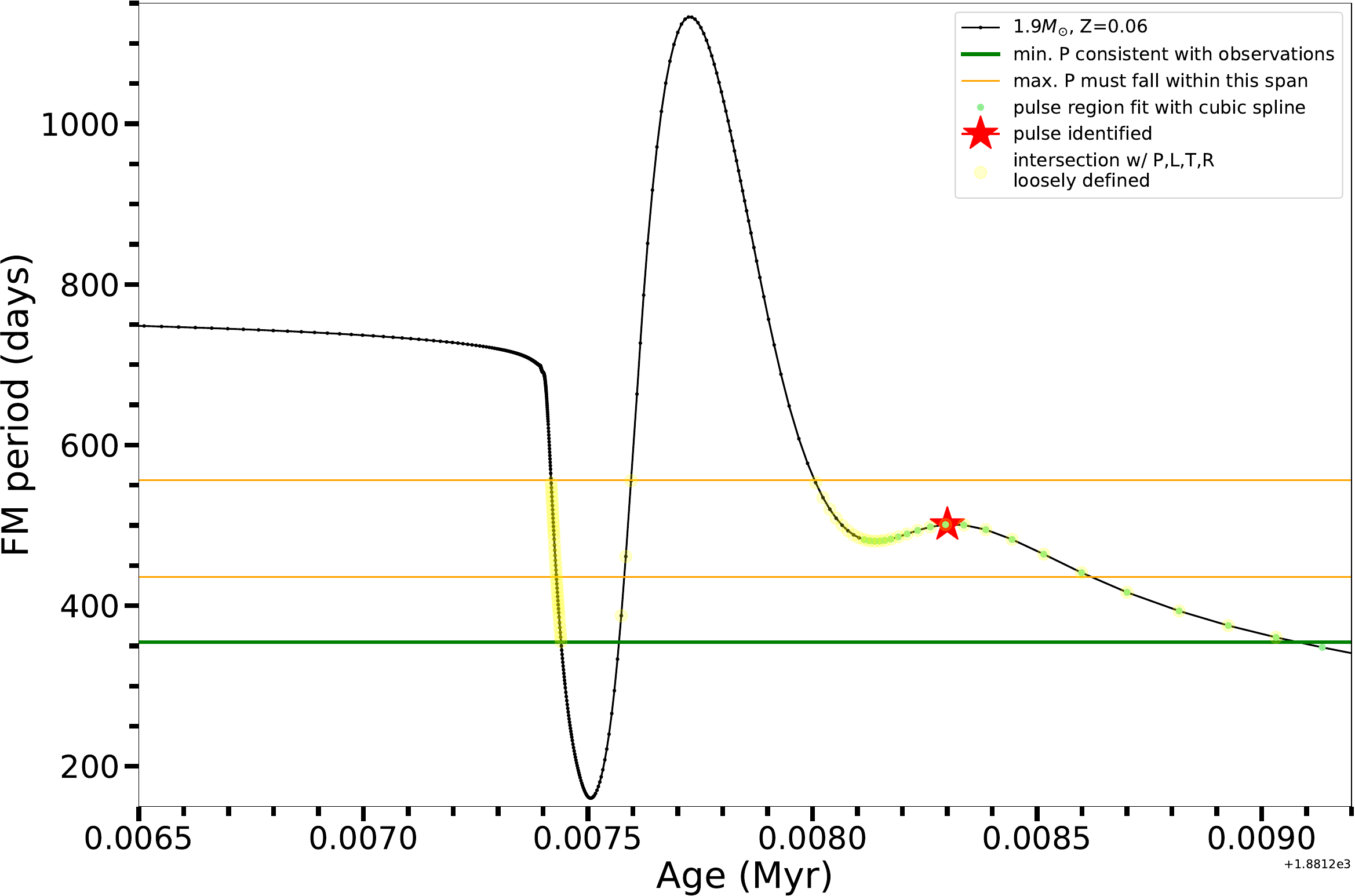}
    \caption{A pulse for which the helium sub-flash is identified as a plausible fit, indicated by the red star on top of the right-most local maximum. Yellow highlighting indicates regions of this curve that are consistent with the period, luminosity, effective temperature, and radius constraints defined in Table \ref{table:obs_domain}. We note that the power-down portion of the pulse (highest maximum, as in the lower-left panel of Figure \ref{fig:slope_hypothesis}) is not a viable fit here given that it begins well outside the boundaries permitted for $P_\text{max}$. 
    }
    \label{fig:subflash}
    \end{figure}

    Because the more recent period observations have much denser sampling and much smaller uncertainties than the earliest measurements, these will inappropriately bias the pulse fit towards the last $\sim$third of the data unless counteracted manually. We use a ``slope hardness'' parameter, $s$, that sets a consistency threshold requiring 
   \begin{equation}
       \dot{P} \approx \Delta P / \Delta t \ge s / \Delta t
    \label{eq:hardness}
   \end{equation} 
   to avoid fitting pulses that are too shallow overall.
   We perform several tests and find that a setting of $s=75$ does the best job of identifying good fits and excluding poor ones. 
   We caution, however, that the value chosen for $s$ can have a significant impact on the solution space. We computed our results for three settings of this parameter: $s=50$, $s=75$, and $s=100$. We focus on solution maps using $s=75$ henceforth, but results for all values are provided in Appendix \ref{appendix:param_var}.
   Overall, this method identifies the correct regions of interest about 95\% of the time.\footnote{Not perfectly--- instances of incorrect pulse identification and indexing are documented in the data visualizer. These do not affect our results.}

    \item For each pulse identified, the observational data are adjusted to center on the pulse, as shown in Figure \ref{fig:pulse_alignment}.
    A data file that replaces the relative ages of the period observations with absolute ages from the models is generated and given a name beginning with \texttt{auto\_age\_fit}. Any given TP-AGB spectrum may host several acceptable pulse solutions (depending on how strictly observational agreement is enforced---see Appendix \ref{appendix:param_var}) but typically there are between zero and $\sim5$ \texttt{auto\_age\_fit}s per model.

    As the age of R~Hya is not well constrained (see Section \ref{sec:age}), and realistic precisions on stellar age determinations are several orders of magnitude larger than the entire duration of a thermal pulse spectrum anyway (e.g. \citealt{Tayar2022, Joyce2023}), we treat absolute age as arbitrary but preserve the relative timing of the period measurements in our fits.

    \item Each period--age relation generated in step 3 is evaluated against its corresponding pulse on the \texttt{MESA} evolutionary track with the same mass and metallicity. In the immediate area of the pulse, the theoretical model is fit with four cubic splines: one curve relating each of period (FM or O1, depending on hypothesis), $T_\text{eff}$, luminosity, and radius to age. 

    The cubic splines are sampled to generate theoretical $L$, $T_\eff$, $R$ and period values corresponding to each measurement. These are compared point-by-point to the observations using a weighted root-mean-square error (WRMSE) framework:
    \begin{equation}
    \text{WRMSE} = \sqrt{\frac{\sum_{i=1}^{n} w_i \cdot (y_i - \hat{y}_i)^2}{\sum_{i=1}^{n} w_i}}
    \label{eq:WRMSE}
    \end{equation}
    where:
    \begin{align*}
    &n \quad \text{is the number of observations or data points,} \\
    &y_i \quad \text{is the observed value for the }i\text{th data point,} \\
    &\hat{y}_i \quad \text{is the predicted value for the }i\text{th data point,} \\
    &w_i \quad \text{is the weight assigned to the }i\text{th data point.}
    \end{align*}
The weightings $w_i$ are given by 
    \begin{equation*}
        w_i = \frac{1}{\sigma_{y,i}^2 } 
    \end{equation*}
    where $\sigma_{y,i}$ is the uncertainty associated to observation $y_i$.  
    
    When goodness-of-fit is determined according to the 210\,d period measurements alone, we call the statistic $P_\text{WRMSE}$, or $P_\text{W}$.

    \item In the case where we wish to factor agreement with $L$, $T_\eff$, and $R$ into our consistency metric along with $P$, we compute Equation \ref{eq:WRMSE} for all quantities and combine the four WRMSE values two different ways. The first is a harmonic mean:
\begin{equation}
    H_\text{WRMSE} = \frac{4}{\frac{1}{x_1} + \frac{1}{x_2} + \frac{1}{x_3} + \frac{1}{x_4}} 
    \label{eq:harmonic_mean}
    \end{equation}
    where $x_1,\ldots,x_4$ are the WRMSE values from \eqref{eq:WRMSE} for $P$, $L$, $T_\eff$ and $R$. We abbreviate this as $H_\text{W}$.

    The second is a pseudo-$\chi^2$ score normalized such that the combination of all classical observational components is weighted equally with a seismic (period-only) score:
\begin{equation}
    S_\text{WRMSE} = \sqrt{ \frac{1}{3}(L_\text{W}^2 + T_\text{W}^2 + R_\text{W}^2) + P_\text{W}^2 }
    \label{eq:pseudo_chisq}
    \end{equation}
    where $P_\text{W}$, $L_\text{W}$, $T_\text{W}$ and $R_\text{W}$ are the values from Equation \ref{eq:WRMSE}, as above. We abbreviate this as $S_\text{W}$.
\end{enumerate}

\begin{figure}
    \centering
    \includegraphics[width=\columnwidth]{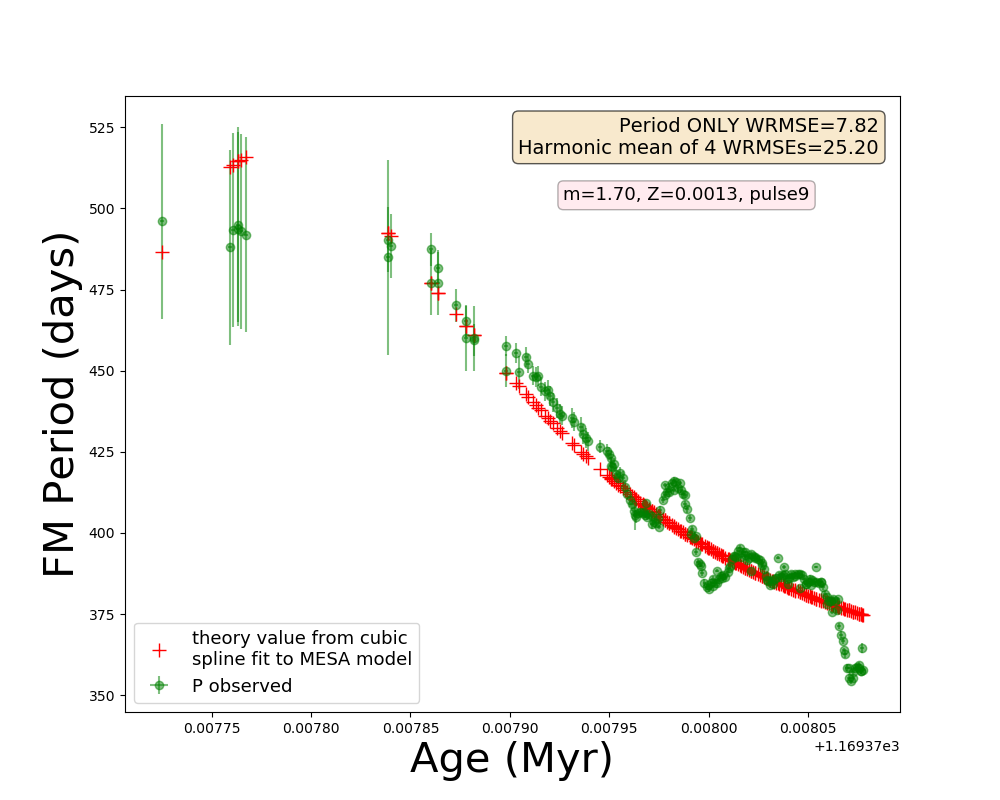}
    \caption{The period measurements (green circles with error bars) are overlaid on predictions from a theoretical pulse (red crosses) computed with \texttt{MESA} (FM, $1.7 M_\odot, Z=0.0013$) and interpolated with a cubic spline. The observations are aligned with theory as described in Step 3 of \ref{sec:modelingprocedure}. Values for two of the five statistics discussed on in Section \ref{sec:fittingprocedure} are shown in the upper right-hand corner.
    }
    \label{fig:pulse_alignment}
\end{figure}

\section{Results}
\label{sec:results}
\begin{figure*}
\centering
\includegraphics[width=\columnwidth]{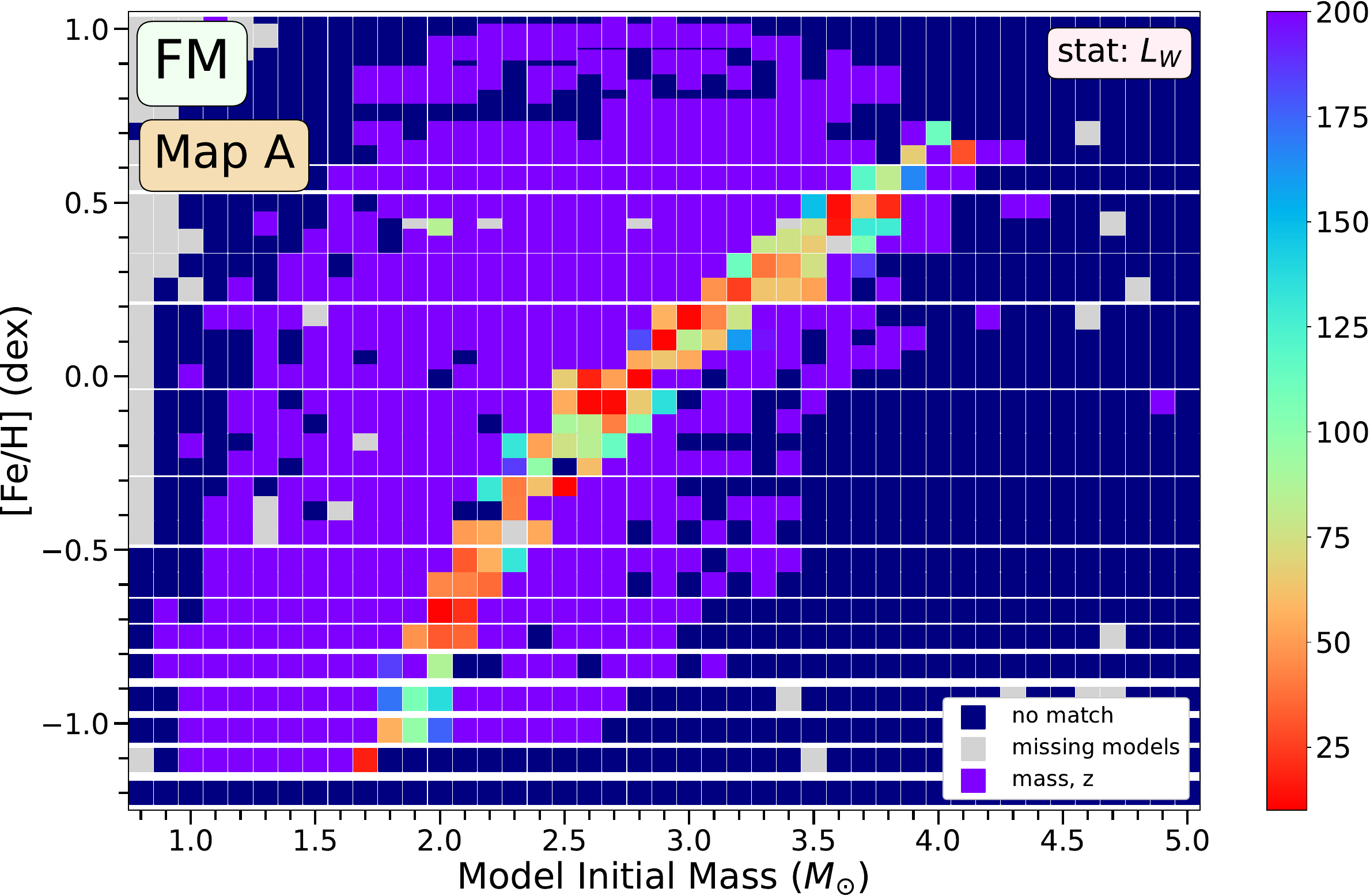}
\includegraphics[width=\columnwidth]{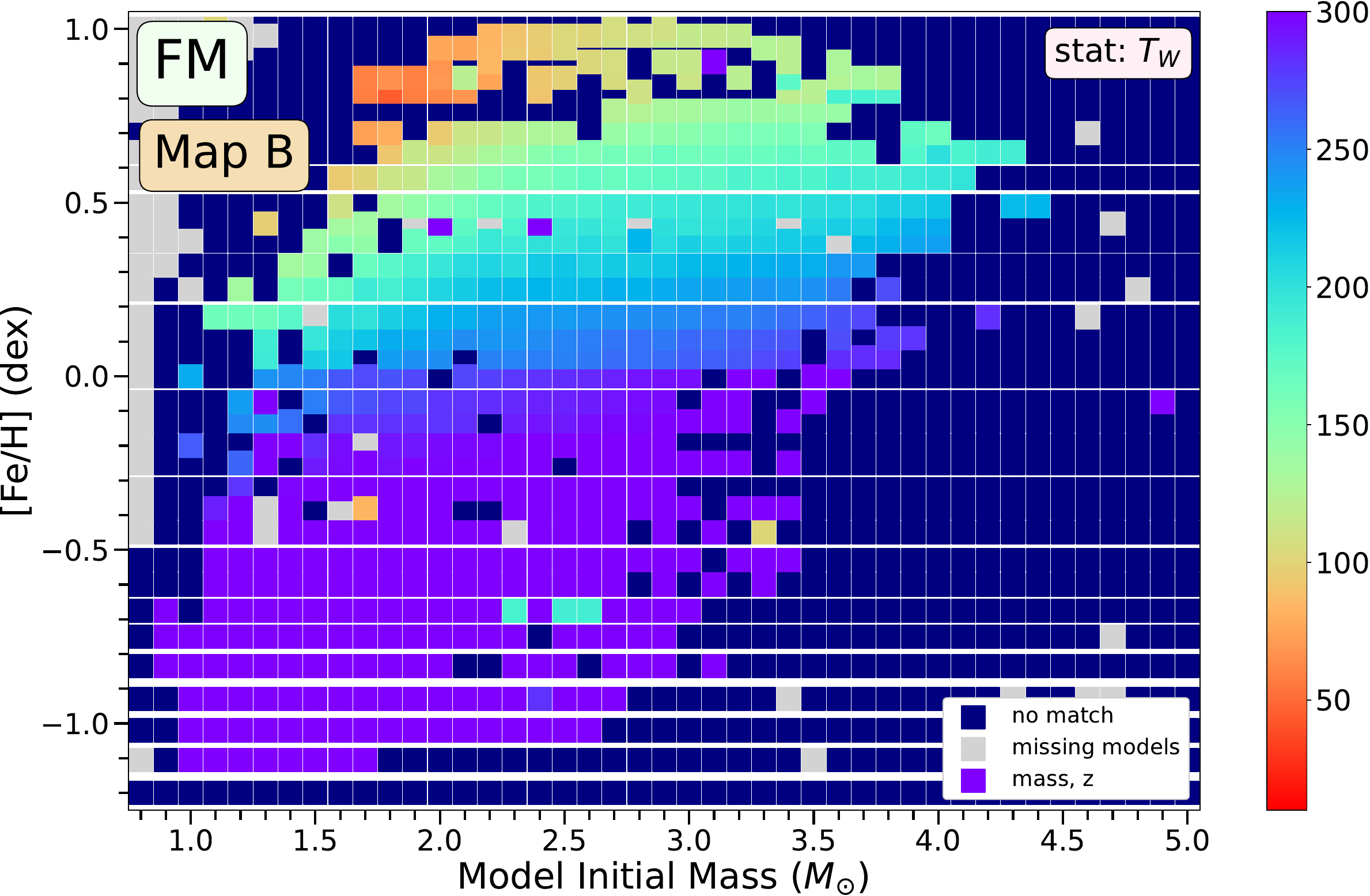}
\includegraphics[width=\columnwidth]{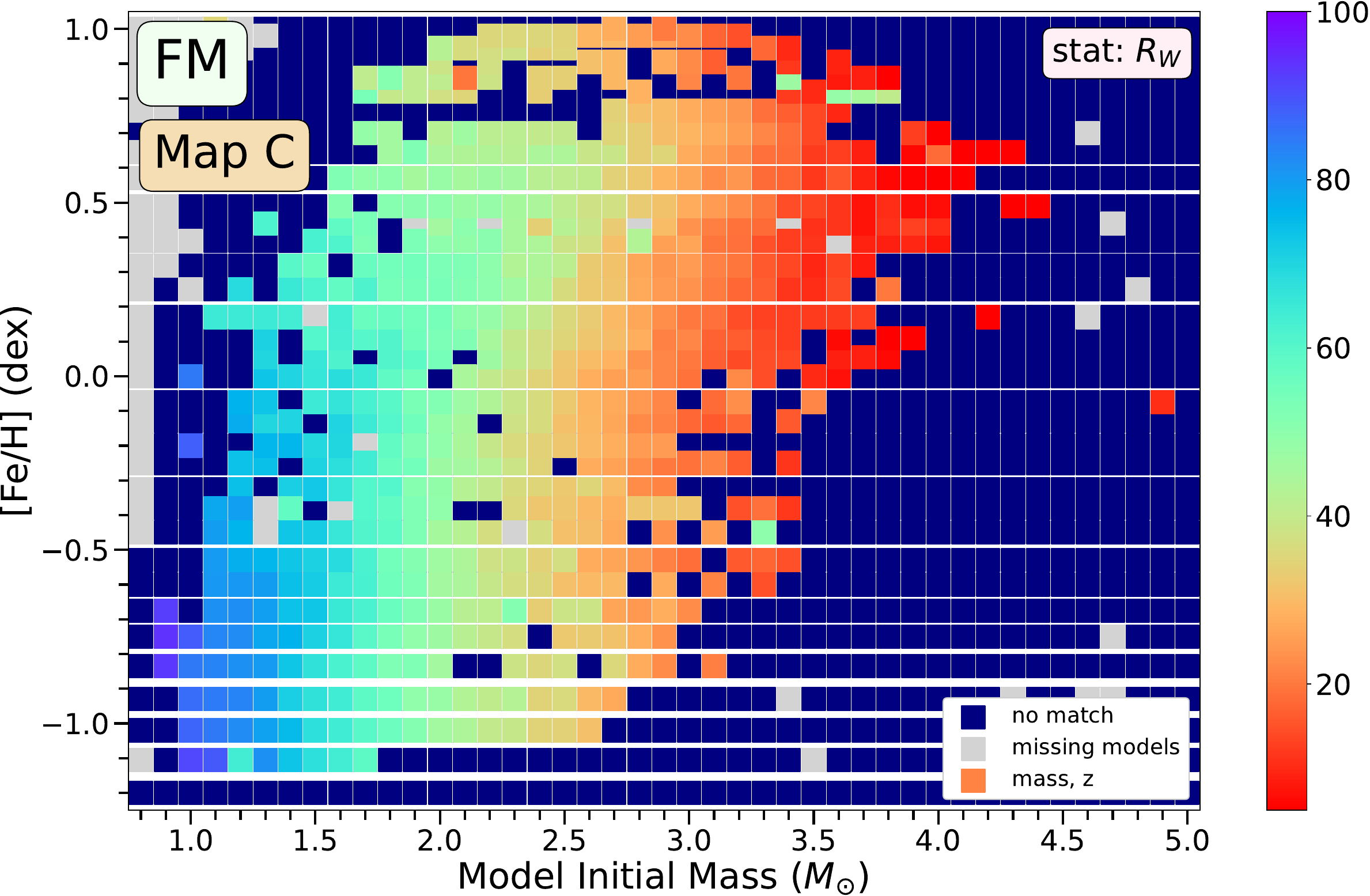}
\includegraphics[width=\columnwidth]{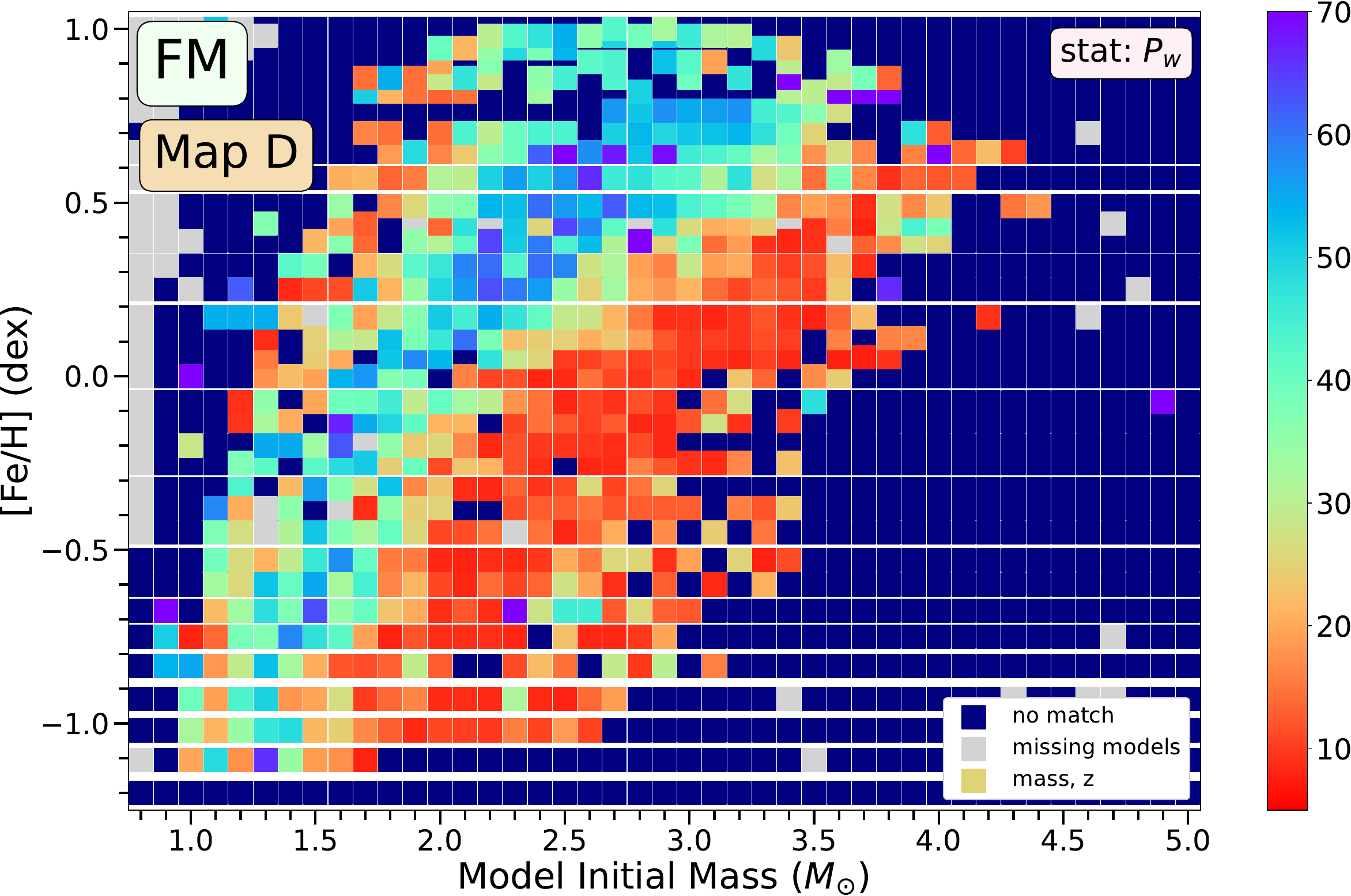}
\includegraphics[width=\columnwidth]{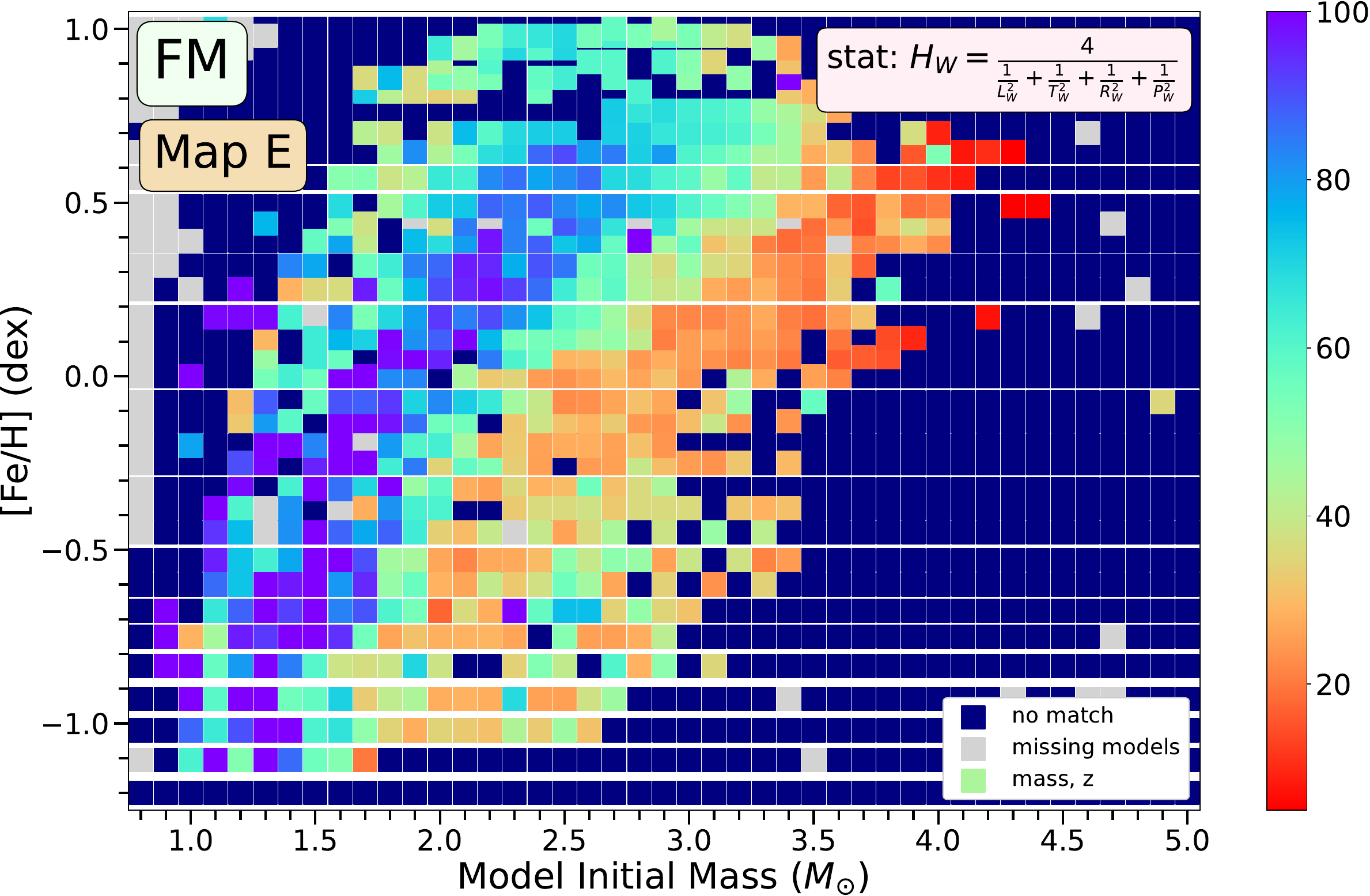}
\includegraphics[width=\columnwidth]{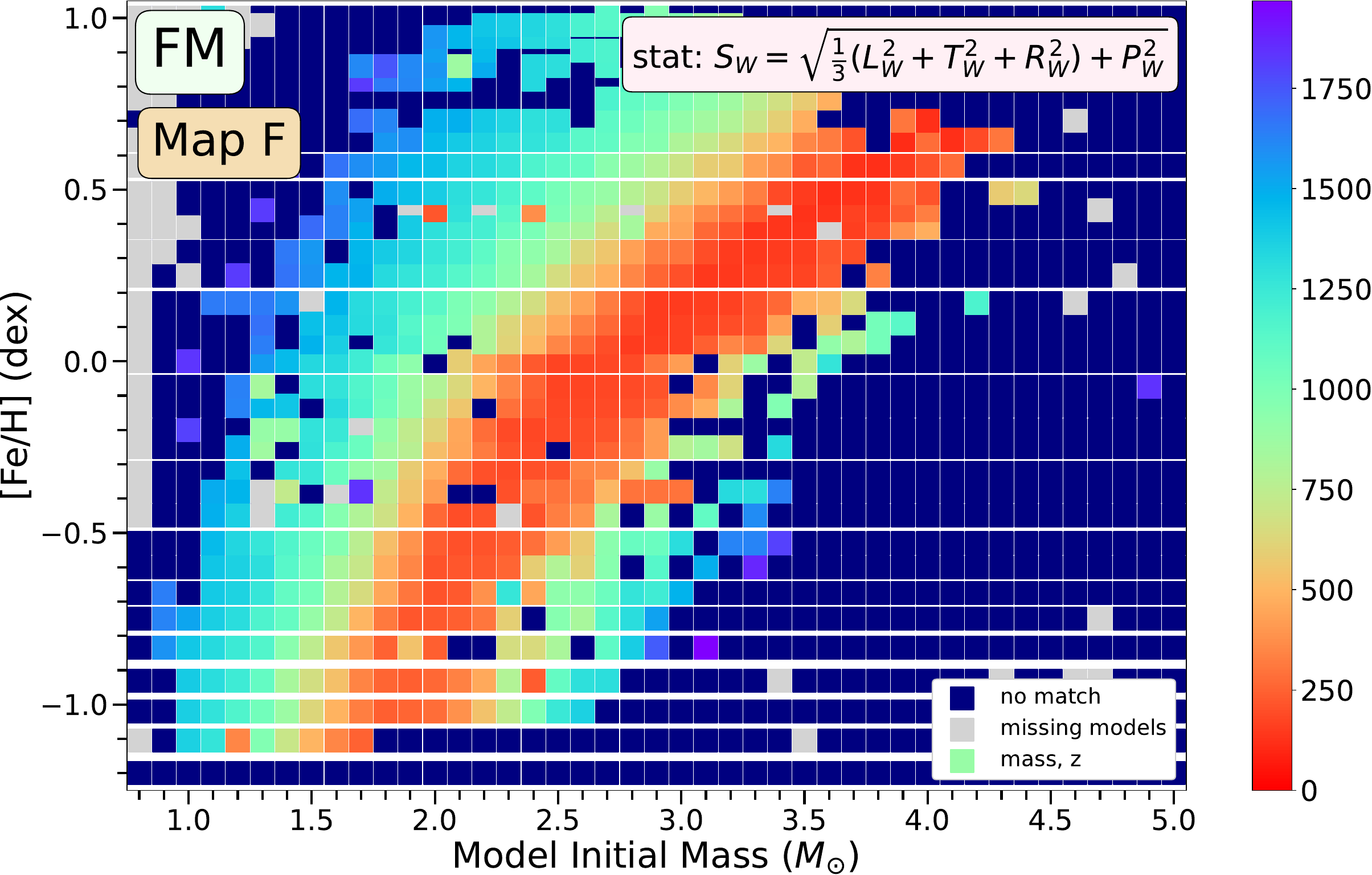}
\caption{Six heat maps showing [Fe/H] vs initial mass are presented for the grid with fixed $Y_\text{init}$ (see Appendix \ref{appendix:param_var} for the equivalent visualization using the helium-scaled grid).
Better fits are coded in redder (warmer) colors, corresponding to a different goodness-of-fit metric in each panel. Map A highlights the FM models that best agree with the luminosity constraints only, and likewise for Map B ($T_\text{eff}$ only) and Map C (radius only). Map D shows the best-fitting models according to agreement with the 210 measured periods.
Maps E and F are statistical composites of the previous four maps, computed two different ways. Map E uses a harmonic mean of the four observational components, and Map F uses a quadrature sum of normalized classical and seismic agreement metrics, as shown in the legends in the upper left-hand corners. Color bar normalization changes from panel to panel, as indicated by the largest value on each panel's color bar, but the direction is constant (better fits in warmer colors). Models that ran successfully but failed to intersect the observational boundaries are shown in navy blue. Models that did not run successfully are shown as grey squares (about $5\%$ of models for the grid with fixed $Y_\text{init} = 0.3$, shown here.)
Boundaries are set according to agreement with the loose observational constraints described in Table \ref{table:obs_domain}.
}
\label{fig:heatmap_FM}
\end{figure*}

\begin{figure*}
\centering
\includegraphics[width=\columnwidth]{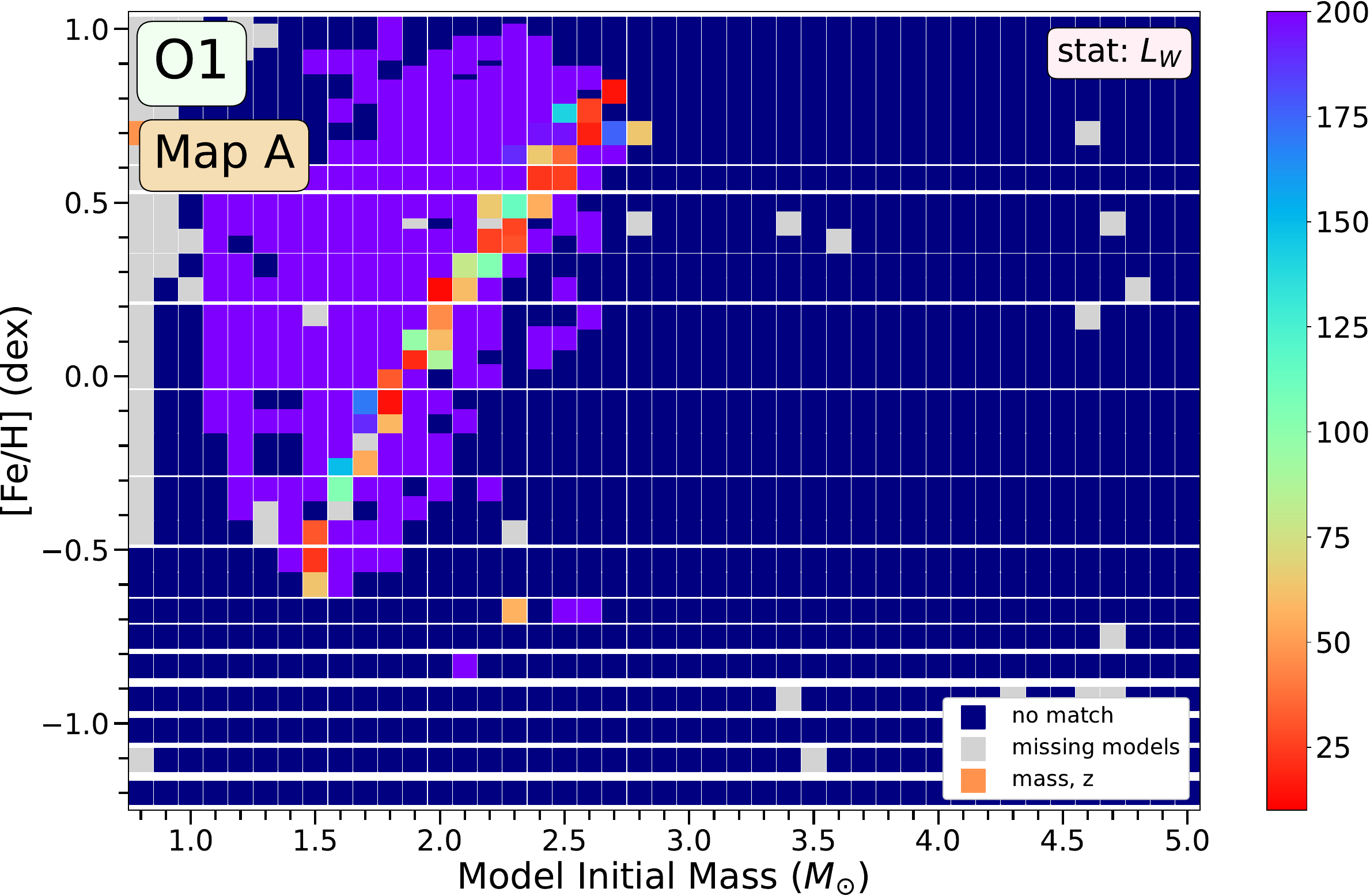}
\includegraphics[width=\columnwidth]{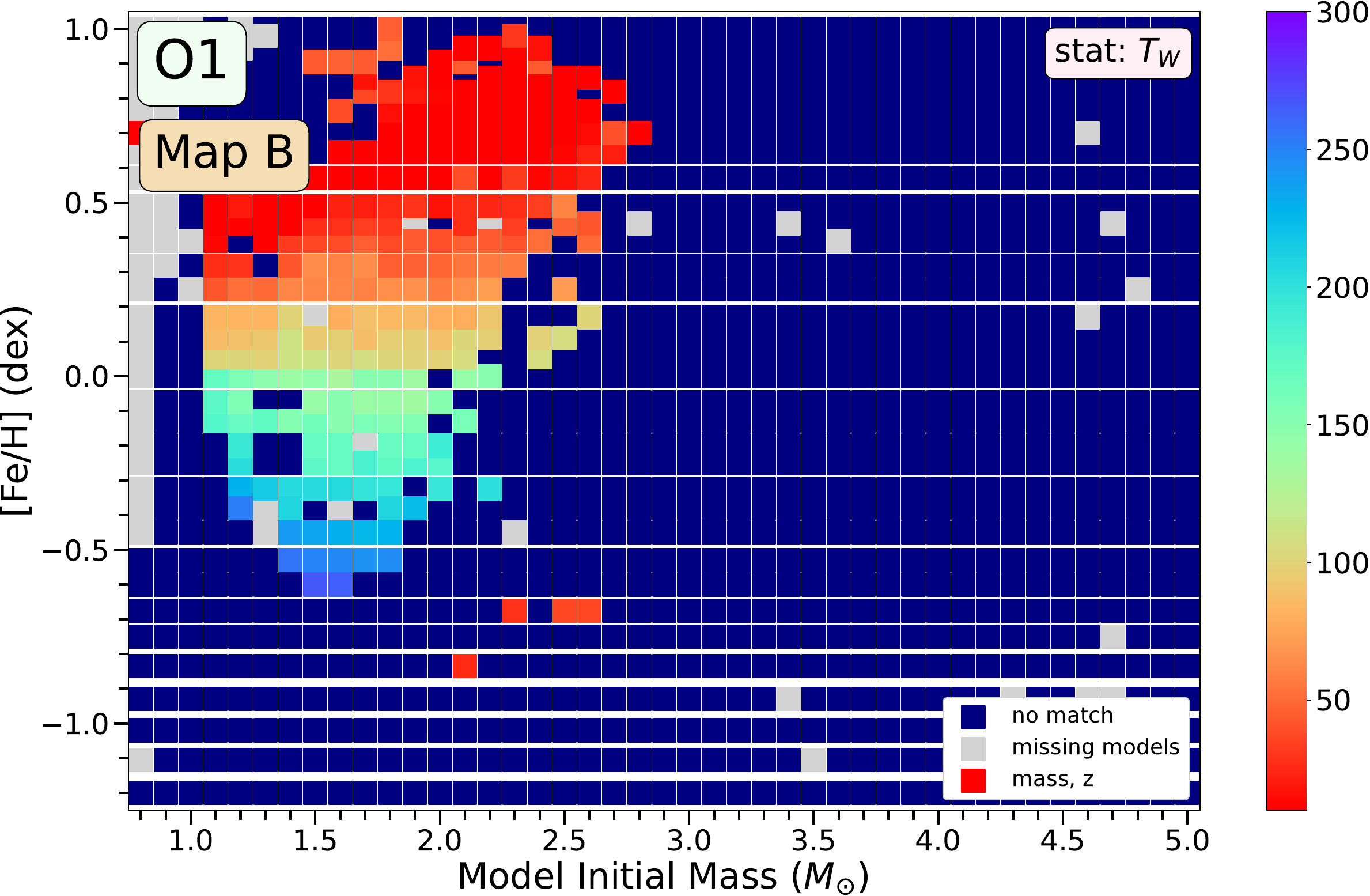}
\includegraphics[width=\columnwidth]{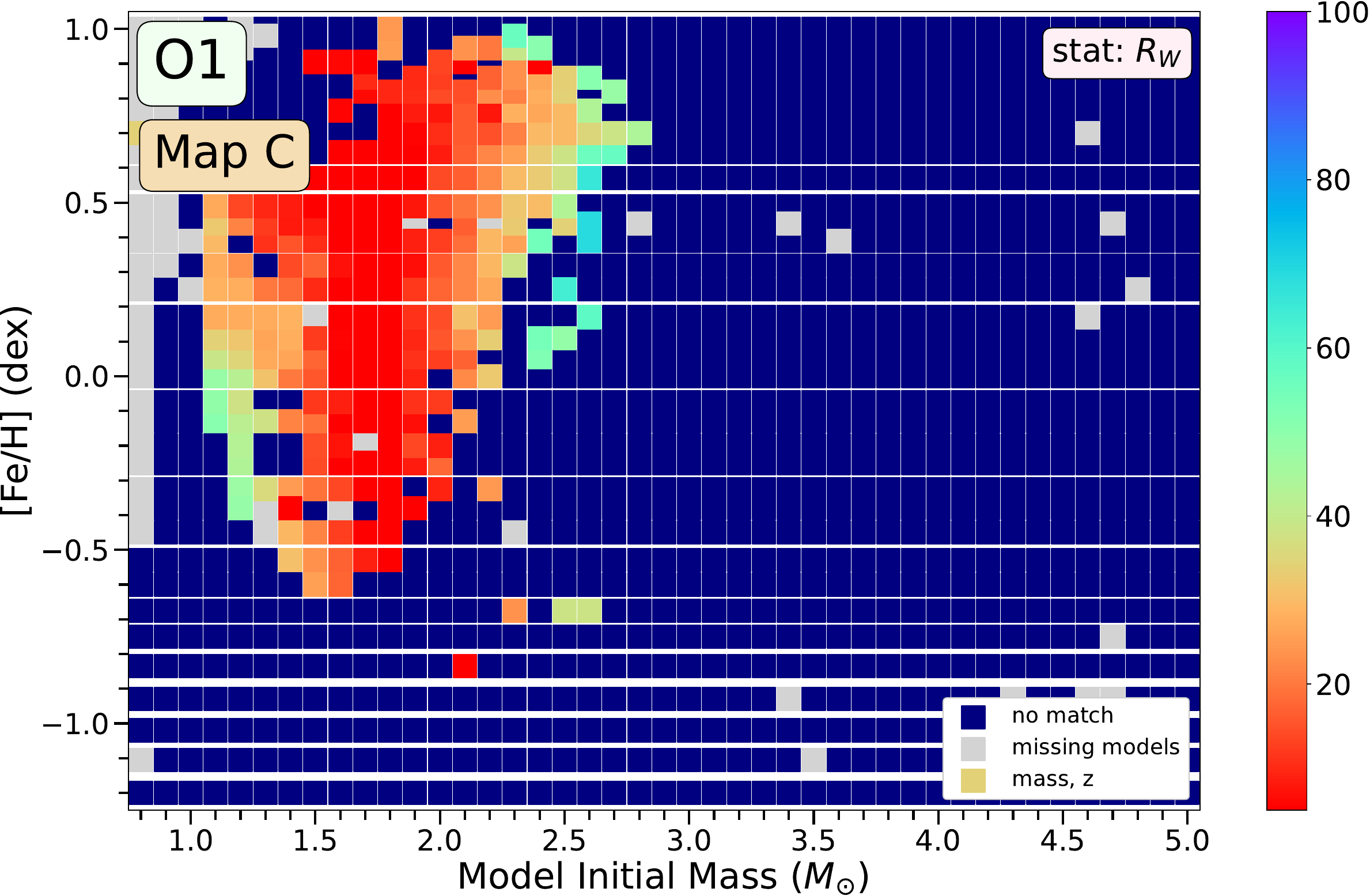}
\includegraphics[width=\columnwidth]{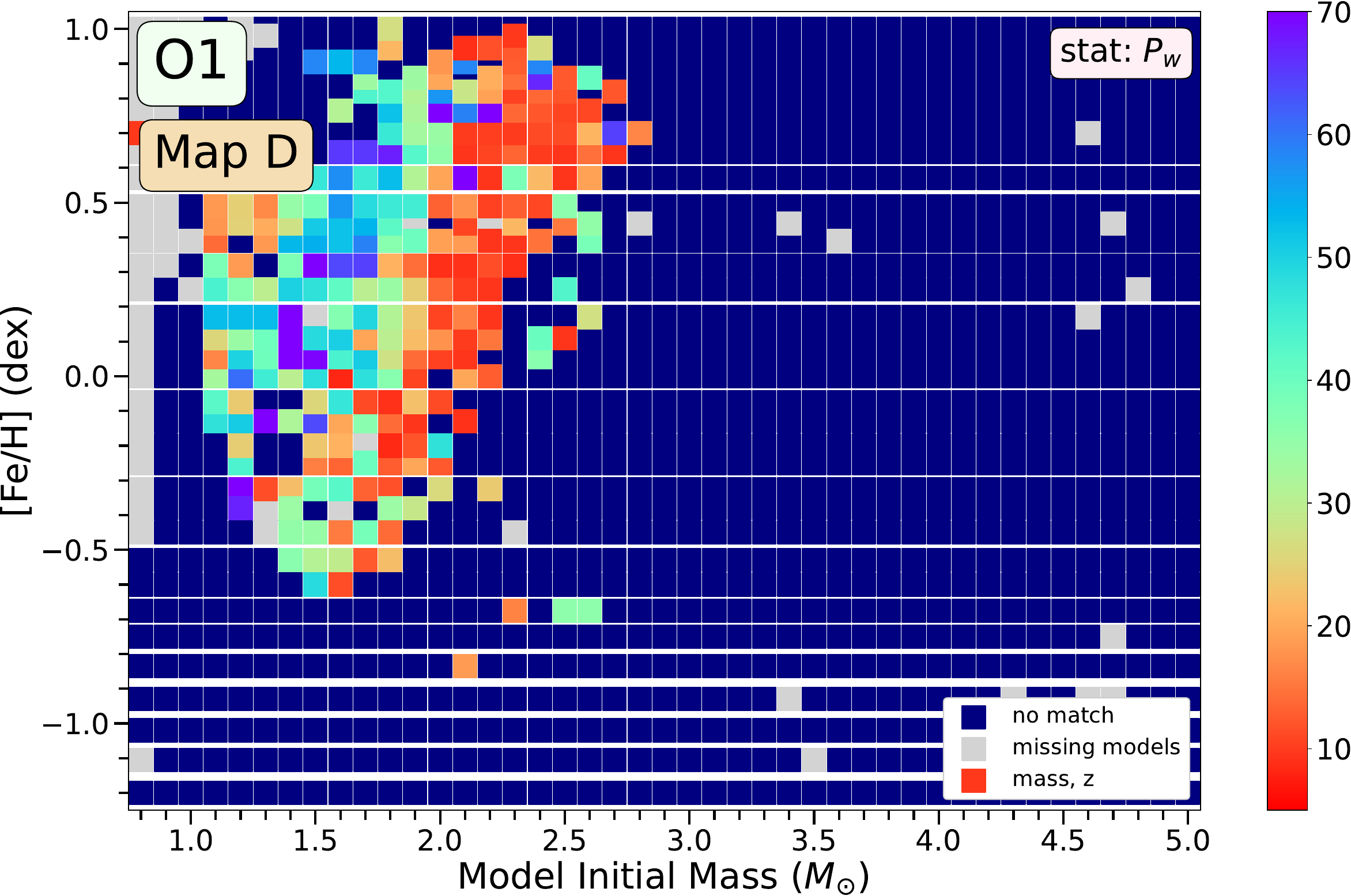}
\includegraphics[width=\columnwidth]{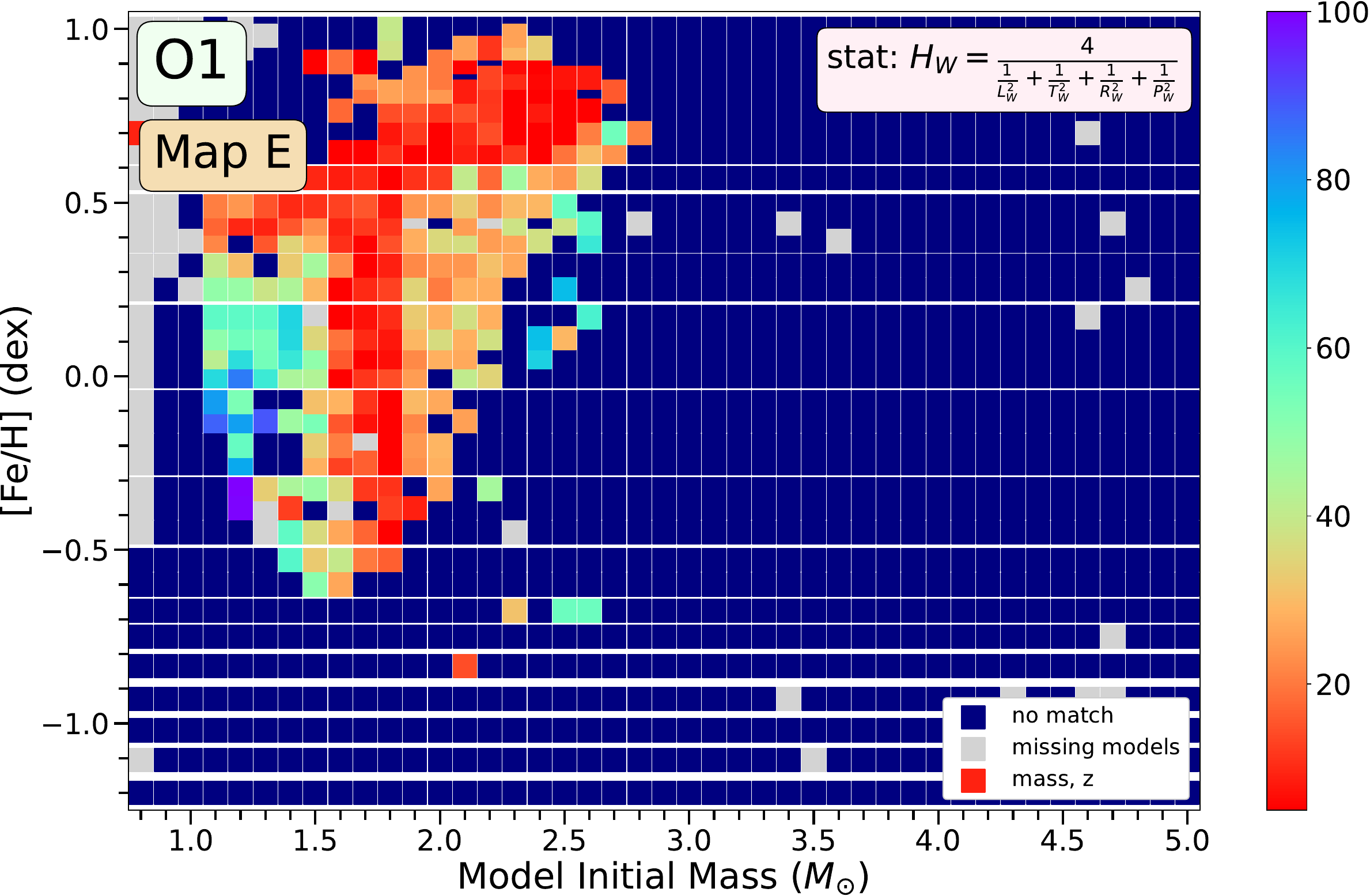}
\includegraphics[width=\columnwidth]{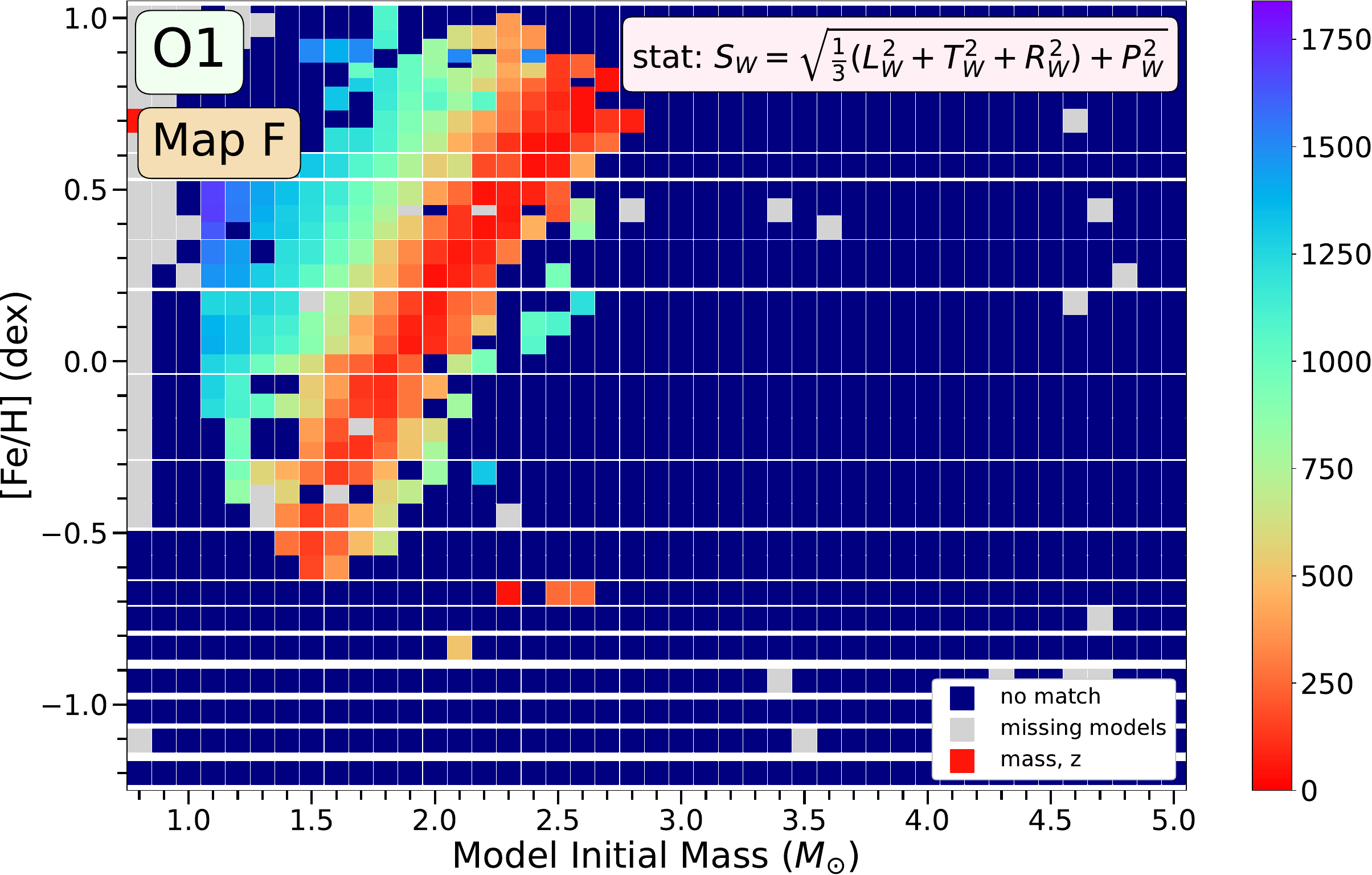}
\caption{Same as Figure \ref{fig:heatmap_FM}, but for the O1 mode assumption. The color normalization in each panel is the same as in its corresponding panel in Figure \ref{fig:heatmap_FM}, so scale is preserved.
}
\label{fig:heatmap_O1}
\end{figure*}
The goodness-of-fit to R Hydrae for 1400 TP-AGB models, per panel, is shown in the form of six mass--[Fe/H]\footnote{$Z$ is converted to [Fe/H] using the \citet{GS98} solar scale. $Z$ values are available for each point using the grid visualizer.} heat maps in Figures \ref{fig:heatmap_FM} (FM) and \ref{fig:heatmap_O1} (O1). Figures show calculations based on the static helium grid ($Y_i=0.3$) unless stated otherwise.

In all panels, hot colors (red) correspond to low WRMSE scores, or better-fitting models, and cool colors (blue) correspond to higher WRMSE scores, or worse-fitting models. The score maximum and minimum shown on the color bars differ, with values chosen for similar visual scale. All models with scores greater (worse) than the maximum given on the plot are shown in purple. Navy blue squares indicate models that were run successfully but did not intersect with the observations as specified in the first step of the fitting procedure (Step 1, Section \ref{sec:data_processing}). 
Grey squares indicate models that were launched but failed (roughly 3\% of cases). Colored squares show the minimum WRMSE score among all pulse fits in the \texttt{MESA} track for that mass and metallicity, thus representing the fit of only the best pulse associated to that model. 

All maps reflect the looser interpretation of the observational constraints described in Table \ref{table:obs_domain} and have a slope hardness parameter of $s=75$. The agreement statistic adopted in each panel of Figures \ref{fig:heatmap_FM} and \ref{fig:heatmap_O1} is indicated in the upper right corner: Map A shows $L_\text{W}$, Map B shows $T_\text{W}$, Map C shows $R_\text{W}$, and Map D shows $P_\text{W}$ (Equation \ref{eq:WRMSE}). Maps E and F show composites of these four statistics: Map E shows the harmonic mean, $H_\text{W}$ (Equation \ref{eq:harmonic_mean}) and Map F shows $S_\text{W}$ (Equation \ref{eq:pseudo_chisq}).

Each of the statistics based on a single classical parameter (Maps A, B, C) shows the sensitivity of the models to one component of the $L$, $T_\eff$, $R$ observations. We discuss the FM maps in Figure \ref{fig:heatmap_FM} first. Map A shows that the luminosity is highly constraining along a single ridge starting at (1.7 $M_\odot$, [Fe/H] $= -1.1$) and extending to $(4.1, +0.65)$. Map B indicates that low-mass, high-metallicity models do the best job of reproducing the observed effective temperature, but the sensitivity is not particularly strong anywhere. Map C shows that agreement with radial constraints is best among higher-mass models.

Map D shows the best-fitting models according to $P_\text{W}$, which is an arguably valid way to assess the results overall. Since the $L$, $T_\eff$, $R$ constraints are folded in to the fit determinations by setting the boundaries of the domain (see step 1 in Section \ref{sec:fittingprocedure}), Map D shows a composite of classical and seismic agreement weighted strongly, but not exclusively, towards seismic consistency. By this measure, there are two ridges of good solutions: the primary FM ridge, extending from roughly (1.7 $M_\odot$, [Fe/H] $= -1.1$) to $(3.8, +0.6)$, similar to the luminosity ridge in Map A; and a secondary FM ridge, extending from $(1.1, -0.13)$ to $(1.9, +0.8)$. These ridges are less smooth and show much more scatter than the luminosity ridge in Map A, but they follow an arc of similar slope in both cases.

The $P_\text{W}$ statistic roughly traces contours of mean density, a quantity to which radial period variations are sensitive but scaled by different factors in the FM vs O1 cases. We found that the models on the primary FM ridge have best-fitting pulse indices around the 15th TP, whereas models along the secondary, lower-mass, higher-metallicity ridge have best-fitting pulse indices around the 4th TP (see Section \ref{sec:Monash}). 

Map E shows goodness-of-fit according to $H_\text{W}$, the harmonic mean of $L_\text{W}, T_\text{W}, R_\text{W}$ and $P_\text{W}$, which incorporates the classical observational constraints in the domain as well as in the statistic itself. By this metric, agreement with radius plays a more prominent role in the solution space, but the ridges shown in Map D are reflected here as well. 

Map F shows the solution space according to $S_\text{W}=$~$ \sqrt{ \frac{1}{3}(L_\text{W}^2 + T_\text{W}^2 + R_\text{W}^2) + P_\text{W}^2 }$, which treats overall agreement with the classical observational constraints and agreement with the individual period measurements equally. Because the individual WRMSE terms are weighted by the measurement uncertainties and normalized according to number of observations, this agreement statistic is similar to a weighted $\chi^2$ score.\footnote{It is not a true $\chi^2$ score for several reasons, including that $L$, $T_\eff$, $R$, and $P$ are not independent.}

We take the $S_\text{W}$ statistic to be the best indicator of goodness-of-fit out of all those considered. Using this metric, Map F shows one clear ridge of preferred solutions, with smooth tapering into less well-fitting regimes. The ``hot'' ridge has roughly the same lower and upper coordinates as Maps A, D and E, covering about $(1.6, -1.1)$ to $(4.1, +0.7)$ in an arc about $M_\odot$ wide with the shape of the thin luminosity ridge in Map A. Because there is no mathematically rigorous way to quantify how much more probable a model with $S_\text{W} = 250$ is compared to a model with $S_\text{W} = 1500$, we consider this measure to be an indicator of relative likelihood and treat all models along this red FM ridge to be equally good fits.

Moving now to Figure \ref{fig:heatmap_O1}, we see many of the same features. Map A hosts a thin ridge of high sensitivity to luminosity. Map B shows comparatively strong sensitivity to effective temperature relative to Map B in Figure \ref{fig:heatmap_FM}, but with scaling in the same direction: higher-metallicity models are a better fit to the temperature constraints. Map C shows a somewhat different sensitivity to the radial constraints, with a large patch of well-fitting models centered at $1.5M_\odot$ regardless of metallicity and tapering in both mass directions. 

The $P_\text{W}$ statistic in Map D reveals a ridge similar to the one in Map D of Figure \ref{fig:heatmap_FM}, but with a steeper slope (see Section \ref{sec:global_features} for more discussion of this). Map E shows greater radius and temperature sensitivities among O1 models compared to FM models.

As in the FM case, we take Figure \ref{fig:heatmap_O1}'s Map F to be the best indication of agreement of the O1 models with all observations. Here, the arc of Map D appears more sharply and smoothly, ranging from (1.5$M_\odot$,[Fe/H]$=-0.6$) to (2.8 +0.8).

Interpretation of the solution space continues in Section \ref{sec:discussion}. 

\subsection{Data visualization tool}
\label{sec:visualizer}
An interactive data visualization application is provided with this grid. We provide data files containing the calculations necessary to reproduce all of the realizations in Figures \ref{fig:heatmap_FM} and \ref{fig:heatmap_O1} and their counterparts for the helium-varied grid, as well as additional parameter variations explored in Appendix \ref{appendix:param_var}.

Some of the visualizer's capabilities include mouse-over information, pulse identification plots, and links to the evolutionary tracks associated to each model. These are demonstrated in Figure \ref{fig:data_visualizer} in Appendix \ref{appendix:visualizer}. However, the documentation in the Github repository itself is most comprehensive: \url{https://github.com/mjoyceGR/AGB_grid_visualizer}.

\subsection{Caveats}
Given that a substantial fraction of models were terminated based on run time limitations (72 or 120 hours maximum for the fixed helium or helium-varied grids, respectively) or convergence difficulties (\texttt{max\_number\_retries $= 500$}) rather than by virtue of reaching their physically motivated stopping conditions, it is possible that any such model was stopped before encountering the thermal pulse that would provide the best fit to observations. 
Run incompleteness could therefore impact the score shown for a model in the heat map, and non-uniformity of incompleteness (e.g., models with very super-solar abundances are more likely to encounter convergence failure than solar-like models) could impact trends in the heat maps overall. 

That said, premature model termination is not a concern for R~Hya, which is typically best fit by one of the first 15 TPs regardless of helium variation. Where models do max out their run time, they have evolved far beyond the observational constraints for R Hya. This can be confirmed by inspecting the pulse identification figures available for every model in the grid visualizer.

There are, however, more significant limitations to this grid in terms of its utility as a general AGB fitting tool. Chief among these is the lack of variation in mass loss. The number of thermal pulses an AGB model undergoes is highly sensitive to the choice of mass loss efficiency, $\eta$, and mass loss prescription. The strength of the TPs and occurrence of third dredge-up depend on the mass of the envelope, and how the envelope mass changes over the course of the TP-AGB phase is likewise sensitive to these mass loss settings (see, e.g., \citealt{Ventura2005b, StancliffeJeffery2007, Marigo2020, Karakas2022}.)

Similarly, this grid does not consider variations in $\alpha_\text{MLT}$ or choice of opacity tables, the impacts of which have been documented in \citet{Cinquegrana22solarcal} and \citet{Cinquegrana2023}. We do not explore variations in solar scale. We also note that the quantity $\Delta Y/\Delta Z$ remains fixed to 2.1 for the grid that scales $Y_\text{init}$ with $Z$. Not only is there considerable tension in the literature regarding the appropriate choice for $\Delta Y / \Delta Z$, but it may vary depending on the region of the Galaxy, local chemistry, number of enrichment sources (e.g., supernovae) in the vicinity, and time. 

It is likely that scores would change in response to modifying any one of these stellar modeling assumptions, and possible (though less likely, based on explorations done here) that overall trends could shift as well. Such questions are worth exploring in future investigations. Nonetheless, this grid is an excellent resource for determining (preliminary) fundamental parameters for AGB stars and provides a much more accessible alternative to performing one's own AGB calculations.

Finally, the grid calculations presented here are linear and adiabatic, and neither of these simplifications applies in earnest to AGB stars and their pulsation modes. We explore non-linear and non-adiabatic considerations further in Section \ref{sec:advanced_tactics}.

\section{Discussion}
\label{sec:discussion}
We now discuss the key features of the F panels of Figures \ref{fig:heatmap_FM} and \ref{fig:heatmap_O1}---the preferred statistical composite maps---and fundamental parameter determinations for R Hydrae, including comparisons to other fundamental parameter determinations in the literature. 

\subsection{Global features of solution spaces}
\label{sec:global_features}
The FM and O1 composite maps show very similar features in their solution spaces. Both maps feature a diagonal ridge tracing the preferred solutions, shown in red (warmer colors). The ridge in the FM map runs from around $1.9\,M_\odot$ up to almost $4.0\,M_\odot$ with increasing metallicity up to [Fe/H] = 0.7. Along this ridge, solutions in the upper region starting from $2.5\,M_\odot$ and nearly solar metallicity and upwards have lower scores than those below this.  The preferred solutions in the O1 composite heat map (panel F of Fig.~\ref{fig:heatmap_O1}) trace a smaller area and form a more vertical ridge that trends towards lower masses. This ridge runs from $1.5\,M_\odot$ to $2.6\,M_\odot$, between [Fe/H] indices of $-0.6$ and $+0.8$. In both modes, the ridges are truncated where models cease to intersect the observational boundaries for any pulse.

The difference in slope between the ridges in Figures~\ref{fig:heatmap_FM} and  \ref{fig:heatmap_O1} reflects variation in the O1/FM period ratio, which depends on the physical parameters of the models. As we have shown in previous works on variable evolved stars, the strong dependence of the O1/FM period ratio on quantities like luminosity can be used to further constrain the model fits. This technique was used to characterize the TP-AGB star T~UMi and the red supergiant Betelgeuse \citep{TUMi,Betelboi}, but unfortunately cannot be applied to single-mode stars like R~Hya.

\subsection{Fundamental Parameters of R Hya}
\label{subsec:fundpar}
\begin{figure*}
\centering
\includegraphics[width=\columnwidth]{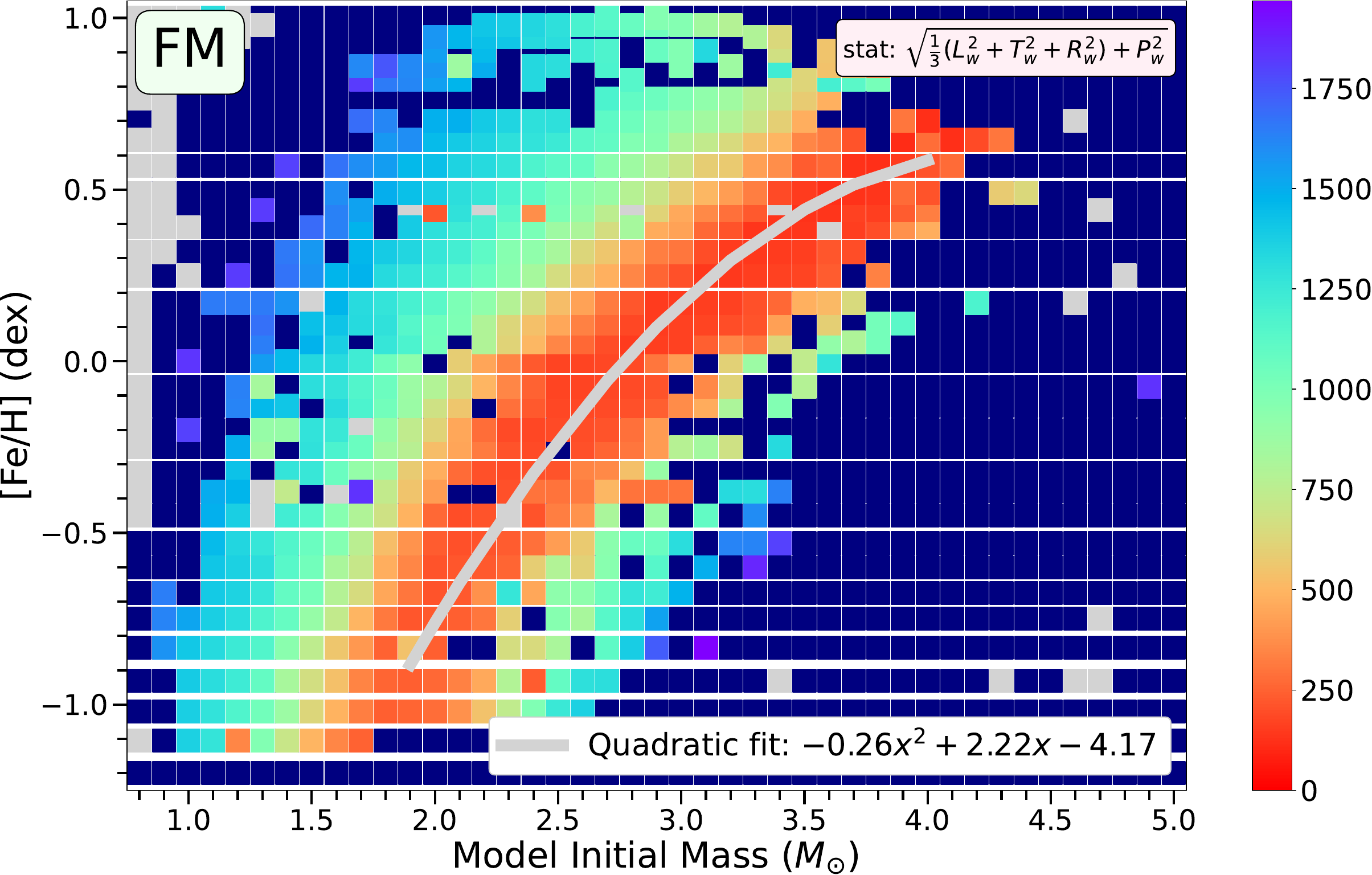}
\includegraphics[width=\columnwidth]{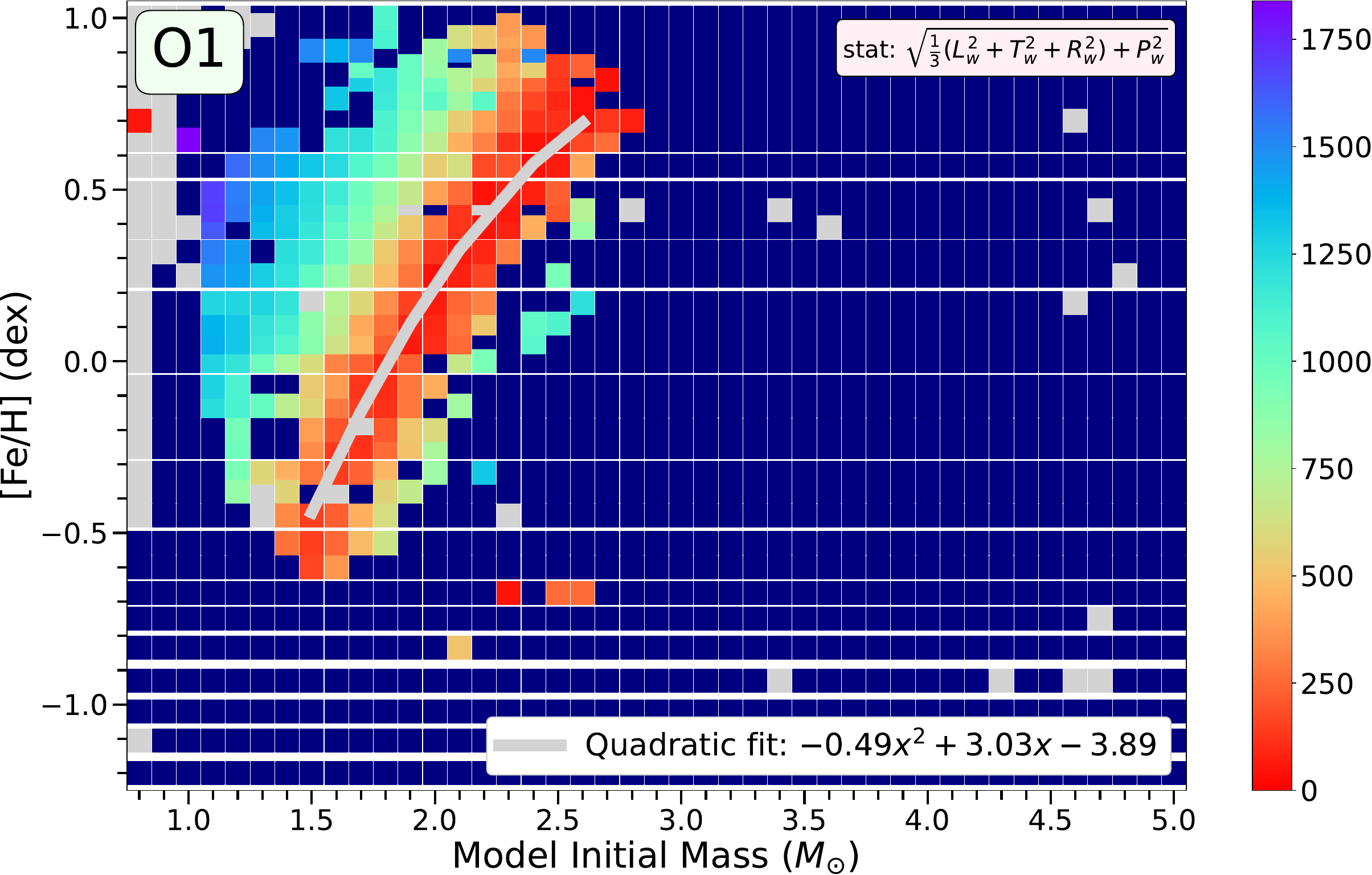}
\caption{Quadratic fits to the ridges of best-fitting solutions for the FM (left) and O1 (right) are overlaid on the heat maps adopting the $S_\text{W}$ statistic. See also: Equations \ref{eq:FM_scaling} and \ref{eq:O1_scaling}.}
\label{fig:quadratics}
\end{figure*}

Treating all solutions constituting the primary ridges in the composite maps (F panels, Figures \ref{fig:heatmap_FM} and \ref{fig:heatmap_O1}) as equally valid, the best-fitting parameters of R~Hya follow well-behaved quadratic scaling relations between mass and metallicity, depending on the mode assumed.
For the FM, this relation is approximately
\begin{equation}
 {\rm [Fe/H]} = -0.26 M_\text{init}^2 + 2.22 M_\text{init} - 4.17;
\label{eq:FM_scaling}
\end{equation}
and for the O1,
\begin{equation}
 {\rm [Fe/H]} = -0.49 M_\text{init}^2 + 3.03 M_\text{init} - 3.89.
\label{eq:O1_scaling}
\end{equation} 
These are shown overlaid on the heat maps in Figure \ref{fig:quadratics}.
\indent Good solutions for initial mass under the FM assumption are greater than $1.5M_{\odot}$ and extend up to roughly $4 M_{\odot}$. This range covers almost all metallicites considered ($-1.2 \le$ [Fe/H] $\le +0.7$), with the lower bound on [Fe/H] and the upper bound on mass enforced by agreement with the $L$, $T_\eff$, $R$ boundaries. Good O1 solutions start at higher metallicities ([Fe/H]${}\sim-0.5$) relative to FM solutions and push into metal-enrichment extremes ([Fe/H]${}\ge+0.7$) that can be ruled out by other factors. The range of masses preferred by the O1 is much narrower than the FM, spanning $\sim1.5$--$2.7\,M_{\odot}$ only.

Regardless of mode assumption, the best solutions fall towards the higher-mass edge of the regions permitted by the classical constraints. 
Based on the modeling results alone, we can already surmise that the FM is more likely than the O1: the number of good solutions is larger and covers a wider range of (sensible) initial masses and metallicities in this case. However, we can further discriminate among all good solutions by comparing against independent parameter estimates for R~Hya in the literature and taking into account other sources of physical information, such as the presence of third dredge-up indicators, described in detail in Section~\ref{sec:nucleo}.   

\subsection{Literature Comparison}
\label{subsec:lit_compare}
\begin{figure*}
\centering
cd \includegraphics[width=\columnwidth]{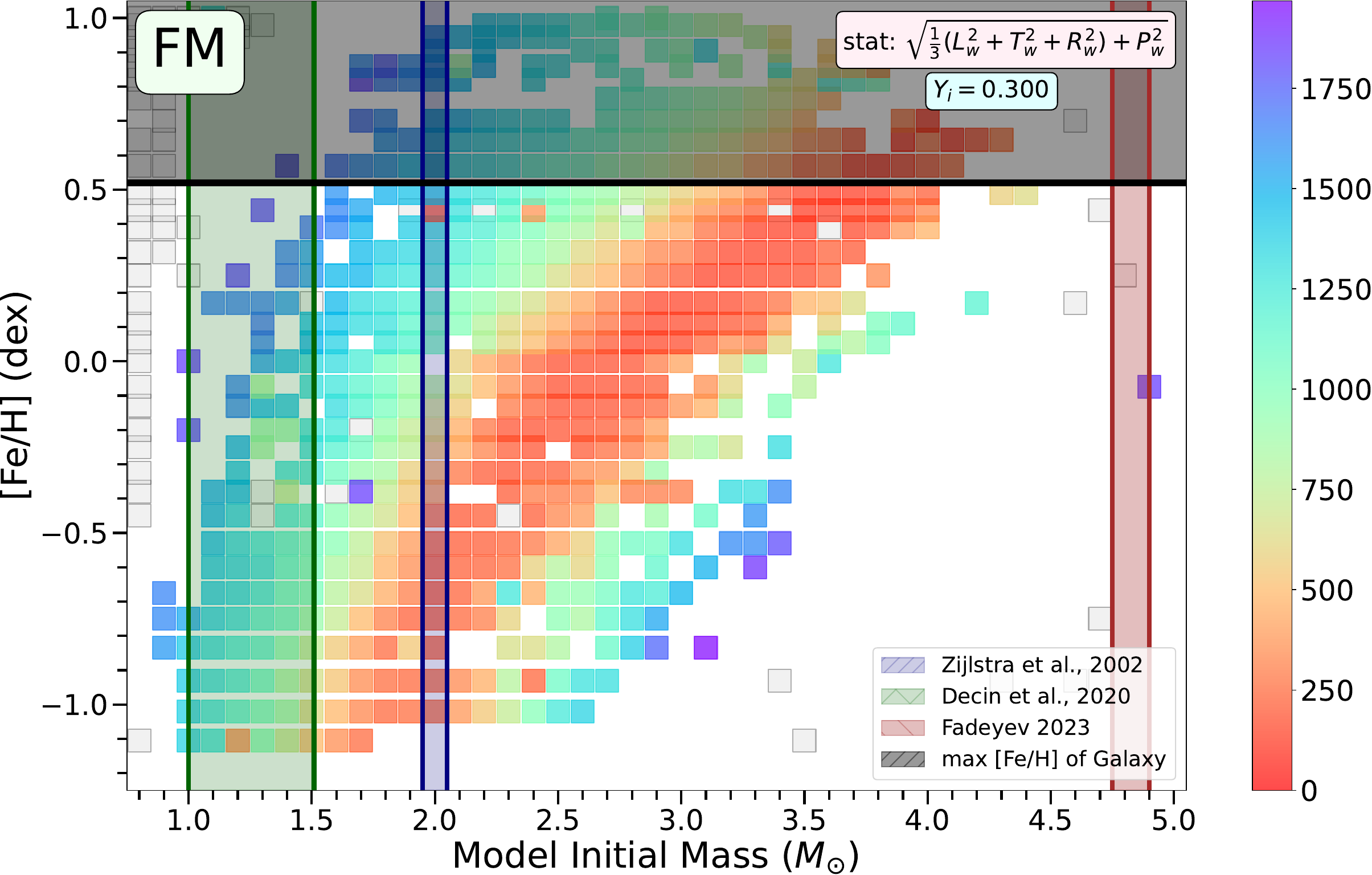}
\includegraphics[width=\columnwidth]{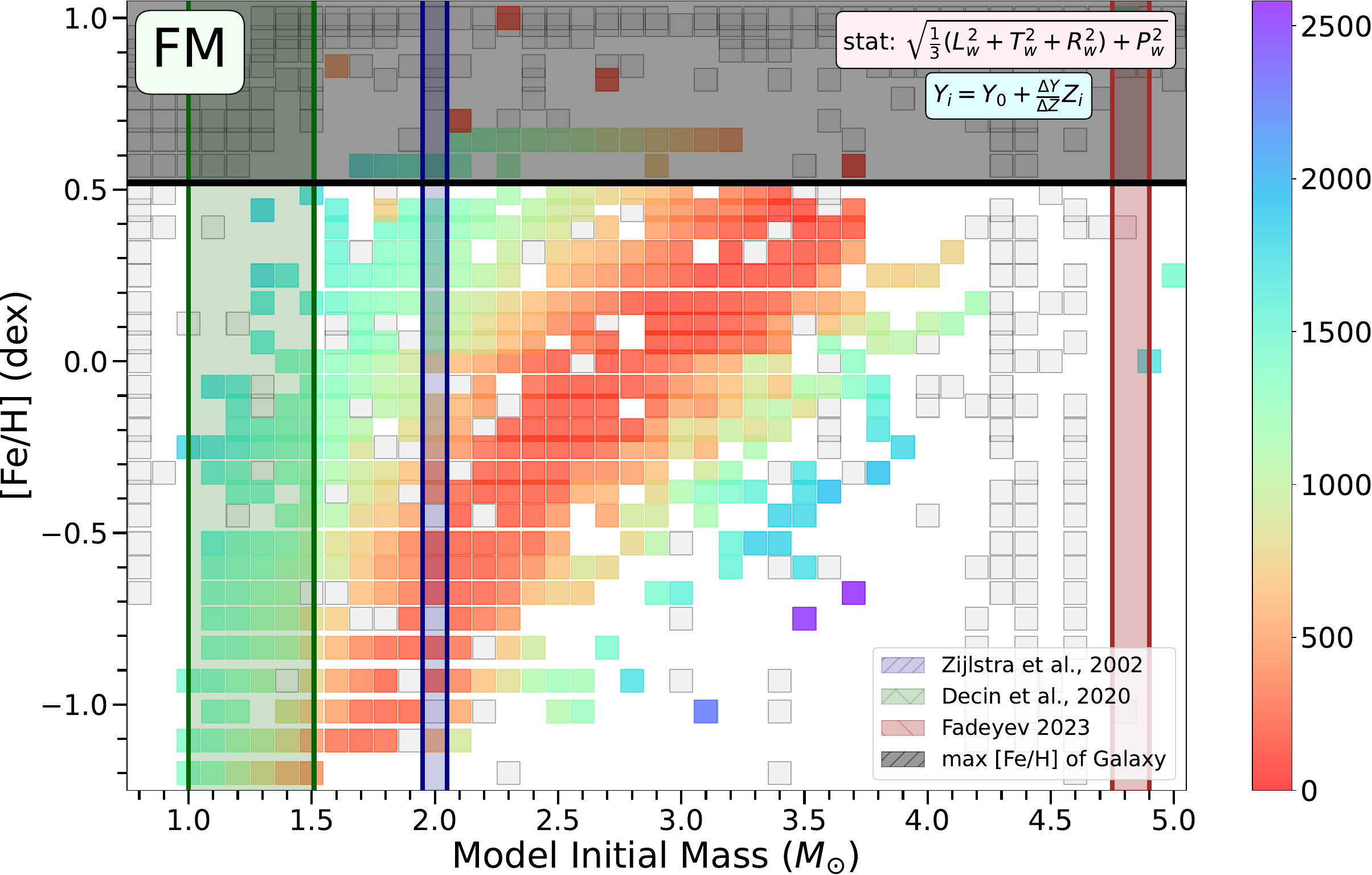}
\includegraphics[width=\columnwidth]{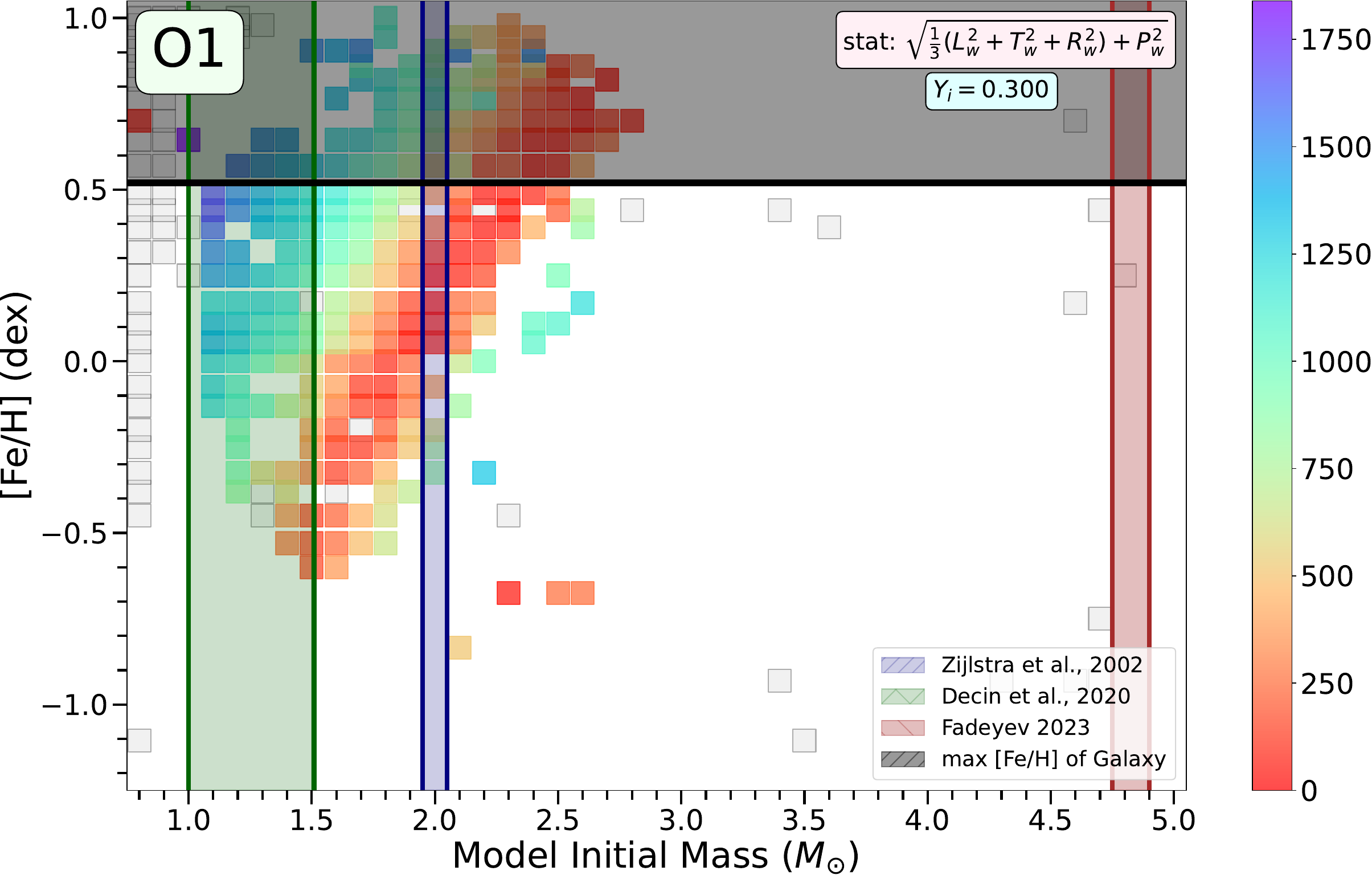}
\includegraphics[width=\columnwidth]{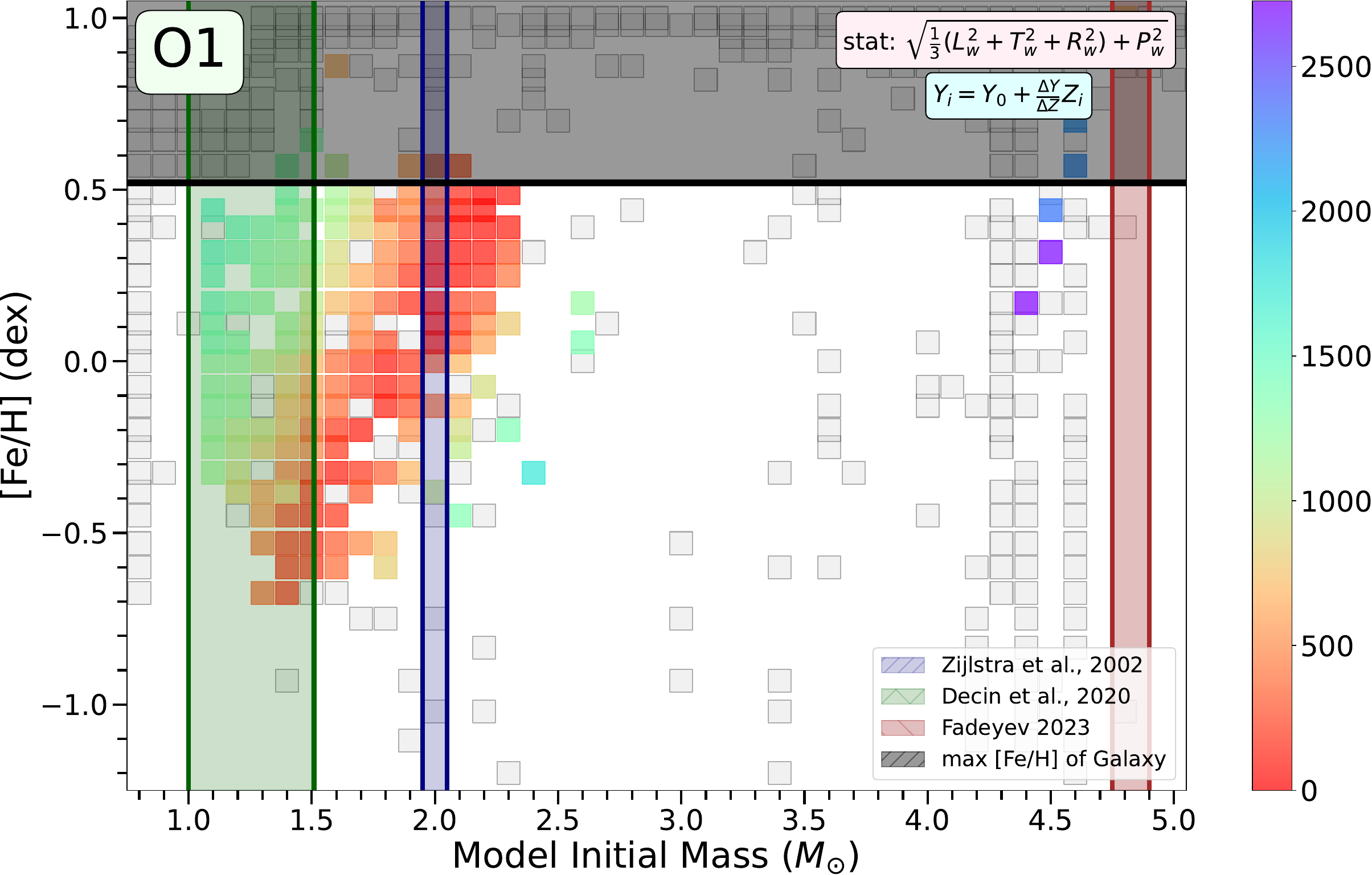}
\caption{The FM and O1 heat maps showing the $S_\text{W}$ composite statistic are shown with mass estimates from other determinations in the literature superimposed, with source as indicated in the legend and listed in more detail in Table \ref{table:proposed_solutions_lit}. Colors are muted for visualization but grey squares preserve their meaning as failed/missing models.
The fixed helium assumption is used in the left-hand panels and helium variation is used on the right, as indicated via text box in the upper right of each panel. 
The black region above [Fe/H]$=0.7$ indicates roughly the maximum metallicity of the Galaxy, suggesting that any solutions with metallicities above this should be excluded.}
\label{fig:heatmap_literature}
\end{figure*}

Table \ref{table:proposed_solutions_lit} summarizes past attempts at determining the fundamental properties of R~Hya, including mass, metallicity, and age. In cases where the parameter determination was made using AGB stellar models, the value is indicated in boldface. Figure \ref{fig:heatmap_literature} shows some of these overlaid on the FM and O1 composite ($S_\text{W}$) maps, with colors muted for visualization.

We first note that there is a huge range of masses and luminosities estimated for R~Hya in the literature, as indicated also in Table \ref{table:proposed_solutions_lit}, and very few estimates for its metallicity or age. 
The mass boundaries derived by \citet{Zijlstra2002} (similarly, \citealt{wood-zarro-1981} and \citealt{Eggen-1985}) intersect with a subset of our best-fitting models between [Fe/H]${}\sim-1.0$ to $-0.5$ under the FM assumption, but considerably higher metallicities, [Fe/H]${}\sim 0$ to $+0.3$, under the O1 assumption. This is indicated by the blue hatched region in Figure \ref{fig:heatmap_literature}.

The higher metallicities required by the O1 map are less likely, given R~Hya's location in the solar neighborhood, so the mass constraints suggested by most of the references in Table \ref{table:proposed_solutions_lit} provide further evidence that the period we have measured corresponds to the FM. One exception is the mass range reported by \citet{Decin-2020} (green hatched region in Figure \ref{fig:heatmap_literature}), which overlaps very little with our preferred FM solutions, barely intersecting them at the lowest masses and metallicities consistent with the observational boundaries.

The O1 solutions show slightly better agreement with \citet{Decin-2020} around [Fe/H]${}=-0.5$, again for the lowest masses. Mass estimates from \citet{Fadeyev-2023} (red hatched region) are significantly higher than the observational boundaries permit in either case. Even assuming that the FM scaling relation (Equation \ref{eq:FM_scaling}) is valid above $4.0 M_\odot$, a mass greater than $4.8 M_\odot$ would require a metallicity exceeding the maximum metallicity of the Galaxy, indicated by the black shaded region at the top of both panels in Figure \ref{fig:heatmap_literature}. Hence, our results are inconsistent with \citet{Fadeyev-2023}. 

\begin{table*} 
\centering
\begin{tabular}{lccccccc}  
\hline \hline
Ref.			&	Birth Mass  & Present-day	&	M$_{\rm core}$		& Luminosity 		&	T$_{\rm eff}$ &	[Fe/H] & Age \\
~				&	(M$_\odot$) & Mass (M$_\odot$)		&	(M$_\odot$)	&	(L$_\odot$)		&	(K)			 & ~  & Gyr		\\
\hline
\citet{wood-zarro-1981}	&	$\sim 2.0 $	         &	--	&	0.61--0.66	& $11500 \pm 1000$	&	--			 &	--	&	--		\\
\citet{Eggen-1985}      &   $\sim 2.0$           &  --  & --  & -- & -- & -- & 0.5--1.0 \\
\citet{Zijlstra2002}	& 	$\sim 2.0 $	         &	--	&	--			&	11600			&	2570--2830	 &	--	&	--		\\
\citet{Decin-2020}		&	$\mathbf{\sim 1.35}$  (1.0--1.5) &	-- &	--			&	7400			&	$2100\pm35$	 &	--	&	--		\\
\citet{Homan-2021}		&    --                  &	$\mathbf{\sim}$ 1.5--2.5	&	--			&	--			&	--	 &	--	&	--		\\
\citet{Trabucchi-2022}  &   2.72  (1.52--3.98) &	--			&	--			&	--	 &	--	&	--	& --	\\ 
\citet{Fadeyev-2023}	&	4.8 (4.75--4.9) &	--		&	--			&	25000			&	--			 &	0.014 &	--		\\
\citet{Fadeyev-2024}	&	4.7  & 4.43--4.50		&	--			&	$18600\pm200$ 			&	--			 &	0.014 &	--		\\
\hline
\end{tabular}
\caption{Summary of proposed solutions for R Hydrae's physical parameters in the literature. Bold values indicate results based on AGB stellar model fits. In the case of \citet{Homan-2021}, the present-day mass estimate refers to the total mass of the central system, which includes two components.
}
\label{table:proposed_solutions_lit}
\end{table*}

\subsection{Age constraints}
\label{sec:age}
\begin{figure*}
\centering
\includegraphics[width=\columnwidth]{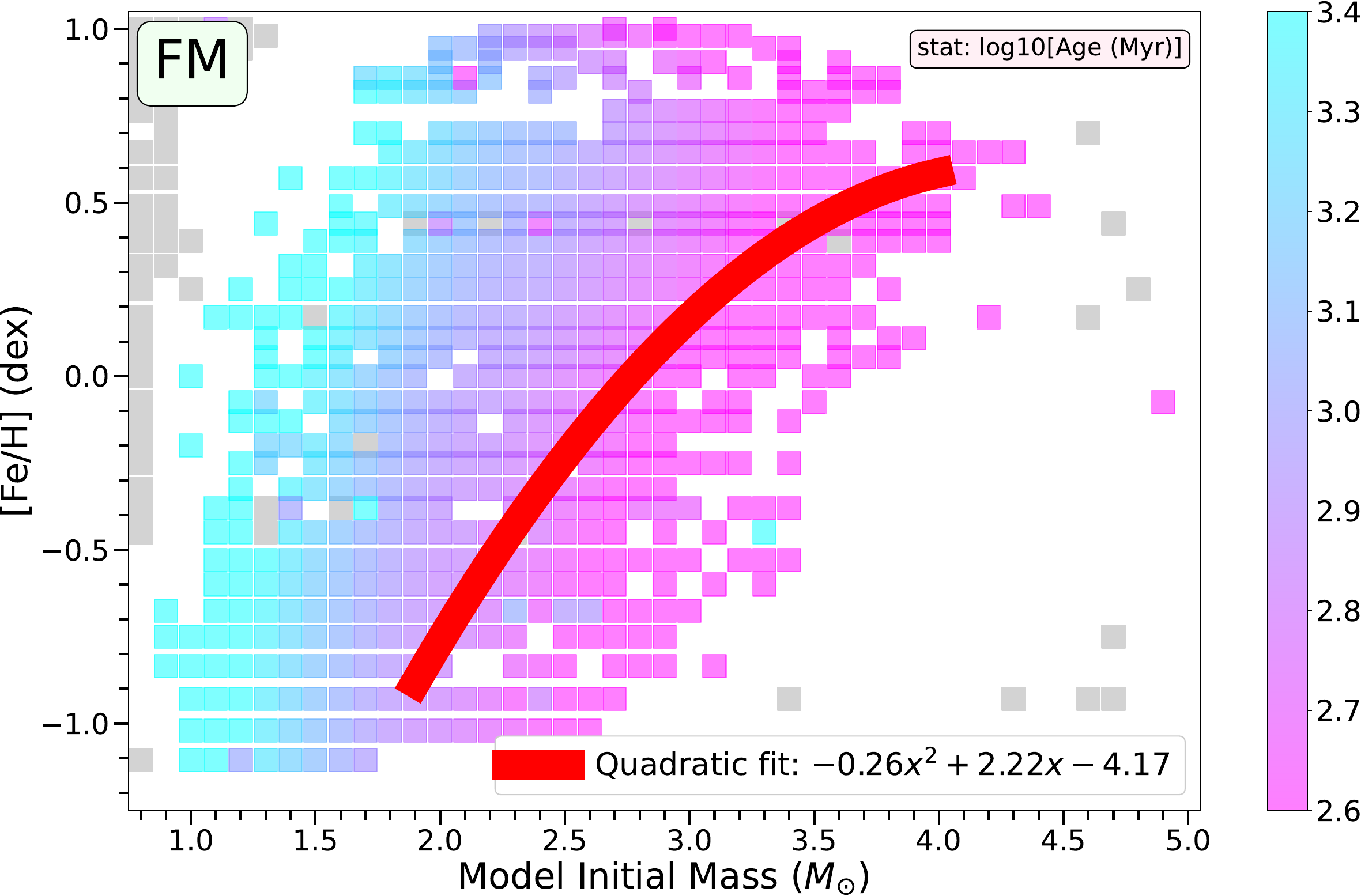}
\includegraphics[width=\columnwidth]{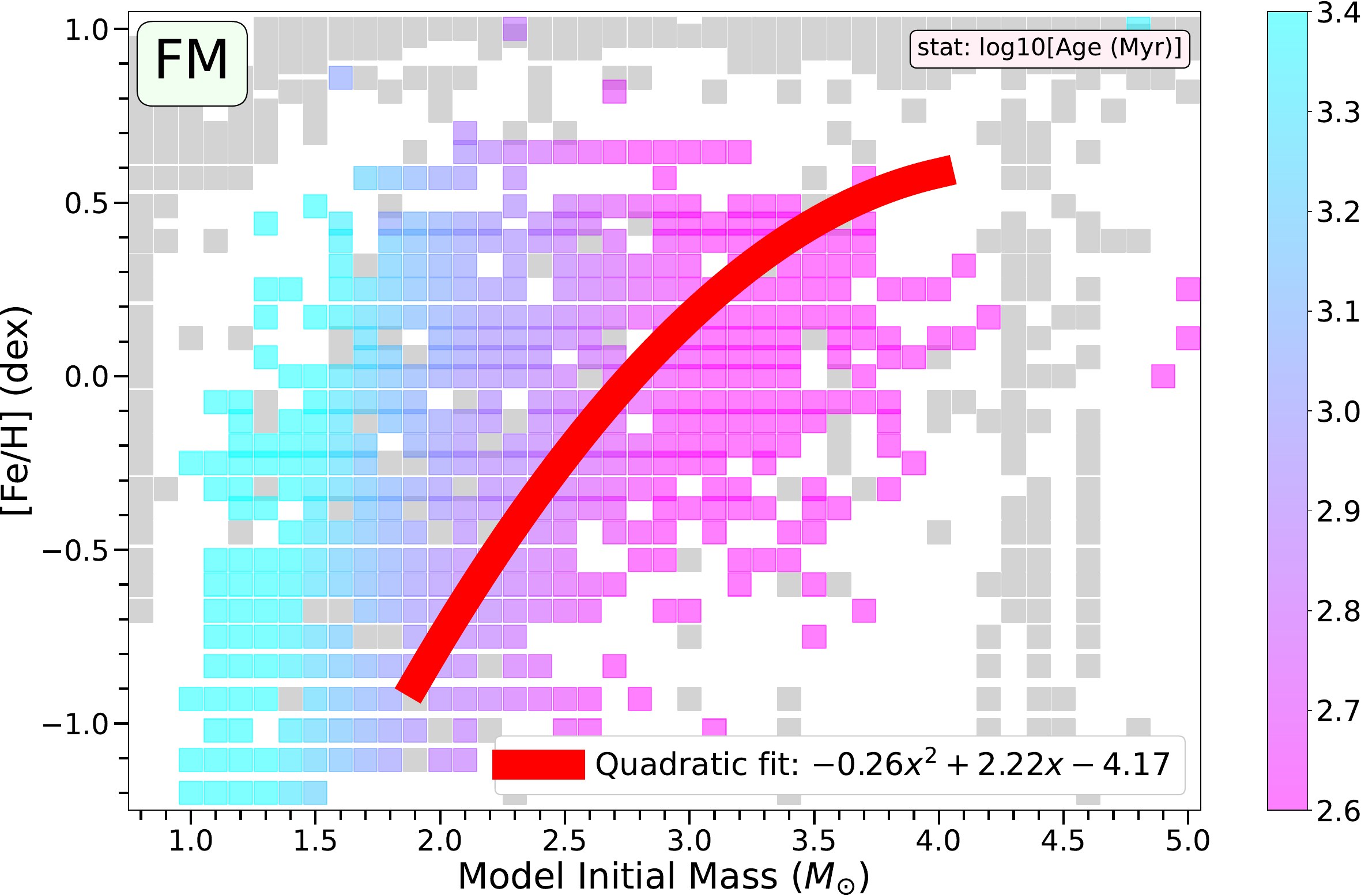}
\includegraphics[width=\columnwidth]{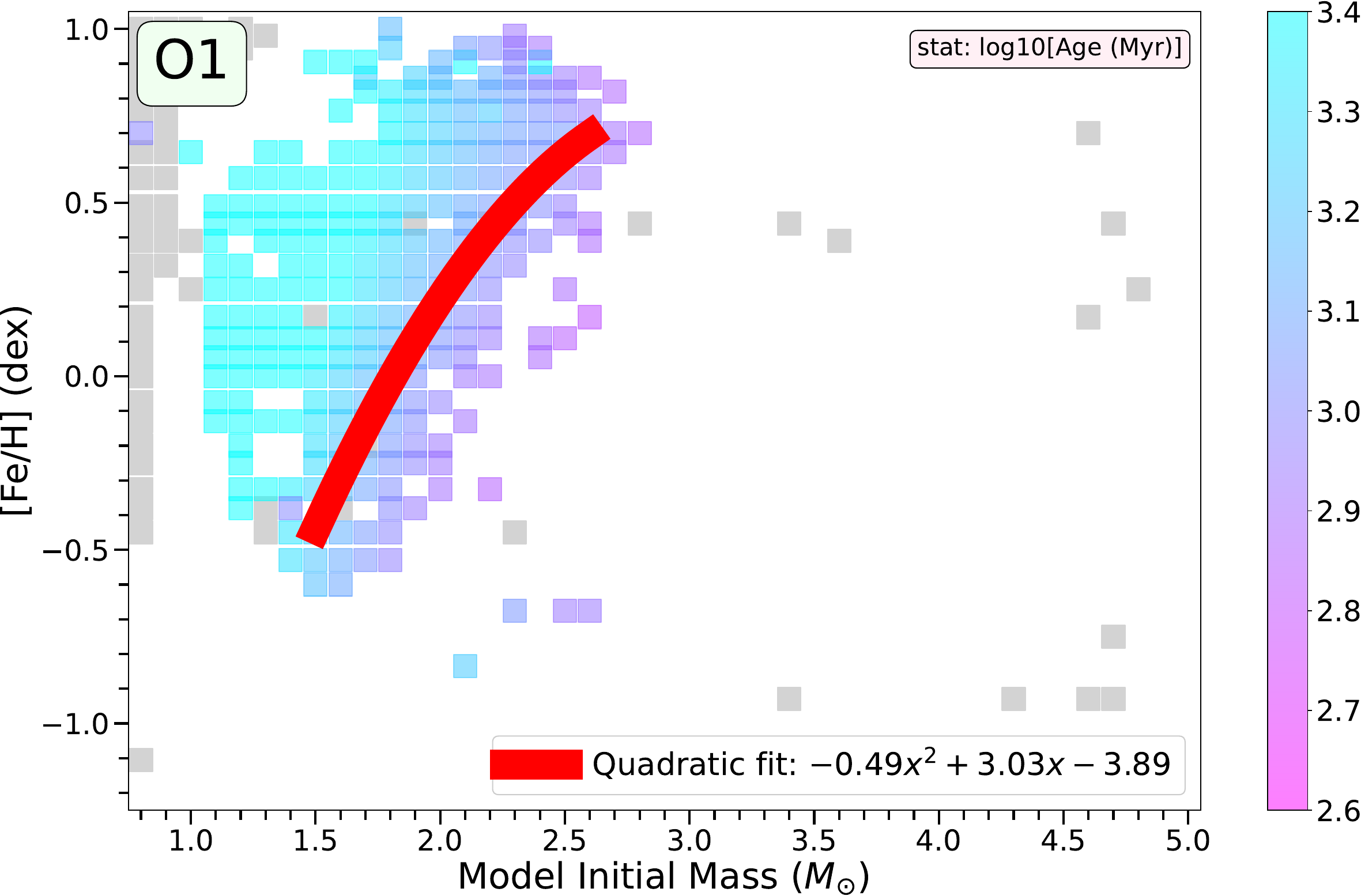}
\includegraphics[width=\columnwidth]{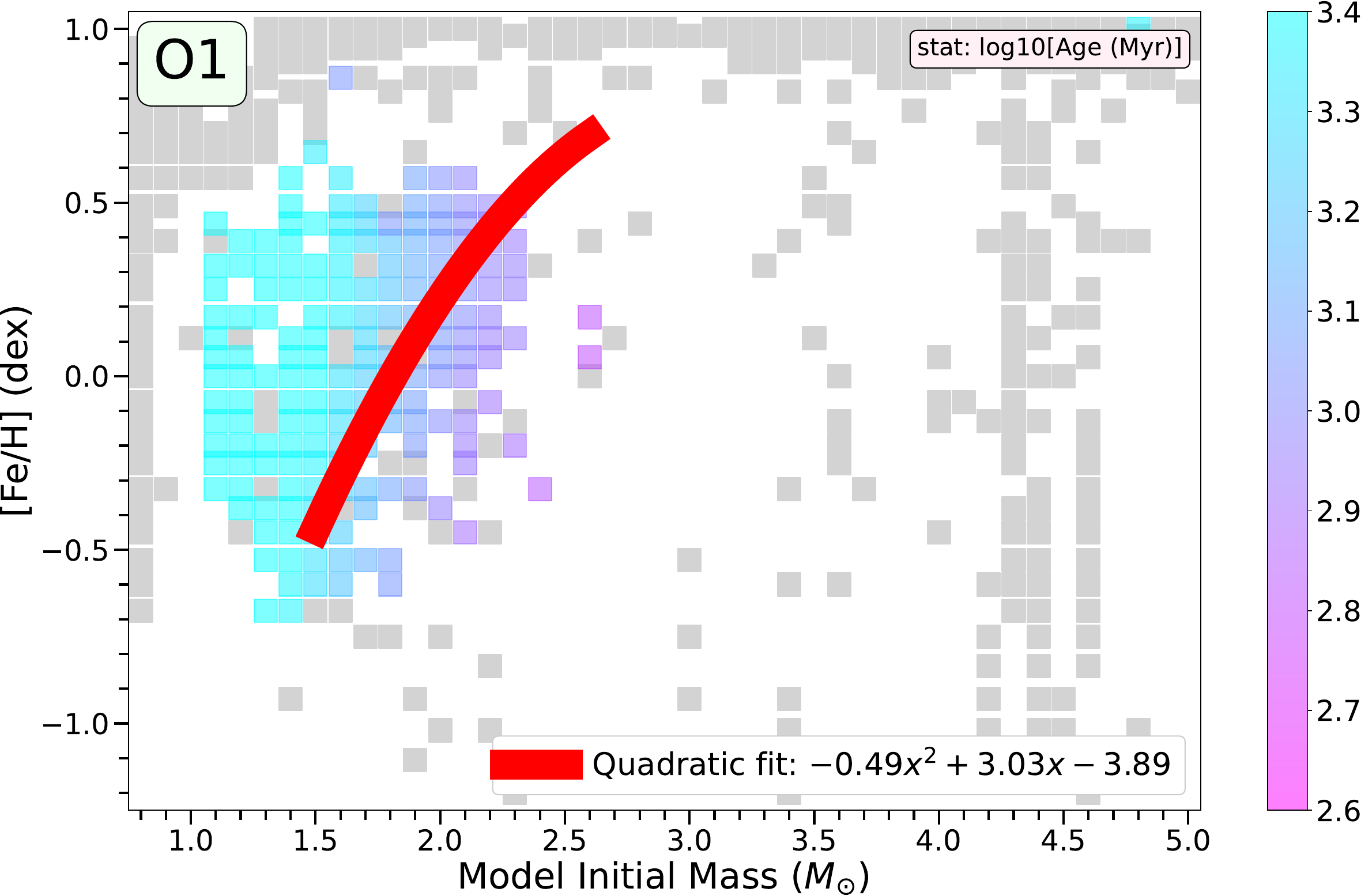}
\caption{The age of the best-fitting thermal pulse is indicated via color bar in units of $\log_{10}$ Myr (2.0 corresponds to 0.1 Gyr, 3.0 corresponds to 1 Gyr, 4.0 corresponds to 10 Gyr, etc) for both mode assumptions and treatments of helium. Ages shown in the left-hand panels are the ages according to the static-$Y_i$ models, and ages shown on the right-hand panels are ages according to the helium-varied models. The quadratic fits to the primary ridges from Figure \ref{fig:quadratics} are reproduced at the same locations.
The treatment of helium does have an impact on the ages: in general, the models that use helium scaling are younger than the helium-fixed models when $Y_i, Z_i$ are super-solar and older when $Y_i, Z_i$ are sub-solar.  
The fixed-helium maps show better coverage than the varied-helium maps due to better convergence and completion rate in the fixed-helium grid. }  
\label{fig:age_heatmap}
\end{figure*}

\citet{Eggen-1985} suggested that R~Hya might be a member of the Hyades moving group, which would constrain its age and mass to be about 0.5--1.0 Gyr and $\sim 2\,M_\odot$, respectively. This is the origin of their age value in Table \ref{table:proposed_solutions_lit}.
However, more recent results indicate that, while the Hyades cluster contains stars that also belong to the tidal streams of Hyades, the moving group itself originates from a resonance with the spiral arms of the Milky Way \citep{Famaey2007,mcmillan2011}. According to \citet{Famaey2008}, about half of the members are not coeval with Hyades, and hence this is not useful as an age constraint for R~Hya. 

While we cannot connect R~Hya to any cluster directly, we can use various period--age relations to estimate its age. However, different age relations give very different answers, especially in the 300 to 500~d period range \citep{Zhang-2023}. Here, we assume that the 350 d period is a good approximation for the out-of-pulse period of R~Hya, but calculate the ages for 300 and 500~d values as well, listing them in parentheses: 
In the 350~d case, the relation of \citet{Eggen-1998} gives an age of 6.9~Gyr (3.6--8.5~Gyr); \citet{Zhang-2023} gives a very similar value of 6.5~Gyr (4.1--7.4~Gyr). In contrast, the relation of \citet{Grady-2019} finds a much younger age of 1.01~Gyr (0.17--1.83~Gyr). The most detailed relations are presented by \citet{Trabucchi-2022}, who provide limits on the parameter space: in this case, the 350~d period corresponds to an age range of $0.5^{+2.6}_{-0.3}$~Gyr. 
Differences in the calibrating samples on which these period-age relations are based could be responsible for the large variance in the solutions they provide: any relation is inherently biased towards the cluster members from which it is calibrated and will therefore not be universally applicable to all Mira and TP-AGB stars.

Another complicating factor is that these calibrations assume a smooth evolution of pulsation period with age, but stars experience large, non-monotonic changes in period over the course of a TP. As stars evolve along the asymptotic giant branch, their radii and thus pulsation periods increase over time, on average. However, this overall trend of period increase with time is decorated with the high-amplitude variations of thermal pulses on much shorter timescales, meaning the star can enter the same period regime at very different points along its AGB evolution, leading to a period--age degeneracy (as discussed in Section \ref{sec:twomodes} and shown in Figure \ref{fig:twomodes}).
As such, period--age relations may not be the appropriate age inference method for 
stars suspected to be actively undergoing a TP.

However, our more sophisticated fitting method breaks the degeneracy present in simple period--age relations. Figure \ref{fig:age_heatmap} shows the age of the best-fitting thermal pulse from each TP-AGB spectrum on the same mass--[Fe/H] plane as Figures \ref{fig:quadratics} and \ref{fig:heatmap_literature}. The fixed-helium assumption applies to the left-hand panels, and the varied-helium treatment applies to the right (with higher model failure rate, as discussed in Section \ref{sec:modelgrid}, indicated by grey squares).
Ages are in Myr, converted to $\log_{10}$ units and indicated by color. The thick red line in both panels is a broadened version of the quadratic fits shown in Figure \ref{fig:quadratics} and indicates the ridge of best agreement for the FM (top) and O1 (bottom) solutions. 

Our best-fitting models indicate a low age for R~Hya: FM solutions have ages below 1.0 Gyr regardless of helium assumption. As shown in Figure \ref{fig:age_heatmap}, FM solutions trace a ridge ranging from 800 Myr (1 Gyr with helium varied) at the lowest masses to about 300 Myr at the highest, whereas overtone solutions give a somewhat higher age range of 1.0 to 1.6 Gyr (extending to 2.5 Gyr when helium is varied). These values are in good agreement with ages inferred from the period--age relations given by \citet{Grady-2019} and \citet{Trabucchi-2022}, but not with the relations of \citet{Eggen-1998} and \citet{Zhang-2023}, which overestimate the ages significantly.  

\subsection{Mass constraints}
\label{sec:mass}
\citet{Trabucchi-2022} also provide period--mass relations based on stellar models.\footnote{The lower mass boundaries are visual estimates based on Figure~B.1 of \citet{Trabucchi-2022}, because the coefficients for the upper edge of the O-rich relation listed in their Table~B.2 do not reproduce the line in the figure.} Using the same set of periods from the previous section for comparison, we find initial masses of $2.72\,M_\odot$ (1.52--3.98\,$M_\odot$), for the 350~d period. The mass range is slightly lower for a 300~d period (1.26--3.28\,$M_\odot$) and considerably higher for 500~d (2.45--6.28\,$M_\odot$). 

An alternative way to estimate masses for TP-AGB stars is through the $^{17}$O/$^{18}$O isotope ratio. According to the model calculations of \citet{DeNutte-2017}, this is a sensitive and  reliable probe of stellar mass throughout the TP-AGB phase. Using the isotope ratios measured by \citet{Hinkle-2016}, \citet{Decin-2020} found a significantly lower mass of $\sim1.35\,M_\odot$. 
Taking the uncertainties in the isotope ratio into account, a mass range of 1.0--1.5 $M_\odot$ can be estimated. We note, however, that \citet{Hinkle-2016} found a ratio of $1.0\pm0.4$ for $\chi$~Cyg, whereas \citet{DeNutte-2017} cites a ratio of $2.0\pm0.5$ for the same star, so the uncertainty in this method or in the isotope ratio could be even higher. 

We list these estimates in Table~\ref{table:proposed_solutions_lit} and mark them on Fig.~\ref{fig:heatmap_literature}. As the maps show, \citet{Decin-2020}'s estimate based on the $^{17}$O/$^{18}$O ratio clearly underestimates the mass of the star compared to the other solutions presented here: while the O1 ridge extends to masses that partially intersect this range, the favored mass ranges are significantly higher for both mode assumptions. 

On the other end of the mass scale, the mass estimated by \citet{Fadeyev-2023} is strongly disfavored by our models. This can be explained by the fact that \citet{Fadeyev-2023} assumed that the earlier ``hovering'' (standstill) in period corresponded to the end of the power-down phase, whereas recent observations show otherwise. 

The generic assumption of a 2\,$M_\odot$ model made by \citet{wood-zarro-1981}, \citet{Eggen-1985} and \citet{Zijlstra2002} falls comfortably within the FM and O1 ridges, but only at relatively low, sub-solar metallicities ([Fe/H] $\le -0.4$) for the FM models.

The masses we calculated from the period--mass relations of \citet{Trabucchi-2022} constitute a broad range that encompasses our best-fitting masses. We find that assuming 350\,d for the fundamental mode gives the most consistent mass values from this relation: 1.52--3.98\,$M_\odot$ (best fit at 2.72\,$M_\odot$), suggesting that the out-of-pulse ``equilibrium'' period for the star is close to 350\,d. The picture is more complicated for the first-overtone masses, since the mass--period and mass--age relations are calibrated by assuming fundamental-mode pulsation.

\citet{Homan-2021} assume that R~Hya is a multiple system, and the mass they provide is an estimate of the present-day mass of a central reservoir containing two components. While their reported mass range is comfortably consistent with many of our preferred models under either mode assumption, the appropriate comparison for \citet{Homan-2021}'s results would be to the end state of binary evolution calculations, which we do not compute.

\subsection{Metallicity limits}
As shown in Section~\ref{sect:kinematics}, the metallicity of R~Hya is possibly super-solar. Metallicities in the disk of the Milky Way extend to about [Fe/H] = 0.5--0.52, based on results from various large-scale spectroscopic surveys such as APOGEE \citep{APOGEE-2015}, GALAH \citep{GALAH-2022} and GES \citep[Gaia-ESO Survey, ][]{GES-2022}. Even higher metallicities have been claimed in the Galactic bulge \citep{Hill11,Bensby13}, but the kinematics suggest that R~Hya is a disk population star. To avoid folding assumptions about R~Hya's membership of particular populations into our analysis, we treat R~Hya's metallicity agnostically and explore a wide range of values centered on the solar [Fe/H], ranging from $-1.0$ to $+1.0$. The latter is the upper limit stars can theoretically reach, according to the calculations of \citet{Cheng-Loeb-2023}. 

For the FM assumption, we find good solutions at almost every value of [Fe/H] ($Z_\text{init}$) shown in Figures \ref{fig:heatmap_FM} through \ref{fig:heatmap_literature}, with the exception of the two lowest metallicities shown. However, calculations were also performed for [Fe/H]${}=-1.53$ and [Fe/H]${}=-2.2$ (not shown) in an exploratory grid (see Section \ref{sec:modelgrid}), and no model with either of these metallicities, regardless of mass, intersected any of the observational constraints. They were thus excluded, and no further efforts to study metallicities between [Fe/H]${}=-2.2$ and [Fe/H]${}=-1.2$ were made on the basis of lack of observational consistency. 

An ``exclusion zone'' set by an upper limit of $+0.52$~dex is indicated in Fig.~\ref{fig:heatmap_literature}. In the static helium case (left column of Figure \ref{fig:heatmap_literature}), the FM solutions extend into this regime for masses above $\sim4.0 M_\odot$, suggesting that such masses are not realistic. Perhaps unsurprisingly, such models generally fail due to convergence difficulties in the helium-varied grid (right panels of Figure \ref{fig:heatmap_literature}) when computing compositions with metallicities above this limit, as helium becomes nearly 50\% of the star's total composition. 

The O1 metallicity range is narrower, beginning at $-0.5$~dex for the lowest masses in the fixed-helium grid but extending well beyond the Galactic upper limit, encompassing models with metallicities as high as $+0.9$~dex. The upper third of the O1 solution space is thus excluded on this basis, ruling out masses $\ge2.6M_\odot$. 

\section{Non-linear and non-adiabatic considerations}
\label{sec:advanced_tactics}
For the sake of computational tractability and achieving the highest level of completeness for the open-source grid, we elected to run our models using the adiabatic assumption in GYRE. However, it has been known for decades that using the adiabatic rather than non-adiabatic assumption can impact pulsation periods for stellar models in this regime \citep{FoxWood1982, Trabucchi-2019, Zinn2023}. Similarly, there exists ample evidence in the literature that the linear approximation is not valid for large-amplitude pulsators (e.g. \citealt{Trabucchi2021}), though there is tension regarding which direction non-linear considerations should push the periods compared to the linear approximation. We address each of these considerations as follows. 

\subsection{Linear non-adiabatic models}
\begin{figure}
\centering
\includegraphics[width=\columnwidth]{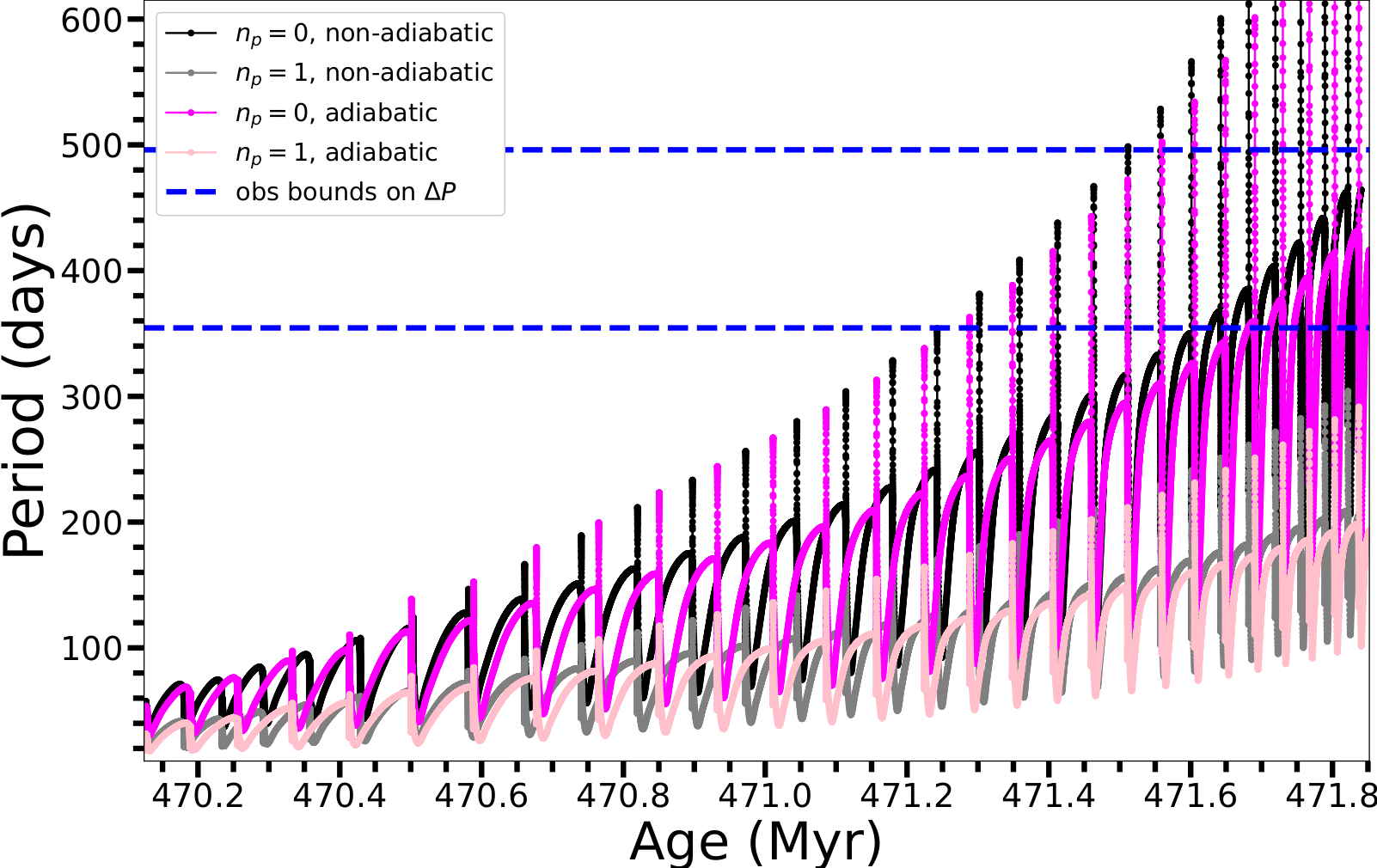}
\includegraphics[width=\columnwidth]{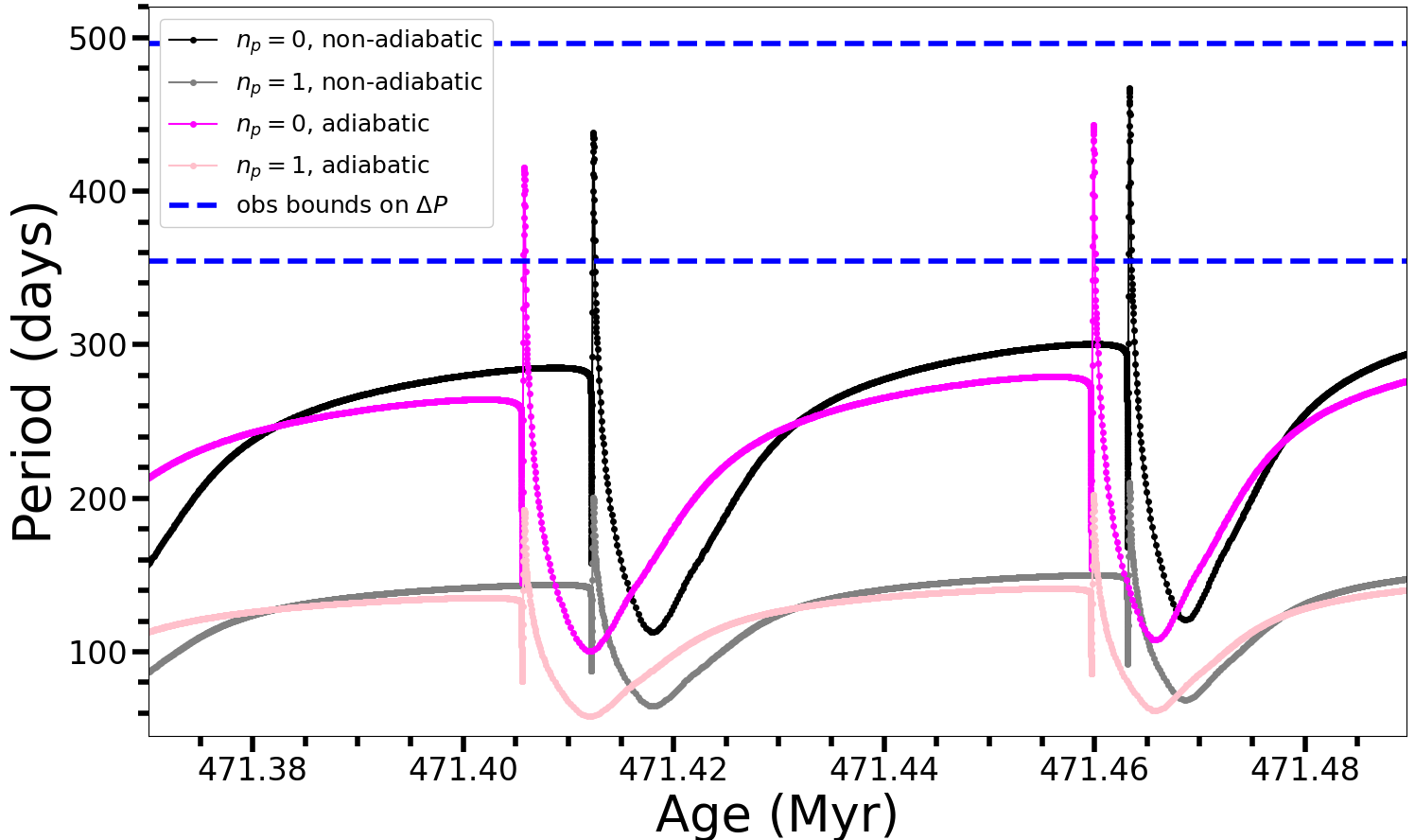}
\caption{A comparison between linear adiabatic and linear non-adiabatic calculations for a helium-varied model with $2.90 M_\odot$ and $Z=0.0216$ ($Y=0.294$) for the FM (magenta/black) and O1 (pink/grey) is shown. \textbf{UPPER:} We note that use of adiabatic (magenta) vs non-adiabatic (black) changes the pulse indices that satisfy the observational constraints, but does not have a strong impact on the predicted periods for the first 20 or more TPs. As the pulse index increases, the discrepancy between periods predicted with adiabatic vs non-adiabatic calculations increases. \textbf{LOWER:} The inter-pulse region is slightly shorter for the non-adiabatic calculations, and the period predictions are slightly larger when using non-adiabatic mode. The period difference is well within the tolerance of our uncertainties, and the change in duration of the inter-pulse region is more likely due to the (significantly) higher structural resolution needed to run GYRE in non-adiabatic mo
de. These temporal shifts only amount to a difference of one TP over the course of $\sim$30 pulses (see Section \ref{sec:Monash} for discussion of pulse index uncertainties.)    
 }
\label{fig:nad_vs_ad}
\end{figure}
In addition to being more physically correct representations of the conditions of AGB stars, non-adiabatic models can calculate the linear growth rates of modes, indicating which modes are unstable to small perturbations. Mode stability for red giant and AGB stars has been investigated by many authors \citep[see, e.g.,][]{Xiong-2007, Takayama-2013, Xiong-2018,Trabucchi-2019}. As we touched upon in Section~\ref{sec:twomodes}, in principle, non-adiabatic models could more definitively constrain R~Hya's pulsation mode. However, mode selection in the non-linear regime (i.e., observing which modes grow to full amplitude) and mode identification based on observations pose theoretical and practical difficulties. As Paper I demonstrated for T~UMi, stars might go through mode transitions over the course of a TP, including entering a double-mode state. Mapping the changes in the stability parameter over the entire grid would increase the computational requirements significantly. Theref
ore we elected not to constrain our grids with a growth rate criterion, but rather to map the full parameter space both for the FM and O1 modes. 

Given that non-adiabatic solutions may differ from adiabatic ones for stars like R~Hya, we recomputed a handful of our preferred models (described in more detail in Section \ref{sec:Monash}) using GYRE in non-adiabatic mode. The results of one representative model ($M=2.9, Z=0.0216, Y=0.294$) are shown in Figure \ref{fig:nad_vs_ad}. We note here that the difference in periods predicted from running GYRE in adiabatic (magenta/pink) mode vs non-adiabatic (black/grey) mode is not significant for the (FM) pulses that are observationally consistent with R Hya. The discrepancy widens as the star evolves, becoming more significant at later pulses, as the star's outer envelope becomes more stratified and the conditions become more strongly non-adiabatic. However, the difference does not, for any of the representative models explored, come close to exceeding the 60-d uncertainty already incorporated into the numerical experiment. The adiabatic vs non-adiabatic discrepancy is roughly t
he same for O1 as for FM periods.

Based on this marginal discrepancy of, typically, $\sim$20 days or fewer, we conclude that the conditions in models that fit R~Hya are only weakly non-adiabatic, and that the adiabatic approximation is valid for this particular star. Though the validity of this approximation does not extend to every pulse in every model over the entire grid, the substantial increase in computation time and structural resolution required to run GYRE in non-adiabatic mode is prohibitive. We therefore do not attempt to run a comprehensive grid in non-adiabatic mode, but we do provide the \texttt{GYRE} control file (\texttt{gyre\_non-ad.in}) used to perform non-adiabatic calculations in the repository.
We leave a more thorough description of these settings to the documentation on GitHub, but note that our non-adiabatic search uses the adiabatic eigenfrequencies  as the trial roots for the non-adiabatic case (as documented in description of the \texttt{`AD'} method for non-adiabatic calculations in the \texttt{GYRE} documentation). In this way, the adiabatic solutions presented in our grid still have utility regardless of how well the adiabatic approximation applies.

A limited alternative to running non-adiabatic calculations directly is provided by \citet{Xiong-2007} for the RGB. However, we found that their analytic approximation was (strongly) invalid for the AGB. \citet{Xiong-2007}'s stated domain of validity for the pulsation constant, $\log_{10} Q$, in their approximation is $\log_{10} Q = -1.30$ to $-0.9$, whereas calculated values of $\log_{10}Q$ for our models ranged from $0$ to $0.25$ along the AGB.

\subsection{A non-linear scaling relation}
\begin{figure}
\centering
\includegraphics[width=\columnwidth]{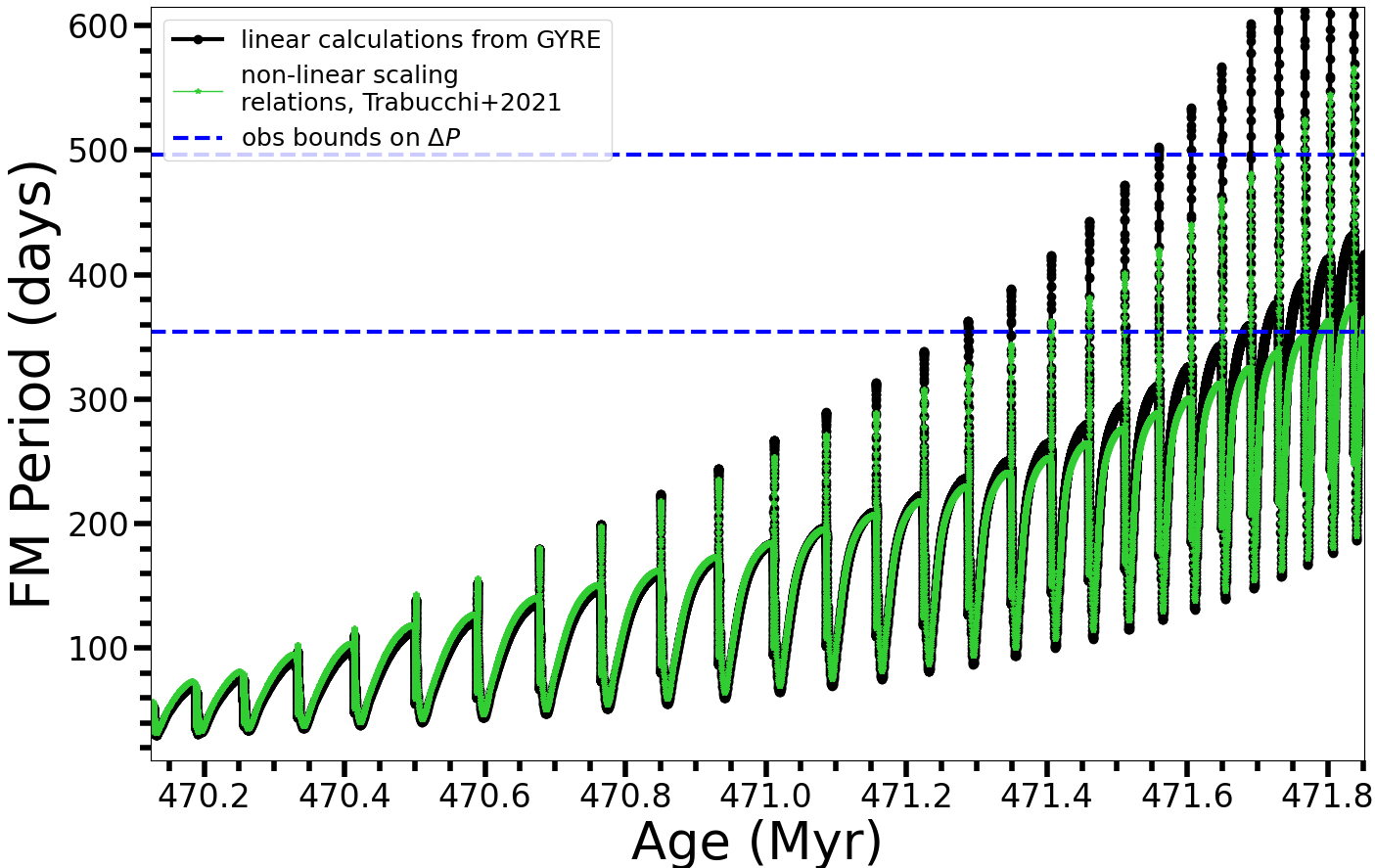}
\includegraphics[width=\columnwidth]{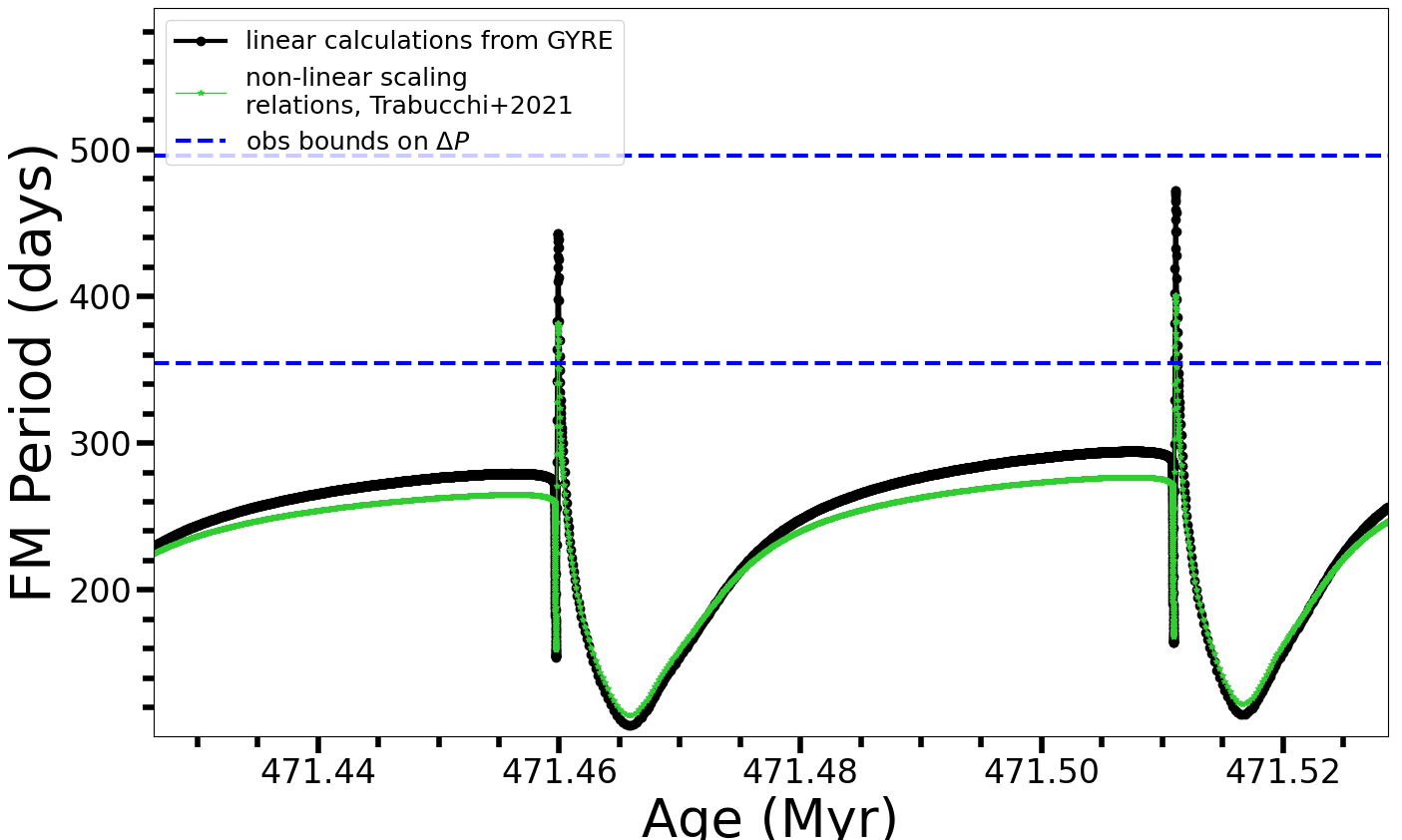}
\caption{A comparison between periods computed directly with linear, adiabatic assumptions in GYRE and periods inferred from the non-linear, adiabatic scaling relations of \citet{Trabucchi2021} for the fundamental mode period of a helium-varied model with $2.90 M_\odot$ and $Z=0.0216$ ($Y=0.294$) is shown. This is the same model presented in Figure \ref{fig:nad_vs_ad}. 
\textbf{UPPER:} The green curve with star-shaped markers shows the periods inferred using scaling relations. The black curve shows linear, direct calculations with GYRE. 
\textbf{LOWER:}  A zoom-in on the same pulses consistent with R~Hya's observed period change shown for the non-adiabatic case in Figure \ref{fig:nad_vs_ad}. 
We note that, as in the adiabatic vs non-adiabatic comparison, the discrepancy worsens as the star evolves. However, in this case, the discrepancy in period is potentially significant enough to affect our results in the early-to-mid pulses found to best fit R~Hya, not just at very late pulses.
The \citet{Trabucchi2021} scaling relations are calibrated to the FM, so we do not investigate the O1.
}
\label{fig:lin_vs_nonlin}
\end{figure}
\begin{figure*}
\centering
\includegraphics[width=\columnwidth]{FM_heatmap_altH_hardness75.pdf}
\includegraphics[width=\columnwidth]{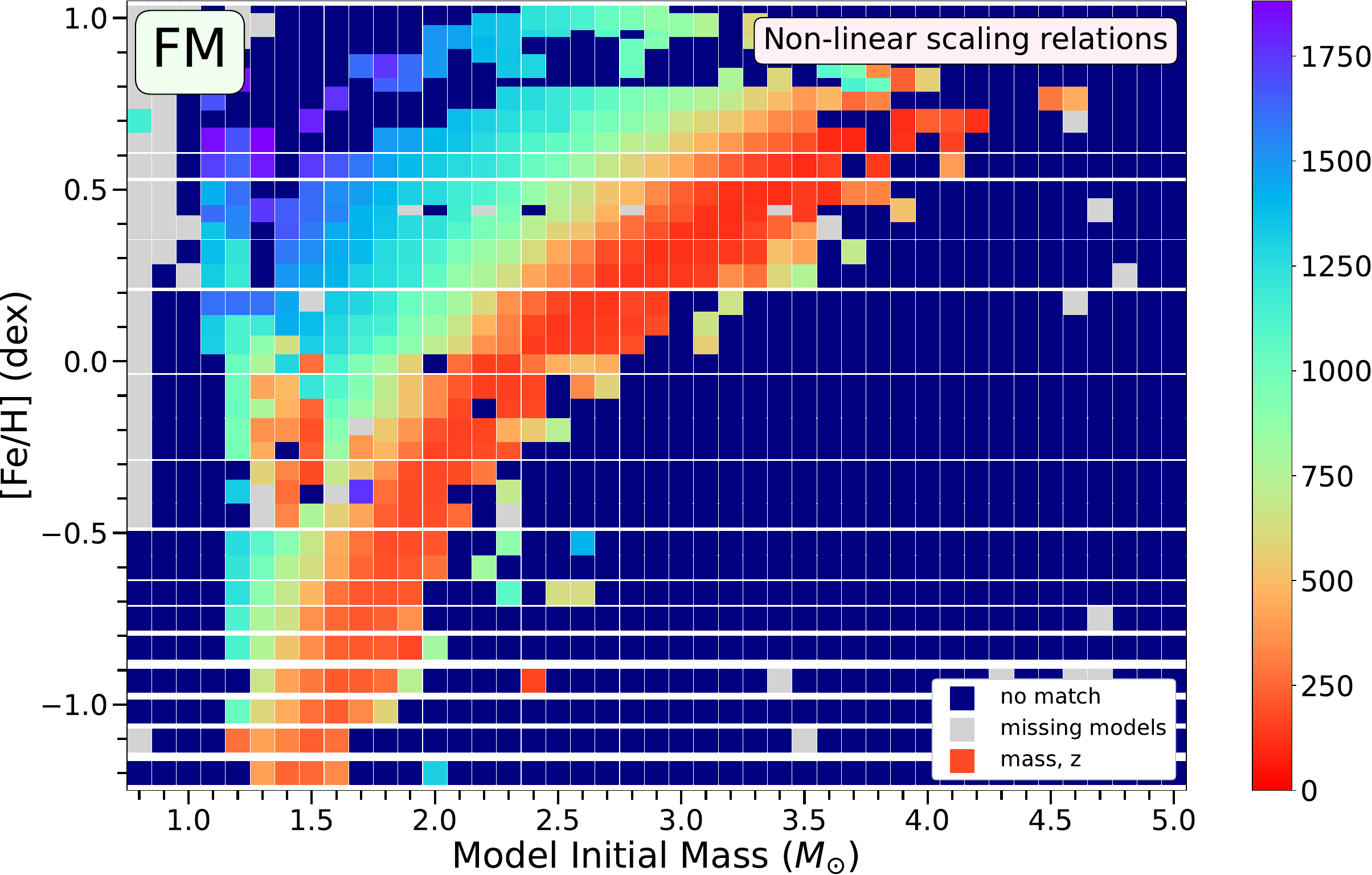}
\caption{The best fit heat map for R~Hya using the $S_W$ statistical realization from Figure \ref{fig:heatmap_FM} (Map F) is reproduced (left) alongside the same map computed using periods from the non-linear scaling relations of \citet{Trabucchi2021} (right), which are a function of present-day mass and radius only, instead of GYRE periods. The difference is most significant for models with lower initial mass and metallicity, but the shape and orientation of the best-fitting parameter regime is similar in both cases. While the non-linear models produce a more restricted fit space for R~Hya, there is a high degree of overlap (but not perfect overlap) between the solution spaces.
}
\label{fig:nonlin_heatmap}
\end{figure*}
As discussed in Paper I, working with linear periods introduces uncertainties in our model fits. While small-amplitude variations in less luminous stars can be approximated well with linear periods, large-amplitude pulsations rearrange the envelope structure of the star, shifting the observed, non-linear periods away from the linear values. However, the degree, or indeed even the direction, of this period shift has not been conclusively determined, and results vary depending on the non-linear pulsation models different authors use. For example, \citet{Yaari-1996} and \citet{lebzelter-wood-2005} found that the periods shorten in the non-linear regime, whereas periods lengthened in the models of \citet{kamath-2010} and \citet{Ireland-2011}. Dependence on initial mass and convective model prescriptions have also been documented \citet{olivier-wood-2005}. While \texttt{MESA} itself has a non-linear pulsation module included, the RSP (Radial Stellar Pulsations; \citealt{Smolec-201
6}) code was developed to reproduce less luminous RR Lyrae and Cepheid-type variations. \texttt{MESA-RSP} runs into dynamical instabilities at around 1000 solar luminosities, which is an order magnitude below the luminosity of R~Hya.

Besides the uncertainties in various parameter choices, the main drawback of non-linear models is that they are very computationally expensive. Because of this, non-linear studies typically rely on only a small set of models, making extrapolations to other physical parameter regimes difficult. More recently, however, \citet{Trabucchi2021} calculated a denser grid of models that made it possible to deduce scaling relations for fundamental mode period differences between linear and non-linear models. 
For these fundamental-mode pulsations, \citet{Trabucchi2021} found that non-linear periods begin to shorten relative to their linear counterparts above a break point in radius (or in period), and eventually saturate at a constant maximum period for the largest stars. This phenomenon would undoubtedly affect our models, in effect lowering peak periods and the hardness of the power down phase, and longer periods will be affected more. For overtone solutions, \citet{Trabucchi2021} found these differences to be negligible, therefore our conclusions for the first overtone model grid are not affected by non-linear shifts.

To investigate the severity of the impact on our results, we implemented the scaling relations of \citet{Trabucchi2021} into our post-processing algorithms and re-ran our analysis using the rescaled period evolution data in lieu of periods computed directly with GYRE. However, since this scaling also introduces many implicit assumptions that are specific to the particular non-linear models used by \citet{Trabucchi2021} (and there is tension in the literature regarding the direction and magnitude of linear vs non-linear trends), we consider these scaled results only as a comparison to the linear model grid. 

In Paper I and in the present study, we 
treated period constraints loosely, both to accommodate possible non-linear shifts and to account for the occurrence of TPs at discrete period values in the models that may or may not match the observed periods perfectly. However, in the example provided in Figure \ref{fig:lin_vs_nonlin}, we see that the discrepancy between linear and non-linear periods is roughly 70 days, which is slightly greater than our uncertainty allowance. The non-linear periods therefore differ enough from the linear calculations that it is worth investigating the impact on the global solution space. A side-by-side, linear vs non-linear comparison for the fundamental mode solution space is shown in Figure \ref{fig:nonlin_heatmap}. 

Globally, our results for R~Hya do not change significantly when using the non-linear scaling relations in lieu of direct calculations with GYRE. It is worth noting that using scaling relations, which predict non-linear periods as a function of present-day mass and radius only, is less computationally expensive than running GYRE. However, we reiterate that these scaling relations are based on case-specific calibrations from one particular non-linear code, and the assumptions adopted in \citet{Trabucchi2021} may or may not be valid across our entire parameter regime.
Rather than repeating the analysis of Section \ref{sec:discussion} for the right-hand panel of Figure \ref{fig:nonlin_heatmap}, it is sufficient to observe that 
(1) using the non-linear scaling relations does have the anticipated impact of pushing R~Hya's best-fitting models towards lower peak periods---especially for lower masses and metallicities; and
(2) it becomes more worthwhile to compute or extrapolate from non-linear models the more evolved (i.e. later in the pulse spectrum) the model is.

We proceed with our analysis using the linear and adiabatic calculations as the \textit{de facto} results, but explore the impact of non-linearity on best-fitting pulse indices in Section \ref{sec:Monash}.

\section{Insights from Nucleosynthesis}
\label{sec:nucleo}
\begin{table*}[ht]
\centering
    \begin{tabular}{cc cc cc c cc cc}
        \toprule
        \multicolumn{1}{c}{\textbf{$M_i$}} &
        \multicolumn{1}{c}{\textbf{$Z_i$}} &
        \multicolumn{1}{c}{\textbf{[Fe/H]}} &
        \multicolumn{1}{c}{\textbf{best pID}} &
        \multicolumn{1}{c}{\textbf{best pID}} &
        \multicolumn{1}{c}{\textbf{1$^\text{st}$ pulse}} &
        \multicolumn{1}{c}{\textbf{$\lambda_{\rm best p}$}} &
        \multicolumn{1}{c}{\textbf{C/O$_{\rm best p}$}} &
        \multicolumn{1}{c}{\textbf{\#TPs}} &
        \multicolumn{1}{c}{\textbf{C/O$_f$}} \\
\multicolumn{1}{c}{($M_\odot$)}
        & 
        & \multicolumn{1}{c}{(dex)}
        &
        \multicolumn{1}{c}{\textbf{$Y_i=0.3$}} &    \multicolumn{1}{c}{\textbf{$Y_i$ varied}} &    \multicolumn{1}{c}{\textbf{w TDU}} &  &
        &
        &
        &
        \\
&
        &
        &
        \multicolumn{1}{c}{\texttt{MESA}} &    \multicolumn{1}{c}{\texttt{MESA}} &    \multicolumn{1}{c}{{Monash}} &  \multicolumn{1}{c}{{Monash}} &  \multicolumn{1}{c}{{Monash}} &  \multicolumn{1}{c}{{Monash}} &  \multicolumn{1}{c}{{Monash}} \\ \hline
        \multicolumn{2}{c}{\textit{Primary FM ridge}} \\ \hline
      1.9 & 0.0018 & -1.02 & 12 & 10   & -&    -   &    -   &   -  & -  \\ \rowcolor{blue!10}[\tabcolsep]    2.0 & 0.0030 & -0.75 & 14  & 12  & 9 &  0.50 & 1.87$^{**}$  &  21 &  5.7 \\
     \rowcolor{green!25}[\tabcolsep]        2.1 & 0.0050 & -0.53 & 15 & 13  & 13 & 0.26 & 0.32  &  24 &  3.5 \\
        \rowcolor{green!25}[\tabcolsep]        2.4 & 0.0095 & -0.25 & 16 & 15  & 15 & 0.05 &  0.25 &  29 &  2.4\\
        2.7 & 0.0135 & -0.075 & 14 & 15  & 17& 0.00  & 0.25  &  33 &  1.91\\
        2.9 & 0.0216 & +0.11 & 15 & 16  & 18& 0.00 &  0.26 & 35 &  1.14 \\
           3.2 & 0.0300 & +0.25 & 14 & 14  & 15 &  0.00  &  0.26 & 35  &  1.02 \\
       \rowcolor{yellow!50}[\tabcolsep] 3.5 & 0.0400 & +0.39 & 12 & 7  & 11 & 0.08  & 0.26 & 32  &  0.68 \\
        \rowcolor{yellow!50}[\tabcolsep]    3.7 & 0.0500 & +0.49 & 9 & 6 & 9 & 0.01 &  0.26 &  28 &  0.44\\
        \hline
        \multicolumn{2}{c}{\textit{Secondary FM ridge }} \\ \hline
        1.2 & 0.0135 & -0.075 & 3 & 3 & none & 0.00  &  0.32 &  14 &   0.31 \\
        1.4 & 0.0300 & +0.25 & 4 & 4 & none & 0.00 &  0.30 & 16  & 0.30\\
        1.8 & 0.0600 & +0.57 & 7 & 2 & none  & 0.00  &  0.27 & 15 &  0.30\\
        \hline
        \multicolumn{1}{c}{\textit{O1}} \\ \hline
        1.5 & 0.0060 & -0.45 & 12 & 11 & 13 & 0.00 & 0.25 & 20 &  1.64\\
        1.7 & 0.0125 & -0.13 & 13 & 13 & none  & 0.00 &  0.27 & 20 &   0.30\\
        1.9 & 0.0216 & +0.11 & 15 & 15 & none & 0.00 &  0.27 & 25 & 0.30\\
        2.1 & 0.0344 & +0.32 & 18 & 18 & none & 0.00 &  0.27 & 30 & 0.30\\
        2.4 & 0.0600 & +0.57 & 20 & no match & none  & 0.00 &  0.27 & 28 & 0.30\\
        2.6 & 0.0800 & +0.70 & 22 & no match & none & - &      -      & -&    -\\  
      \hline
    \end{tabular}
    \caption{The models listed here (shown in Figure \ref{fig:Monash_cloned_models}) are replicated with the Monash stellar evolution code to determine whether third dredge-up is present during the thermal pulse that best fits R~Hya in each case. Where a model undergoes TDU, the pulse at which TDU begins is indicated. Where no TDU occurs in the model at any point, ``none'' is written.
All models assume a Bl\"ocker mass loss efficiency of $\eta = 0.01$ (corresponding to a related ``impact value'' of $\eta = 0.02$ in the Monash code's convention). Helium abundances are always varied per Equation \ref{eq:deltaY_deltaZ} in the Monash models; this assumption is likewise shown for \texttt{MESA}, along with the $Y_i = 0.3$ static assumption. 
We note that the best-fitting pulse index does change depending on the helium treatment used in the \texttt{MESA} grid, but in the majority of cases, not by more than two pulses. There are significant changes in three cases, all with very high metallicities ([Fe/H]$\gtrsim +0.4$): two along the primary FM ridge, highlighted in yellow (see Section \ref{sec:Monash} for more discussion), and the highest mass along the secondary FM ridge. 
Models that exhibit third dredge-up during the best-fitting pulse to R Hya are taken to be those for which the best-fitting pulse index (pID) exceeds the first pulse showing TDU in the corresponding Monash model. These are highlighted in green.
The quantity $\lambda_\text{best p}$ describes the ratio of mass dredged into envelope ($\Delta_{\rm dredge}$) to the change in the mass of the H-exhausted core (see \citealt{KarakasLattanzio2014}) at the best-fitting pulse (for the $Y_i = 0.3$ grid), and similarly for the carbon-to-oxygen ratio at this point (C/O$_\text{best p}$). The total number of TPs the Monash models undergo and their final C/O ratios are provided in the penultimate and last columns, respectively. In two cases, Monash models did not converge for the desired parameters, so no values are provided. 
\textbf{ $^{**}$The model highlighted in blue ($2.0M_\odot, Z=0.003$) undergoes TDU by the time of the best-fitting pulse, but it is also a carbon star by the time of that pulse and therefore not representative of O-rich R Hya.}
    }
\label{table:monash}
\end{table*}
Having determined the preferred parameter ranges for R~Hya using classical and seismic constraints, we turn now to detailed abundances and isotopic ratios. The abundances of AGB stars are sensitive to the convective mixing process known as third dredge-up. We discuss the observational constraints on third dredge-up here. In Section \ref{sec:Monash}, we analyze additional calculations performed with the Monash stellar evolution code, which is optimized and well vetted for AGB and third dredge-up calculations. 
Results for this section and Section \ref{sec:Monash} are summarized in Table \ref{table:monash}.

\begin{table*}[ht]
\centering
    \begin{tabular}{cc cc cc cc}
        \toprule
        \multicolumn{1}{c}{\textbf{$M_i$}} &
        \multicolumn{1}{c}{\textbf{$Z_i$}} &
        \multicolumn{1}{c}{\textbf{[Fe/H]}} &
        \multicolumn{1}{c}{\textbf{best pID}} &
        \multicolumn{1}{c}{\textbf{best pID}} &
        \multicolumn{1}{c}{\textbf{1$^\text{st}$ pulse}} &
        \multicolumn{1}{c}{\textbf{best pID, non-linear}} &
        \multicolumn{1}{c}{\textbf{best pID, non-linear}}  \\
\multicolumn{1}{c}{($M_\odot$)}      &                                    & \multicolumn{1}{c}{(dex)}          &
        \multicolumn{1}{c}{\textbf{$Y_i=0.3$}} &     \multicolumn{1}{c}{\textbf{$Y_i$ varied}} &  \multicolumn{1}{c}{\textbf{w TDU}} &         \multicolumn{1}{c}{\textbf{$Y_i=0.3$}} &     \multicolumn{1}{c}{\textbf{$Y_i$ varied}}    \\
&
        &
        &
        \multicolumn{1}{c}{\texttt{MESA}} &    \multicolumn{1}{c}{\texttt{MESA}} &    \multicolumn{1}{c}{{Monash}} &         \multicolumn{1}{c}{\texttt{MESA}} &    \multicolumn{1}{c}{\texttt{MESA}} \\   \hline
        \multicolumn{2}{c}{\textit{Primary FM ridge}} \\ \hline
                                       1.9 & 0.0018 & -1.02  & 12 & 10  & -  & 16  & 15 \\ 
    \rowcolor{blue!10}[\tabcolsep]   2.0 & 0.0030 & -0.75  & 14 & 12  & 9  & 16  & 15  \\
     \rowcolor{green!25}[\tabcolsep]   2.1 & 0.0050 & -0.53  & 15 & 13  & 13 & 14  & 15  \\
     \rowcolor{green!25}[\tabcolsep]   2.4 & 0.0095 & -0.25  & 16 & 15  & 15 & 17  & 17  \\
     \rowcolor{green!25}[\tabcolsep]   2.7 & 0.0135 & -0.075 & 14 & 15  & 17 & 21  & 22  \\
                                       2.9 & 0.0216 & +0.11  & 15 & 16  & 18 & 16  & 16  \\
     \rowcolor{green!25}[\tabcolsep]   3.2 & 0.0300 & +0.25  & 14 & 14  & 15 & 18  & 18  \\
     \rowcolor{green!25}[\tabcolsep]   3.5 & 0.0400 & +0.39  & 12 & 7   & 11 & 16  & 12  \\
     \rowcolor{green!25}[\tabcolsep]   3.7 & 0.0500 & +0.49  & 9  & 6   & 9  & 12  & -  \\
        \hline
        \multicolumn{2}{c}{\textit{Secondary FM ridge }} \\ \hline
                                       1.2 & 0.0135 & -0.075 & 3  & 3  & none & 10 & 10 \\
                                       1.4 & 0.0300 & +0.25  & 4  & 4  & none & 5  & 5 \\
                                       1.8 & 0.0600 & +0.57  & 7  & 2  & none & 8  & 4 \\
      \hline
    \end{tabular}
    \caption{The first six columns are the same as in Table \ref{table:monash}, but the remaining columns indicate the best-fitting pulse indices according to the non-linear scaling relations of \citet{Trabucchi2021} for the same models. Due to the low degree of completeness in the helium-varied grid and the slightly different mass-metallicity relation among non-linear models, almost all of the test models lacked best-fitting pulse information for their specific mass, Z combination. In this case, the best-fitting pulse index of a model immediately adjacent to a ``missing'' model is substituted. If there were no adjacent models, no value is provided. Note that the scaling relations only apply to the FM, so no O1 pIDs are provided for the non-linear calculations. As in Table \ref{table:monash}, the $2.0M_\odot, Z=0.003$ model (blue) is a carbon star.
    }
\label{table:nonlin_TDU}
\end{table*}

\subsection{Third dredge-up}
\label{subsec:tdu}
The third dredge-up (TDU) can occur following a thermal pulse, when the post-flash expansion results in the envelope cooling, becoming opaque, and allowing convection to develop in regions of the star that were previously radiative (e.g., \citealt{Herwig05}, \citealt{KarakasLattanzio2014}). This includes the now-dormant H-shell, which has been extinguished owing to expansion, and possibly the He-intershell region. If convection reaches into the He-intershell, then the products of partial He-shell burning can be dredged up to the stellar surface. These nucleosynthesis products include $^{12}$C, fluorine, and heavy elements formed by the \textit{s}-process (e.g., Tc, Ba, Pb; see review by \citealt{lugaro2023s}). The efficiency of TDU combined with the mass-loss rate determine the stellar yields of AGB stars and their contribution to Galactic chemical evolution (e.g., \citealt{Kobayashi20}).

The occurrence and efficiency of TDU is sensitive to many factors, including the initial stellar mass and metallicity, the mass of the convective envelope, and the strength of TPs, where the latter two quantities change with time on the AGB (e.g., \citealt{Karakas02}). TDU is also sensitive to the input physics and numerics used within a stellar evolutionary model, in particular on the treatment of convection and convective boundaries  (e.g., \citealt{Frost96}, \citealt{Herwig97}). Theoretical models show that minimum stellar mass for TDU is particularly sensitive to these choices.

There are a few trends in theoretical models that have been consistently corroborated by various authors: the onset of TDU occurs at a lower initial mass in metal-poor AGB models, and the efficiency of TDU increases with decreases in $Z$ and/or increases in $\alpha_{\rm MLT}$, for a given mass (e.g., \citealt{Boothroyd88}, \citealt{Vassiliadis93}, \citealt{Straniero97}, \citealt{Karakas02}, \citealt{weiss2009new}).

There is still debate regarding the efficiency of TDU in intermediate-mass AGB stars over 4$M_{\odot}$ (e.g., \citealt{Karakas12}, \citealt{marigo2022updated}), and the metal-rich regime
was relatively unexplored for AGB stars, until the work of \citet{Cinquegrana2022}
(although see \citealt{Marigo17}).
There may be some evidence that the occurrence of TDU is (at least partly) stochastic in nature, based on observations of post-AGB stars that do not show TDU where expected and vice versa \citep{kamath2017discovery,Kamath2023}.

Unequivocal evidence of TDU is the detection of Tc in a star's spectrum, given that Tc is radioactive with a half-life of $\sim 200,\!000$~years. Most of the AGB stars with detected Tc are S-type or C-type AGB stars that also have enrichment in carbon \citep[e.g.,][]{shetye19,Abia2022}.  However, there are a few M-type (oxygen-rich) AGB stars with Tc detected, such as those described in \citet{Uttenhaler2010}. Most AGB stars with detected Tc are disk AGB stars with metallicities close to solar or a few times below (e.g., \citealt{shetye19}). Few AGB stars have been observed with super-solar metallicities, given the rarity of metal-rich stars in the disk and the short lifetimes of stars in the TP-AGB phase (e.g., \citealt{Karakas14He}). To have a well-characterized, metal-rich AGB star is crucial to constraining our theoretical models (e.g., \citealt{Cinquegrana2022}). 
The ambiguous detection of Tc---and by implication, TDU---in R~Hya thus motivates deeper theoretical investigation, as R~Hya may be just such an example.

\subsection{Isotopic Ratios} 
\label{subsec:isotopic_ratios}
Another source of evidence for TDU is isotopic ratios. During the TDU, the convective envelope moves inwards (towards the stellar interior) and penetrates into regions of prior He-burning. Consequently, changes in surface composition post-TDU reflect an increased concentration of $^{12}$C on the stellar surface. Very minor changes to $^{16}$O and $^{17}$O are expected; any $^{16}$O produced by partial He-shell burning is modest, on the order of 1\% by mass (e.g., \citealt{Boothroyd88}), which is not sufficient to alter the $^{16}$O/$^{18}$O ratio in a solar or metal-rich star (see also \citealt{karakas2010extra}). H burning can produce $^{17}$O, and $^{17}$O is not destroyed by He burning. In contrast, $^{18}$O can be both created and destroyed by He-shell burning. Production of $^{18}$O occurs through $\alpha$-capture onto $^{14}$N via the reaction $^{14}$N($\alpha,\gamma$)$^{18}$F($\beta^{+}$)$^{18}$O, where $^{14}$N is abundant in the intershell region at the beginning of 
a thermal pulse.  Destruction of $^{18}$O primarily occurs via $\alpha$-capture via the reaction $^{18}$O($\alpha,\gamma$)$^{22}$Ne, although if protons are available in the intershell then captures on $^{18}$O can lead to the formation of $^{19}$F via the $^{18}$O($p, \alpha$)$^{15}$N($\alpha, \gamma$)$^{19}$F chain. 

The $\rm ^{12}$C/ $\rm ^{13}C$, $\rm ^{16}$O/$\rm ^{17}O$, and  $\rm ^{16}$O/$\rm ^{18}O$  ratios are particularly important indicators of TDU in TP-AGB stars. 
\citet{Hinkle-2016} measured the following isotopic abundances in R~Hya: $\rm ^{12}$C/ $\rm ^{13}C  = 26 \pm 4$, $\rm ^{16}$O/$\rm ^{17}O  = 635 \pm 150$ and $\rm ^{16}$O/$\rm ^{18}O  = 343 \pm 200$. Both the $\rm ^{12}$C/$\rm ^{13}C$ and $^{16}$O/$^{18}$O ratios indicate that the star may have undergone TDU episodes, but the evidence is not conclusive.

R Hydrae is O-rich \citep{Lockwood-1969, Wallström-2024} and within $-1.0,+0.5$~dex of solar metallicity, so modelling the star based on a solar-scaled initial abundance is a reasonable assumption. Doing so for metallicities between $Z=0.014$ and $Z=0.08$ leads to an initial $\rm ^{12}$C/$\rm ^{13}C$ of 21 at the first thermal pulse. TDU can mix primary $\rm ^{12}$C to the stellar surface, increasing the surface $\rm ^{12}$C/$\rm ^{13}C$ ratio from its value at the first TP. A modest increase, e.g., 21 to 26 over $\sim$ 10 pulses, can occur if the initial gas is substantially enhanced in metals. However, the uncertainty on the C ratio is $\pm 4$, so this is not strong evidence that TDU has occurred. 

The $^{16}$O/$^{18}$O ratio in R Hydrae is quite low at $343 \pm 200$. Solar-scaled theoretical models with an initial metallicity close to solar (and 2--4$M_{\odot}$) will produce an $^{16}$O/$^{18}$O surface ratio of $\sim 600$--$700$ at the first thermal pulse.
Minor decreases in this ratio occur with TDU; we can see more substantial decreases in the ratio in some models if the $^{18}$O undergoes proton capture to form $^{19}$F via $^{18}$O($p,\alpha$)$^{15}$N($\alpha,\gamma$)$^{19}$F (see \citealt{Lugaro04reaction, Cristallo09, Jorissen92}). 

Significant $^{19}$F production does occur in our most metal-rich models, which leads to an $^{16}$O/$^{18}$O ratio of the appropriate magnitude. However, to the best of our knowledge, F has not been measured in R Hydrae. While F-bearing molecules have been observed in the atmospheres of similar carbon stars \citep{Jorissen92,Abia2015,Abia19}, R~Hya is not thought to be a carbon star, which our calculations corroborate (see Section \ref{sec:Monash}). Likewise, \citet{Wallström-2024} reports no detection of the molecule AlF in R~Hya in their recent study.   

Third dredge-up is not the sole factor in determining the abundance of \textit{s}-process elements on the stellar surface. More metal-rich gas contains a greater number of Fe seed nuclei on average, and the total neutron exposure is proportional to the number of neutrons per Fe seed \citep{clayton83}. Consequently, there are fewer free neutrons available to participate in the \textit{s}-process and produce $^{99}$Tc. Although the presence of $^{99}$Tc indicates that TDU has occurred recently, the absence of it does not imply the opposite. In short, R~Hya's abundances provide some evidence that it has undergone TDU, but we cannot make a definitive claim of TDU based on this information alone. 

\section{Comparison with the Monash Code}
\label{sec:Monash} 
\begin{figure*}
\centering
\includegraphics[width=\columnwidth]{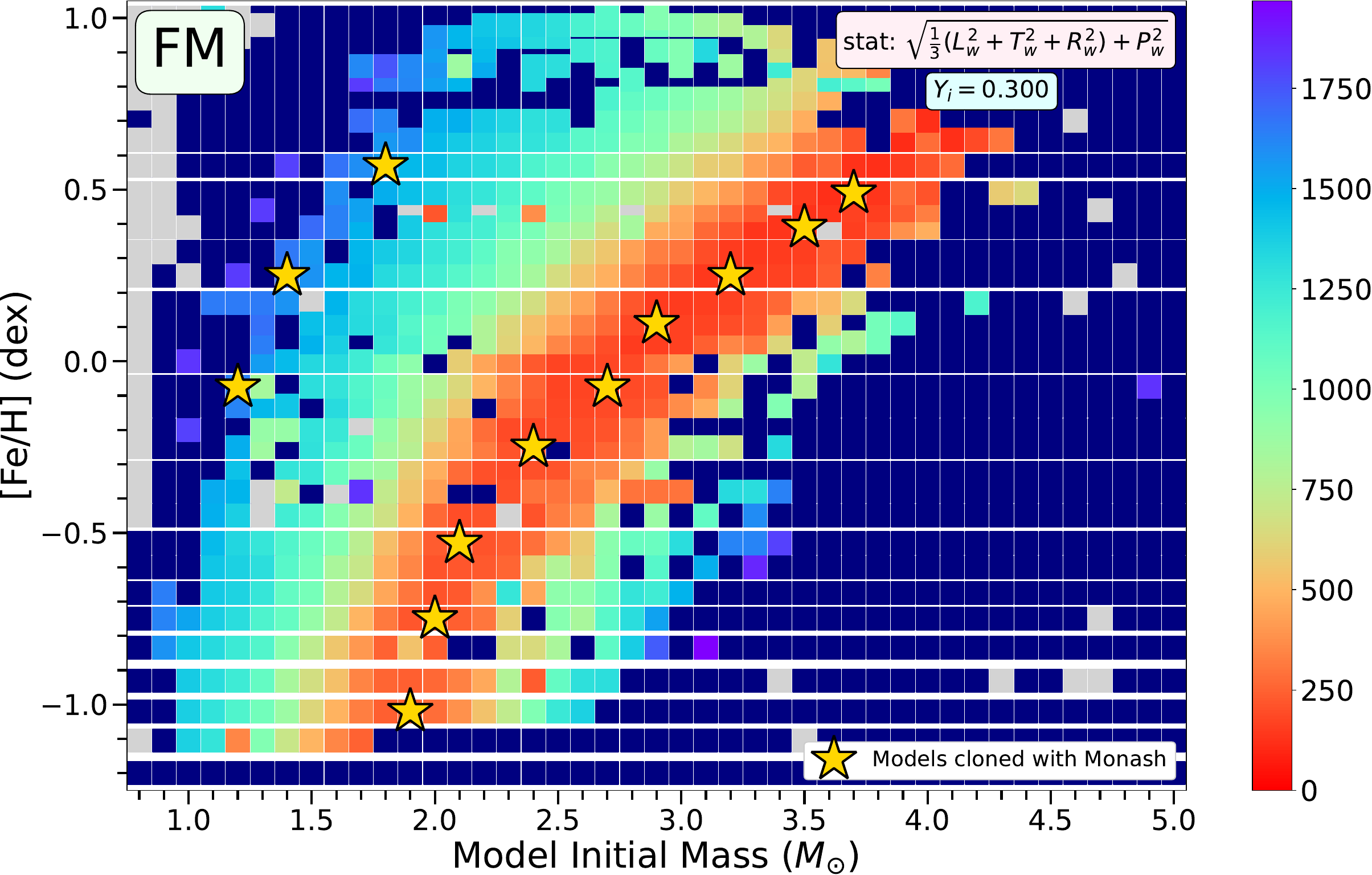}
\includegraphics[width=\columnwidth]{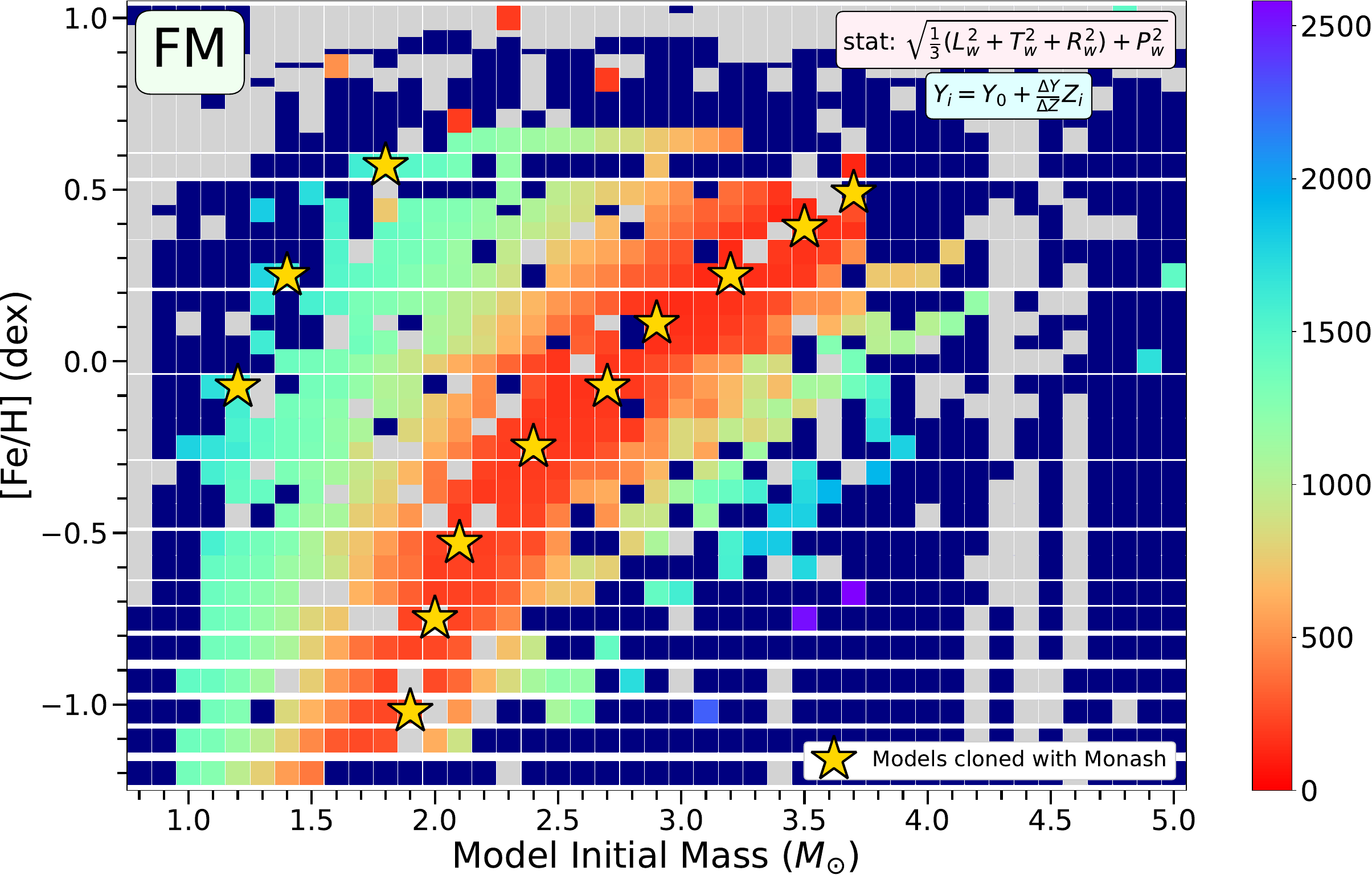}
\includegraphics[width=\columnwidth]{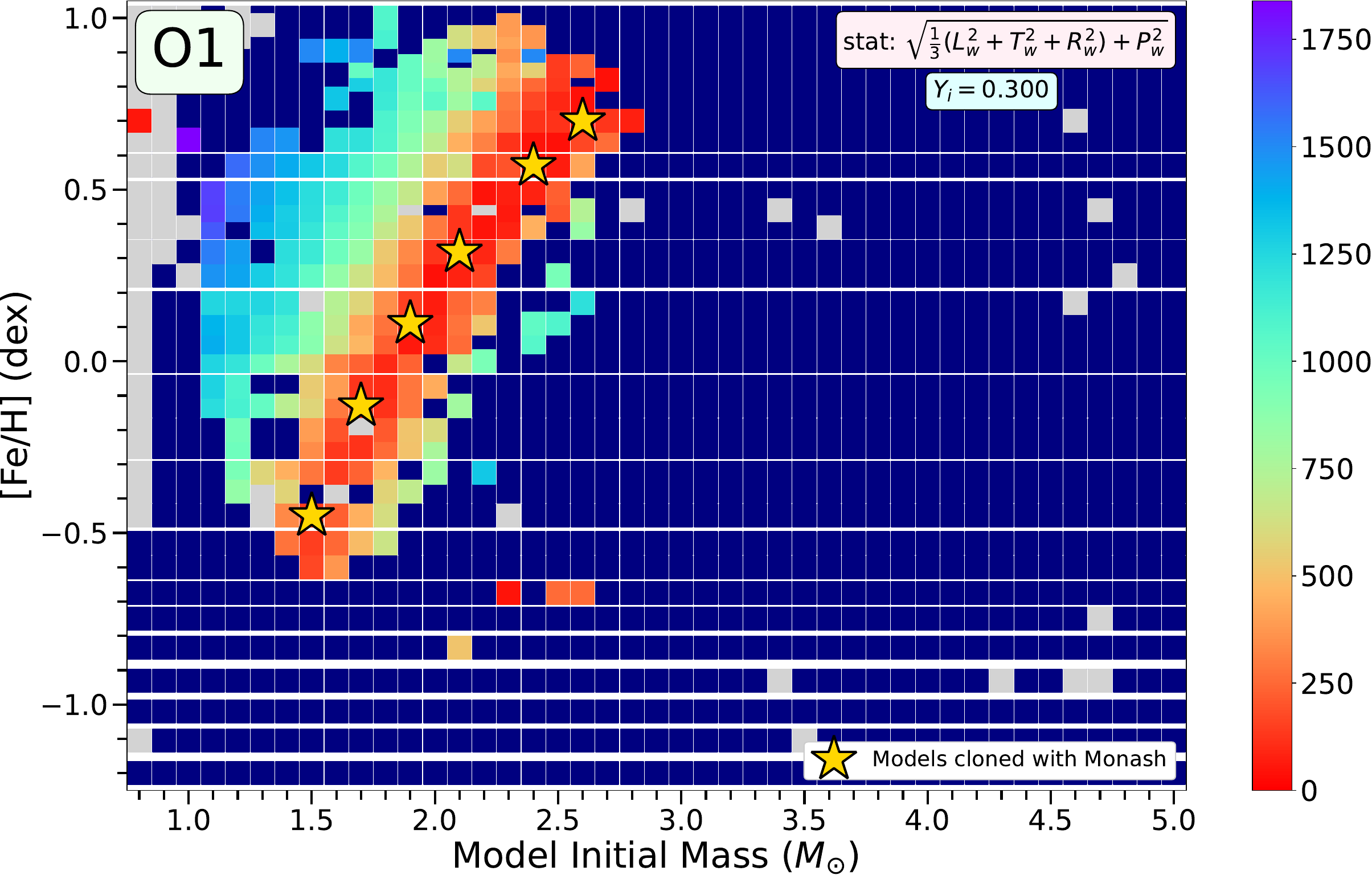}
\includegraphics[width=\columnwidth]{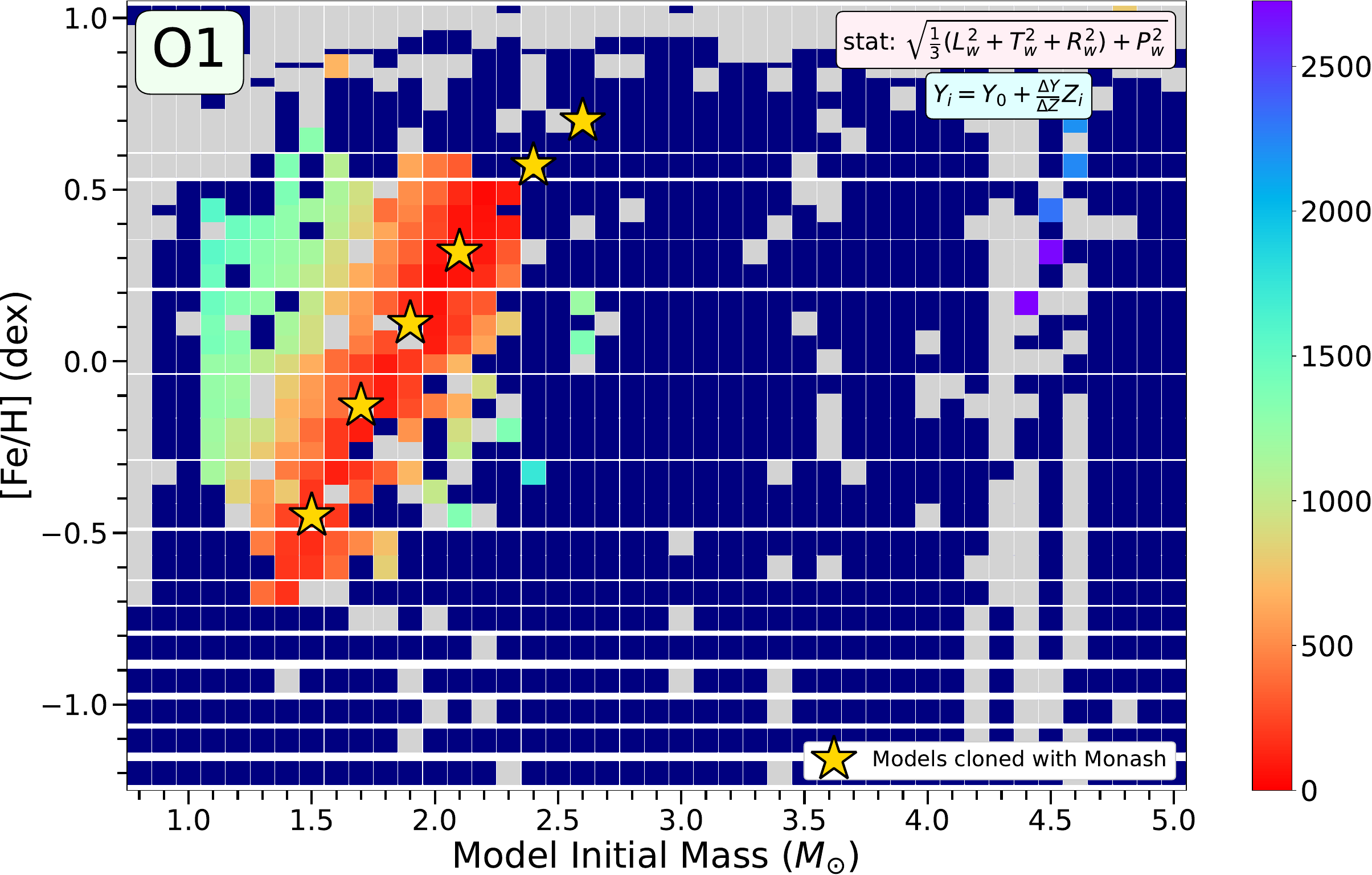}
\caption{ Gold stars indicate parameter combinations for which Monash ``clone'' models were run and results analyzed. On the left, the selection is shown overlaid on the $S_\text{W}$ (F) heat maps of Figures \ref{fig:heatmap_FM} (FM) and \ref{fig:heatmap_O1} (O1). On the right, the same is shown for equivalent heat maps generated from the $Y_i \sim Z_i$ grid.  
}
\label{fig:Monash_cloned_models}
\end{figure*}
To explore the question of TDU in R~Hya further, we turn now to additional theoretical calculations. We recompute (or ``clone'') a selection of the preferred models for R~Hya with the Monash stellar evolution code (an adaptation of the Mount Stromlo Stellar Evolution code; \citealt{LattanzioThesis, Lattanzio86, Frost96, Karakas07}), which has been extensively tested in the AGB star regime and produces detailed information on TDU and related parameters. \textbf{While modeling TDU in \texttt{MESA} is possible, \texttt{MESA} is not optimized for these calculations and does not by default provide reliable predictions of TDU efficiency (see \citealt{Rees2024}).} The models cloned with Monash are described in Table \ref{table:monash} and indicated with gold stars on Figure \ref{fig:Monash_cloned_models}. Nine FM and six O1 models were chosen to represent the ridges described by Equations \ref{eq:FM_scaling} and \ref{eq:O1_scaling}, and three additional FM models were chosen to trac
e the secondary FM ridge present in the $P_\text{W}$ and $H_\text{W}$ realizations shown in Figure \ref{fig:heatmap_FM} (Maps D and E, respectively).

The input physics for these calculations is the same as described in \citet{Cinquegrana2023}. Briefly: we adopt the initial masses and metallicities outlined in Table~\ref{sec:Monash} and the helium scaling rule described in Equation~\ref{eq:deltaY_deltaZ}: $Y_i = Y_0 + \frac{\Delta Y}{\Delta Z} \times Z_i$. We use solar-scaled initial abundances based on \citet{Lodders03}, the OPAL high-temperature opacities \citep{IglesiasRogers1996} and custom low-temperature opacities generated with the \AE SOPUS tool \citep{MarigoAringer2009, marigo2022updated}. Convection is treated with the mixing length theory of convection \citep{Prandtl25, Vitense53}. Boundaries between convective and radiative regions are defined using the method of \citet{Lattanzio86} and \citet{Lattanzio96}. We adopt the mass loss rates of \citet{Reimers1975} for the first giant branch and \citet{Bloecker1995} for the AGB. In \citet{Cinquegrana2023}, we calibrated the \texttt{MESA} inlists (also used in this work
) to mimic the physical output of the Monash code. Models are run from the zero-age main sequence to the tip of the TP-AGB or where convergence issues halted evolution. 

Table \ref{table:monash} provides the indices of the best-fitting pulses for a selection of \texttt{MESA} models that trace the ridges of good solutions, shown in Figure \ref{fig:Monash_cloned_models}. Results for both helium assumptions are included. The table also provides detailed TDU results from the Monash clone models.

The key indicator for understanding the likelihood of recent TDU in R~Hya is whether TDU in the Monash model begins at an earlier or later pulse than the pulse found to best fit R~Hya with \texttt{MESA}: third dredge-up is assumed to be ongoing after it has commenced. For three mass and metallicity combinations, highlighted in green in Table \ref{table:monash}, the best-fitting pulses are later than the first pulse with third dredge-up, regardless of helium assumption:
($2.0 M_\odot$, [Fe/H] $=-0.75$), ($2.1 M_\odot$, [Fe/H] $=-0.75$) and ($2.4 M_\odot$, [Fe/H] $=-0.25$). 

When the $Y_i\sim Z_i$ helium treatment is used, there are two cases in which \texttt{MESA}'s best-fitting pulse switches from following to preceding the pulse at which TDU begins: ($3.5 M_\odot$, [Fe/H] $=+0.39$) and ($3.7 M_\odot$, [Fe/H] $=+0.49$). These are highlighted in yellow in Table \ref{table:monash}. We note that these cases involve very super-solar metallicities, which increases their uncertainty. Nonetheless, we take the best pulse IDs for these models to be the ones extracted from the helium-varied grid, as this grid uses the same treatment as the Monash models. 

A general trend demonstrated by Table \ref{table:monash} is that all of the models selected from the primary FM ridge will eventually undergo TDU, and nearly all of the rest will not (with the exception of 1.5 $M_\odot$, [Fe/H]$=-0.45$). Given the well-known sensitivity of TDU to modeling choices (as discussed in Section \ref{subsec:tdu}), and the large uncertainties in AGB calculations, we caution against reading too much into specific pulse indices. As the models highlighted in yellow in Table \ref{table:monash} demonstrate, there are many things that could shift the pulse index found to best fit R~Hya. 

The precariousness of individual pulse identification is further underscored by Table \ref{table:nonlin_TDU}, which shows the best-fitting pulse indices when using the non-linear period scaling relations of \citet{Trabucchi2021} instead of GYRE periods. In general, the best-fitting pulses shift later in the non-linear case compared to the linear case. In this scenario, all but one preferred model shows the onset of TDU occurring before the best-fitting pulse.
If the assumptions made by \citet{Trabucchi2021} do apply across this regime of the parameter space, such calculations provide even stronger evidence of recent TDU in R~Hya.

In addition to the fact that the non-linear results clearly favor a recent TDU scenario, we emphasize that, along the primary FM ridge, all of the best-fitting pulse indices for R~Hya inferred under our standard assumptions are within 4 pulses of undergoing third dredge-up, regardless of helium prescription. The majority are within two. When considering the large uncertainties in AGB models and the fact that TDU is achieved somewhere---and always close to the best-fitting pulse---for all of the parameter combinations along this ridge, a claim of recent TDU in R~Hya is certainly consistent with our results. Our conclusion based on our models computed under a wide range of assumptions is thus that R~Hya has undergone TDU in its most recent pulse.

\section{The Past and Future of R~Hya}
\label{sec:moneysec}
\begin{figure*}
\centering
\includegraphics[width=\textwidth]{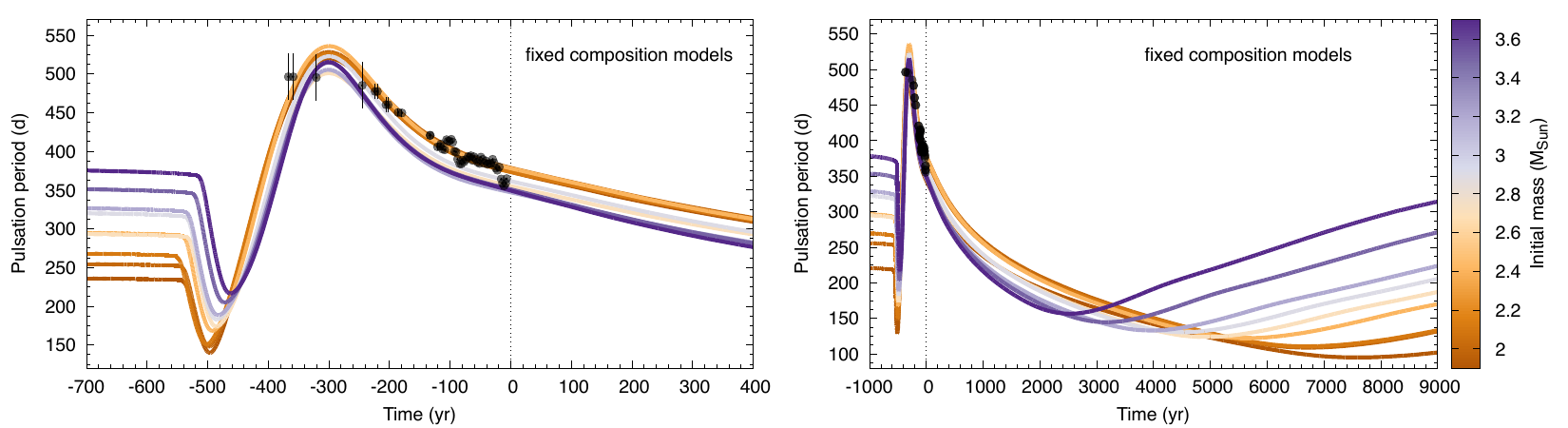}
\includegraphics[width=\textwidth]{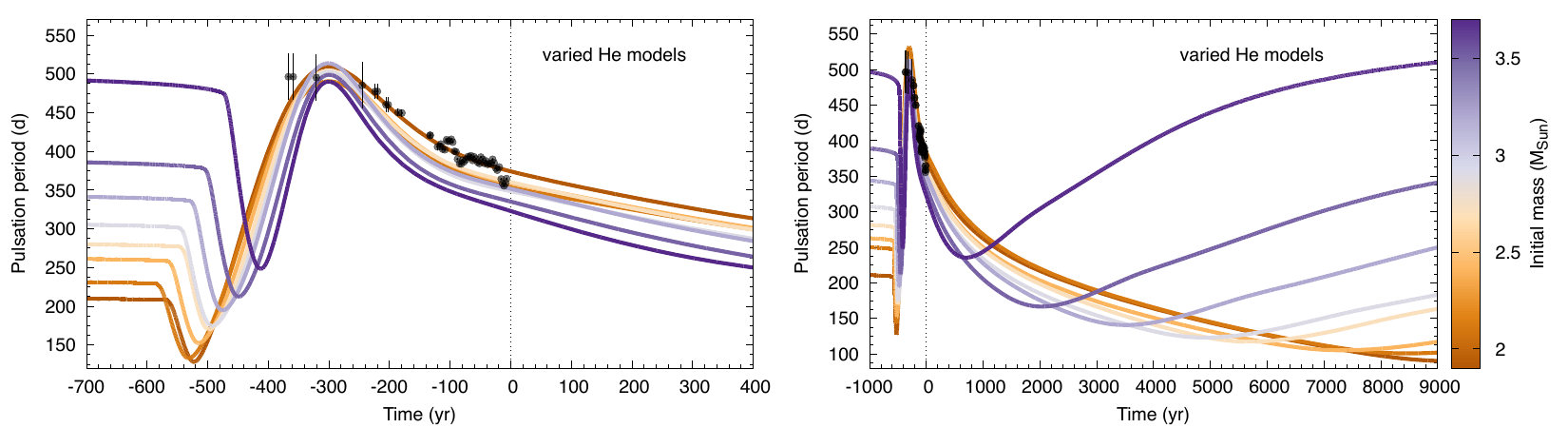}
\caption{Selected TP models (lines) that match the observed period evolution of R~Hya (black dots) best. Colors indicate initial mass. All models are shifted such that maximum period occurs at --300 yr, whereas zero corresponds to the current year, 2024, relative to the observations. Upper panels show models with fixed composition, whereas lower panels show models with varied $Y_i$ (He abundance). Left panels show the initial phase of the TP while right panels show the power-down phase in its entirety.  
}
\label{fig:money_plot}
\end{figure*}
To give a more direct comparison of the preferred solutions with the observations, we plot the period evolution of the FM modes from the nine primary ridge models listed in Table~\ref{table:monash} and overlay the observed period evolution of R~Hya in Fig.~\ref{fig:money_plot}. Here, we show pulses that match the period evolution best (i.e., $P_\text{W}$ rather than $S_\text{W}$). Pulses are shifted in time so that the maximum expansion phases are aligned, as in Figure \ref{fig:pulse_alignment}. Zero marks the current year, 2024. The top row corresponds to the fixed composition models, and the bottom row shows the same for models with adjusted He enrichment.  

Figure \ref{fig:money_plot} confirms that pulses that fit the observations very well can be found over a wide parameter range, between 1.9 to 3.7 M$_\odot$. Higher-mass models offer either poor matches, in accordance with the higher $P_\text{W}$ scores at the high-mass, high-metallicity end in Figure~\ref{fig:heatmap_FM}, or failed to converge (for the He-varied models). Overall, the He-varied models in the lower panels show greater variance than the fixed-composition models, especially before and after the power-down phase. Based on the left-hand panels of Figure~\ref{fig:money_plot}, the TP started about 460--500~yr ago (470--580 for~yr for He-varied), in agreement with the predictions of \citet{wood-zarro-1981}. The initial period strongly depends both on mass and composition, ranging between 230--380 (210--490)~d. This is in agreement with the assumption made in Section~\ref{sec:mass} regarding the out-of-TP period for the period--mass relation calculations. 

The onset (knee) phase ends about 45--50 (40--60) years after the shell flash in every model. Our calculations thus confirm that maximum expansion happened during the 18th century, when data on this star were sparse. 
R~Hya's maximum expansion therefore coincides with or happened shortly before the generation of the dust that now forms a detached dust shell around the star \citep{ZhaoGeisler2012}. 

Looking forward, we expect the period of the star to continue to decrease, although at a slower rate. For the next few hundred years, we estimate the long-term period change, $\dot{P}$, to be around $-0.17$ to $-0.20$~d/yr. This decrease will continue for a long time: as the lower panel of Figure~\ref{fig:money_plot} illustrates, it will be at least another millennium, but more likely several thousand years, before the pulsation period of the star reaches its minimum of around 100--150 (90--230)~d. This, of course, assumes that the star does not go through a mode switch. Nevertheless, the time scale of the power-down in our simulations is an order of magnitude longer than the values predicted by \citet{Fadeyev-2023} based on their model fits. It will also take an order of magnitude longer for future astronomers to confirm our predictions for R~Hya than to confirm our predictions for T UMi from Paper I.

\subsection{Comparison to T Ursae Minoris}
\label{subsec:TUMi}
This paper is the conceptual successor to the seismic--evolutionary modeling of T Ursae Minoris pioneered by \citet{TUMi}. Here, we briefly discuss the similarities and differences between the two studies.

In Paper I, we detected the onset of double-mode pulsation induced by the structural changes T UMi experienced due to a TP. This allowed us to model not only the period evolution of a single mode (as was done here), but the evolution of the mode period ratio as well. This was highly constraining for T UMi and enabled the derivation of very precise estimates for its fundamental parameters.

Furthermore, T UMi was at the onset of the TP, going through the ``knee'' phase (rapid radial decline) over the last few decades. The knee is where the most drastic changes in stellar parameters take place within a TP, and the (severe) slope of the period change was another constraint we could exploit. Within a few decades, the knee phase will provide another clue: the pulsation period is expected to reach its minimum value in the next 10--60 years, with the time span being sensitive to the physical parameters---especially the initial mass---of the star.  

For R~Hya, the knee phase is neither captured nor constrained by observations. As Figure~\ref{fig:money_plot} shows, while the onset of the TP is confined to about 50 years, and the duration of the knee is about a century, for every mass, the initial periods at the beginning (top) of the knee span a wide range.
Yet, the different periods still lead to very similar power-down phases. We can thus conclude that stars in the knee of a thermal pulse are easier to characterize with our method than stars in later sub-phases of the pulse. 

The rapid evolution of T~UMi allowed us to put forth predictions for its period evolution that are verifiable within a human lifetime. In contrast, all signs point to R~Hya's having already entered a slower phase of the TP, for which our period predictions can only be tested thoroughly in a few millennia. We can, however, make another prediction: the majority of our models suggest that the star already entered the TDU phase of the TP-AGB. This can and should be tested with more sensitive spectroscopic observations to determine, conclusively, whether $^{99}$Tc is present at the surface of R~Hya. 

\section{Conclusions}
\label{sec:conclusions}
The observational, asteroseismic, theoretical, and chemical characterization of R~Hydrae is summarized in Table \ref{table:fundpar}. We enumerate our primary findings as follows:
\begin{enumerate}
\item We presented new period measurements and found that the pulsation period and amplitude of R~Hya have continued to decrease since the late 2000s, after experiencing an extensive plateau caused by period meandering (Section \ref{sect:per-evo}).  

\item We have assessed the goodness-of-fit to R~Hya of more than 3000 asteroseismic, TP-AGB stellar models according to six different statistical metrics, two assumptions for its observed pulsation mode (Section \ref{sec:twomodes}), and using two different treatments of initial helium (Sections \ref{sec:fittingprocedure}). 
We showed that R~Hya's best-fitting initial parameters are nicely described by a quadratic scaling relation 
${\rm [Fe/H]} = -0.26 M_\text{init}^2 + 2.22 M_\text{init} - 4.17$
over the range $1.6$--$3.7M_\odot$ for the fundamental mode (Section \ref{subsec:fundpar}). This relation applies to models computed under linear, adiabatic assumptions, which we demonstrate to be reasonable for R~Hya. 

\item We explored the impact on our results of using GYRE in non-adiabatic mode and using non-linear period scaling relations from \citet{Trabucchi2021} to predict pulsation periods (Section \ref{sec:advanced_tactics}). In the former case, a small selection of preferred models was investigated, and the period discrepancy compared to the adiabatic assumption was below our uncertainty threshold, suggesting that the conditions of R~Hya's outer layers are only weakly non-adiabatic. In the latter case, we recomputed the entire solution space and found a slight shift towards lower mass, lower metallicity solutions, affecting especially the lowest mass and metallicity models. The non-linear calculations produced somewhat higher best-fitting pulse indices (Section \ref{sec:Monash}, Table \ref{table:nonlin_TDU}), which provides evidence in support of the conclusion that third dredge-up has taken place in R~Hya's most recent TP.

\item We have shown that the FM is preferred over the O1 for R~Hya's pulsation mode (Section \ref{sec:discussion}) and released a grid containing exact solutions for both of these modes as well as the next 8 radial orders ($n=2,\ldots,9$) across the entire parameter space studied here, for use by the community (Section \ref{sec:visualizer}; Appendix \ref{appendix:visualizer}).  These results are computed using the linear and adiabatic assumptions. While we have shown that both of these simplifications are reasonably valid for R~Hya, we recognize their limitations, especially for later TPs, as conditions become increasingly non-adiabatic and the structure becomes more complicated, thereby enabling more complex convection--pulsation interactions and higher mode amplitudes in the envelope. We provide the necessary infrastructure and evolutionary data to compute non-adiabatic and approximate non-linear periods for all models in the grid as part of the repository accompanying this
 study.

\item R Hya is in the power-down phase (Section \ref{sec:shape_of_pulse}) of a middle-index TP, most likely somewhere between the 9th and 16th pulses. The star is expected to undergo between 20 to 35 TPs overall and most likely become a Carbon star (Section \ref{sec:Monash}), but it is not currently a Carbon star (Section \ref{subsec:phys_par}). 

\item There is some ambiguity regarding the detection of $^{99}$Tc in R Hya's spectrum (Section \ref{sec:observations}). While this question has had few investigators, the current literature consensus is that there is Tc in the UVES and HERMES spectra \citep{Lebzelter2003,Uttenhaler2010,tumi-uttenthaler2011,Uttenthaler-2019-Tc}. Constraints from isotopic ratios are similarly ambiguous, but provide weak evidence in support of R~Hya having recently undergone TDU (Section~\ref{sec:nucleo}).
These results are consistent with our deduction that the star is likely to have undergone third dredge-up in its most recent pulse.

\item In the parameter regime where the best-fitting models overlap with literature determinations---namely, masses between $1.5 - 2.7M_\odot$---our models consistently exhibit TDU. 
Further, when using non-linear periods based on the scaling relations of \citet{Trabucchi2021}, every preferred model but one exhibits TDU (Section \ref{sec:Monash}, Table \ref{table:nonlin_TDU}).
Therefore, when we interpret our solution space in the context of the broader literature and the uncertainties introduced by assumptions of linearity, TDU in R~Hya becomes even more likely (Section \ref{subsec:lit_compare}). 

\item[] Although indications for TDU in R~Hya based on any one source considered here may be inconclusive, we assess the evidence in aggregate. 
The UVES and HERMES spectra, isotopic ratios, previous literature parameter determinations, and modeling results all point to R~Hya having undergone TDU in its most recent thermal pulse. 

\item Using our best models, we demonstrated that the TP in R~Hya started about five centuries ago. R~Hya's maximum expansion coincided with, or happened shortly before, a recent dust production event \citep{ZhaoGeisler2012}. The power-down phase is expected to continue for a few millennia, with a gradually slowing period decrease, before reverting to expansion again (Section \ref{sec:moneysec}). 
\end{enumerate}

\textbf{Using order-of-magnitude estimates for the durations of inter-pulse ($\sim10^5$ years) and thermal pulse ($\sim10^2 -10^3$ years) phases, one should expect to find roughly one star in the midst of a thermal pulse among 100 AGBs. A decent probability of finding one specifically in the power-down phase of a thermal pulse requires a sample closer to 1000 AGBs. The significance of our results is thus underscored by the rarity of such an identification.}

\begin{deluxetable*}{l l l}
\tablecaption{Characteristics of R Hydrae}
\tablecolumns{3}
\tablehead{
\colhead{Parameter} &
\colhead{Value} &
\colhead{Notes}
}
\startdata
\hline 
\textbf{Assumed in \texttt{MESA}/\texttt{GYRE}} & & \\ \hline
$M_\text{init}$ & $1.0$ to $5.0 M_{\odot}$, $\delta M=0.1$ &  $M=0.8,0.9M_\odot$ also attempted but failed to converge \\
$Z_\text{init}$  & 0.0010 to 0.1500 &  0.0001 and 0.0005 also attempted, found no observational overlap \\
$\text{[Fe/H]}$  & $-1.2$ to $+1.0$ dex &  based on \citet{GS98} solar scale \\ 
Helium assumption, fixed & $Y_i = 0.3$ & fixed \\
Helium assumption, $Y_i \sim Z_i$ & $Y_i = Y_\text{BBN} + \frac{\Delta Y}{\Delta Z} \times Z_i$ & $Y_\text{BBN} = 0.2485$; $\frac{\Delta Y}{\Delta Z} = 2.1$ \\
$\alpha_{\text{MLT}}$ & 1.931 & fixed \\
$\eta_{\text{Bl\"ocker}}$ & 0.01 & fixed\\ 
\hline
\textbf{Inferred from \texttt{MESA}/\texttt{GYRE}} & & \\ \hline
Initial mass   & $1.6$ to $3.7$ $M_\odot$  & FM ridge, excluding models too metal-rich for Galaxy \\
Initial Z      & $0.001$ to $0.05$ &  \\
Initial [Fe/H] & $-1.2$ to $+0.5$ dex & following the relation $\text{[Fe/H]}=-0.26M_\odot^2 + 2.22M_\odot - 4.17$ \\
Age ($Y_i = 0.3$)           & 300 to 800 Myr & 300 Myr to 1 Gyr when $Y_i \sim Z_i$ \\
Pulse index  ($Y_i = 0.3$)  & $9^\text{th}$ - $16^\text{th}$ & within $\pm 3$ pulses of TDU onset in most cases \\
Pulse index  ($Y_i \sim Z_i$)  & $6^\text{th}$ - $16^\text{th}$ & low best pulse indices for two very metal-rich models \\
\hline
\multicolumn{3}{l}{\textbf{Inferred from Monash Code \& Nucleosynthesis Considerations } } \\ \hline
Third Dredge-Up? &  Probable, not definite  & becomes very likely if R~Hya has sub-solar initial metallicity\\
Carbon star?     &  No        & C/O $\ge 1$ for only $2.0M_\odot$, [Fe/H]$=-0.75$ model \\
\hline
\multicolumn{2}{l}{\textbf{Inferred from Overall Analysis }}   & \\ \hline
Pulsation Mode & FM & on the basis of larger solution space and literature analysis \\
Time since He-shell flash & 470 to 580 yr &  based on best-fitting pulses, varied $Y_i$  \\
Predicted $\dot{P}$ & --0.17 to --0.20 d/yr & based on best-fitting pulses  \\
\enddata
\tablecomments{See Table \ref{table:obs_constraints} for a summary of all observational parameters and literature-derived constraints and estimates.}
\label{table:fundpar}
\end{deluxetable*}

\begin{figure}
\centering
\includegraphics[width=\columnwidth]{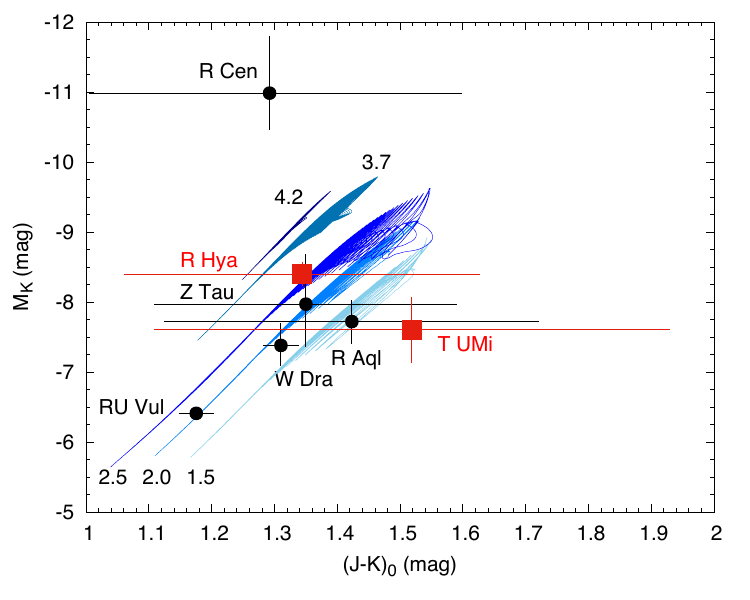}
\caption{Potential TP-AGB target stars positioned in a (J--K)$_0$ vs $M_K$ CMD. We highlight the two stars that we analyzed already in red. Lines are [Fe/H] = 0.11 (Z = 0.0216) TP-AGB stellar tracks from our study, with their initial masses, in $M_\odot$ units, indicated with numbers. }
\label{fig:tpagbtargets}
\end{figure}

\section{Future work}
\label{sec:future}
AGB stars represent one of the largest sources of uncertainty in the modeling of stellar populations, dust formation, and galactic chemical evolution. The amount and strength of mixing in their interiors, as well as their mass-loss histories, have historically been poorly constrained. The existence of thermal pulses and the third dredge-up are established, but understanding their numbers and strength, as well as the types of stars that actually experience them, has been challenging.

In the past, the lack of detailed constraints on any individual star has made it hard to support or refute any particular model, and the uncertainties on any model have been unfortunately large. As we begin to be able to place constraints on the physical interiors of individual thermally pulsing stars, including on their initial masses and compositions, their locations on the AGB, and the number and strength of the pulses they have experienced, we begin to substantively alter our broader understanding of the physics of AGB stars. In turn, we begin to better our understanding of this consistently challenging but astrophysically significant phase of stellar evolution. 

Beyond R~Hya and T~UMi, there are other Miras exhibiting clear long-term period changes. RU~Vul shows signs of a sudden transition from a relatively stable period to an ongoing decrease, reminiscent of T~UMi, which suggests the possibility that we are observing yet another AGB star entering a thermal pulse \citep{Uttenthaler-2016}. If RU~Vul is indeed going through the ``knee" phase, it will provide a good opportunity to determine the physical parameters of another TP-AGB star from seismic constraints. 

We searched for additional targets that offer the possibility of more detailed seismic modeling using our model grid. Based on the O--C data of \citet{Karlsson-2013}, some Mira stars feature a long-term, monotonic component in their period evolution that exceeds the faster changes associated with meandering. Stars like this include R~Aql, R~Cen, W~Dra, and Z~Tau. At least some of these could be experiencing a TP, going through one of the ascending or descending branches between the inversion points. \citet{wood-zarro-1981} proposed that R~Aql is in a descending phase similar to that of R~Hya. With the aid of classical observational constraints, we can use our grid to determine preferred parameter ranges for these stars as well.

We plot these targets on a near-infrared ($M_K$ vs \textit{J--K}) CMD in Figure~\ref{fig:tpagbtargets}. To construct the plot, we calculated the extinction-corrected $M_J$ and $M_K$ magnitudes of the stars using the SeismoLab python package \citep{bodi-seismolab}. The package queries the Gaia distances and 2MASS photometry of the stars and calculates the extinction coefficients from the \citet{green-2015} dust maps. The stellar tracks indicated in the plot come from our grid: they differ in initial mass (1.5, 2.0, 2.5, 3.7, 4.2~$M_\odot$, respectively), but have a near-solar metallicity of $Z = 0.0216$ in all cases. We converted the luminosity of the model into \textit{J} and \textit{K} brightness using the bolometric correction of \citet{Choi2016}. 

The names of the stars are annotated on Figure~\ref{fig:tpagbtargets}. Despite the apparently large distance between the test models plotted and the seemingly outlying candidates RU~Vul and R~Cen, it has already been suggested that both stars are undergoing thermal pulses \citep{Hawkins-2001,Uttenthaler-2016}. Confirming or contradicting this finding is a natural follow-up to the present analysis.

Figure~\ref{fig:tpagbtargets} further illustrates the difficulty of positioning these stars relative to TP-AGB models based on classical constraints alone, thus underscoring the importance of new and better approaches to characterizing evolved, variable stars. 
Hybrid seismic-evolutionary analysis will allow us to identify preferred models and put detailed physical constraints on these and more targets individually---a task that will be accelerated by the publicly available grid we have computed here. This will increase the number of well-characterized AGB stars significantly, which in turn will drive our knowledge of this fundamental, yet in many ways still mysterious, stage of stellar evolution forward.

\section*{Acknowledgements}
M.J. gratefully acknowledges funding of MATISSE: \textit{Measuring Ages Through Isochrones, Seismology, and Stellar Evolution}, awarded through the European 
Commission's Widening Fellowship.  
This project has received funding from the European Union's Horizon 2020 research and innovation programme.
This research was supported by the `SeismoLab' KKP-137523 \'Elvonal grant of the Hungarian Research, Development and Innovation Office (NKFIH). M.J. and J.T. acknowledge NASA ATP grant 80NSSC22K0812.
The authors wish to thank Earl Bellinger and Ebraheem Farag for useful discussion, advice on modeling choices, and contributions to software infrastructure. 
M. Joyce and L. Moln\'ar wish to thank Yuri Fadeyev and Stefan Uttenthaler for discussion during revision on approaches to non-linear modeling and spectroscopic observations, respectively.
M.J. wishes to thank John Bourke for discussion and typesetting.
Our models were run on the OzSTAR national facility at Swinburne University of Technology. The OzSTAR program receives funding in part from the Astronomy National Collaborative Research Infrastructure Strategy (NCRIS) allocation provided by the Australian Government.
The authors also wish to thank the OzSTAR supercomputing team for resources and instruction.
We acknowledge with thanks the variable star observations from the AAVSO International Database contributed by observers worldwide and used in this research.
This work made use of Astropy:\footnote{\url {http://www.astropy.org}} a community-developed core Python package and an ecosystem of tools and resources for astronomy \citep{astropy:2013, astropy:2018, astropy:2022}. 
This research made use of NASA’s Astrophysics Data System Bibliographic Services, and of the SIMBAD database operated at CDS, Strasbourg, France.
We acknowledge the use of the GPT-3.5 language model developed by OpenAI in this research.

\bibliography{new.ms}
\bibliographystyle{aasjournal}

\pagebreak
\appendix
\section{Data tables}
\label{appendix:data}

We calculated the longitudinal evolution of the pulsation period and amplitude from the DASCH and AAVSO data, as described in Section~\ref{sect:per-evo}. Tables~\ref{table:daschperiods} and \ref{tab:aavsoperiods} list the data presented in Figures~\ref{fig:period_ev} and \ref{fig:period_ev_dense} and used in our model fitting process. We also list here the archival timing data we presented by \citet{Zijlstra2002} converted into periods and the uncertainties we assigned to them in Table~\ref{table:historicperiods}.

\begin{table}[ht!]
\centering
\begin{tabular}{ccccc}
\hline
JD & P & $\sigma$P & $A_1$ & $\sigma A_1$ \\
(d) & (d) & (d) & (mag) & (mag) \\
\hline
2416746.3 & 405.65 & 0.97 & 2.032 & 0.048 \\
2419904.3 & 404.72 & 0.74 & 1.890 & 0.030 \\
2422415.7 & 405.65 & 0.62 & 2.032 & 0.048 \\
2425785.2 & 412.50 & 0.69 & 1.854 & 0.054 \\
2428834.5 & 397.78 & 0.83 & 1.700 & 0.045 \\
2431643.1 & 380.81 & 0.56 & 1.541 & 0.087 \\
2446445.5 & 390.79 & 3.72 & 1.436 & 0.149 \\
\hline
\end{tabular}
\phantom{\textbf{Table 6. }}
\caption{Evolution of the pulsation period calculated from the DASCH observations. }
  \label{table:daschperiods}
\end{table}

\begin{table}[ht!]
\centering
\begin{tabular}{ccc}
\hline
JD & P & $\sigma$P \\
(d) & (d) & (d) \\
\hline
2328201 & 496.03 & 30 \\
2331120 & 496.03 & 30 \\
2345033 & 495.05 & 30 \\
2372677 & 484.97 & 30 \\
2380446 & 477.10 & 10 \\
2381876 & 477.10 & 10 \\
2387003 & 459.98 & 10 \\
2388386 & 459.98 & 10 \\
2394350 & 450.05 & 5 \\
\hline
\end{tabular}
\phantom{\textbf{Table 7. }}
\caption{Approximate times of maxima based on the historical records from \citet{Zijlstra2002}, with our estimated uncertainties. }
  \label{table:historicperiods}
\end{table}

\begin{deluxetable}{ccccc}
\tablecaption{Evolution of the pulsation period calculated from the AAVSO observations. \label{tab:aavsoperiods}}
\tablecolumns{5}
\tablehead{
\colhead{JD} & \colhead{P} & \colhead{$\sigma$P} & \colhead{$A_1$} & \colhead{$\sigma A_1$} \\
\colhead{(d)} & \colhead{(d)} & \colhead{(d)} & \colhead{ (mag)} & \colhead{(mag)}
}
\startdata
2413513.7 & 420.55 & 2.85 & 2.226 & 0.195 \\
2418068.7 & 405.62 & 4.68 & 2.070 & 0.138 \\
2419695.5 & 406.33 & 0.97 & 2.213 & 0.041 \\
2420092.3 & 409.27 & 0.72 & 2.213 & 0.041 \\
2421119.7 & 402.72 & 1.42 & 2.183 & 0.054 \\
2422419.7 & 401.84 & 0.67 & 2.387 & 0.036 \\
2423510.1 & 414.79 & 0.74 & 2.221 & 0.038 \\
2424242.3 & 412.73 & 0.90 & 2.015 & 0.035 \\
2425231.4 & 413.21 & 0.61 & 2.230 & 0.030 \\
2426343.7 & 415.34 & 0.67 & 2.266 & 0.032 \\
2427263.9 & 411.62 & 0.77 & 2.071 & 0.035 \\
2428275.3 & 399.81 & 0.57 & 2.187 & 0.031 \\
2429215.0 & 399.06 & 0.51 & 2.210 & 0.022 \\
2430138.4 & 389.76 & 0.38 & 2.062 & 0.028 \\
2431381.9 & 383.86 & 0.42 & 1.761 & 0.028 \\
2432390.3 & 383.56 & 0.36 & 1.806 & 0.025 \\
2433143.9 & 388.40 & 0.41 & 1.720 & 0.024 \\
2434058.9 & 386.23 & 0.49 & 1.875 & 0.025 \\
2435187.9 & 389.16 & 0.62 & 1.772 & 0.029 \\
2436447.0 & 392.93 & 0.49 & 1.830 & 0.020 \\
2437385.7 & 392.42 & 0.37 & 1.781 & 0.020 \\
2438152.5 & 394.16 & 0.32 & 1.831 & 0.020 \\
2439259.4 & 388.12 & 0.45 & 1.656 & 0.024 \\
2440087.0 & 392.72 & 0.60 & 1.656 & 0.030 \\
2441356.7 & 390.06 & 0.54 & 1.707 & 0.018 \\
2442264.1 & 385.62 & 0.42 & 1.563 & 0.037 \\
2443337.8 & 384.17 & 0.44 & 1.690 & 0.031 \\
2444210.3 & 392.35 & 0.40 & 1.605 & 0.032 \\
2445312.5 & 389.56 & 0.61 & 1.540 & 0.026 \\
2446191.9 & 384.39 & 0.42 & 1.524 & 0.025 \\
2447345.8 & 387.36 & 0.44 & 1.576 & 0.022 \\
2448243.3 & 382.97 & 0.36 & 1.719 & 0.023 \\
2449354.1 & 384.18 & 0.38 & 1.454 & 0.018 \\
2450251.4 & 384.05 & 0.31 & 1.671 & 0.023 \\
2451302.1 & 389.57 & 0.35 & 1.890 & 0.027 \\
2452222.4 & 384.96 & 0.34 & 1.799 & 0.023 \\
2453186.8 & 377.99 & 0.33 & 1.785 & 0.019 \\
2454129.9 & 375.50 & 0.42 & 1.832 & 0.017 \\
2455278.4 & 379.59 & 0.51 & 1.460 & 0.024 \\
2456423.3 & 364.04 & 0.39 & 1.425 & 0.028 \\
2457275.1 & 358.25 & 0.34 & 1.275 & 0.022 \\
2458135.0 & 355.26 & 0.58 & 1.051 & 0.018 \\
2459175.4 & 358.40 & 0.77 & 0.909 & 0.029 \\
2459704.4 & 364.55 & 1.54 & 1.046 & 0.031 \\
\enddata
\end{deluxetable}

\break
\section{Effects of other fitting parameters}
\label{appendix:param_var}

\begin{figure*}[hb!]
\centering
\includegraphics[width=0.48\columnwidth]{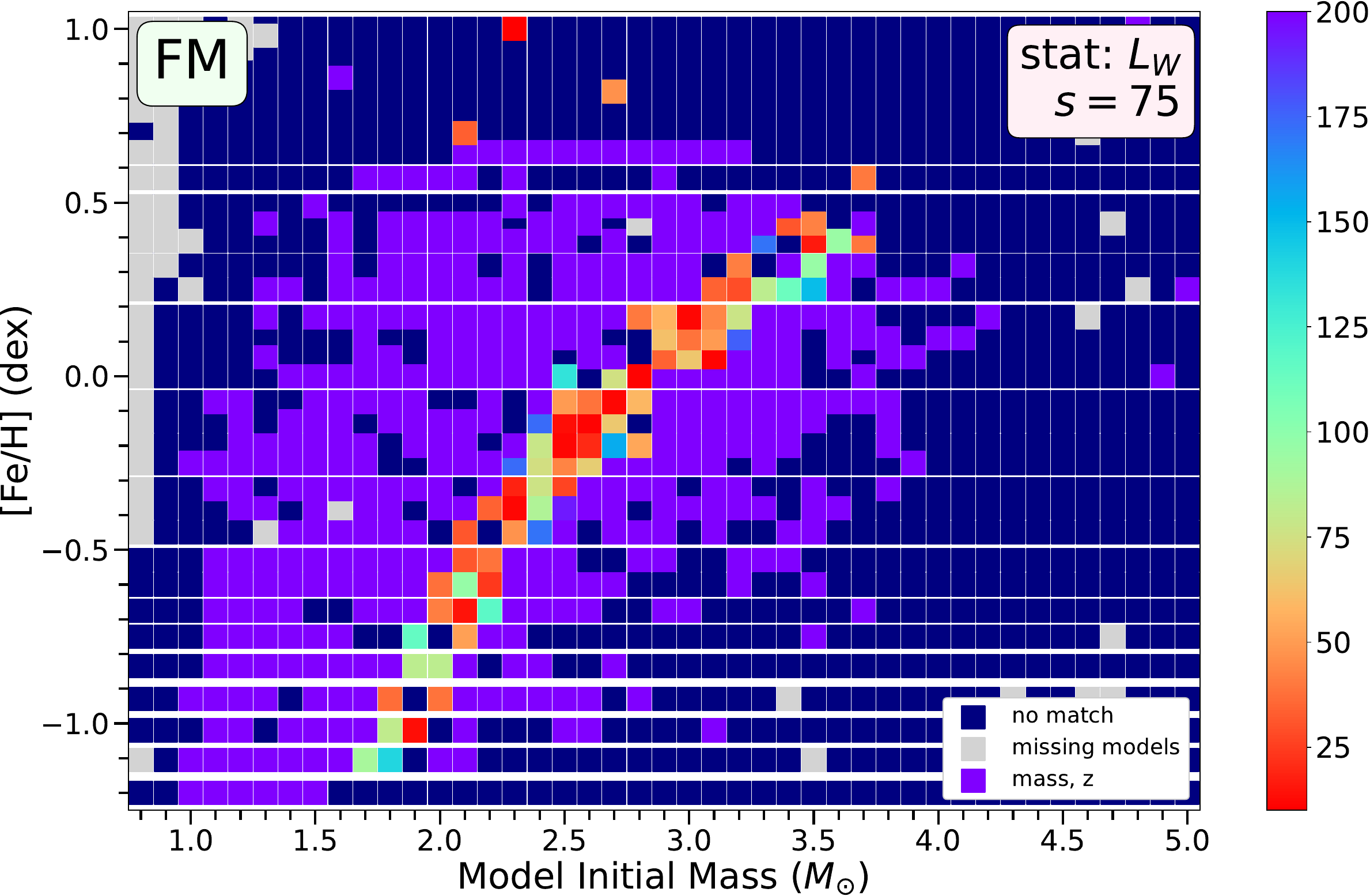}
\includegraphics[width=0.48\columnwidth]{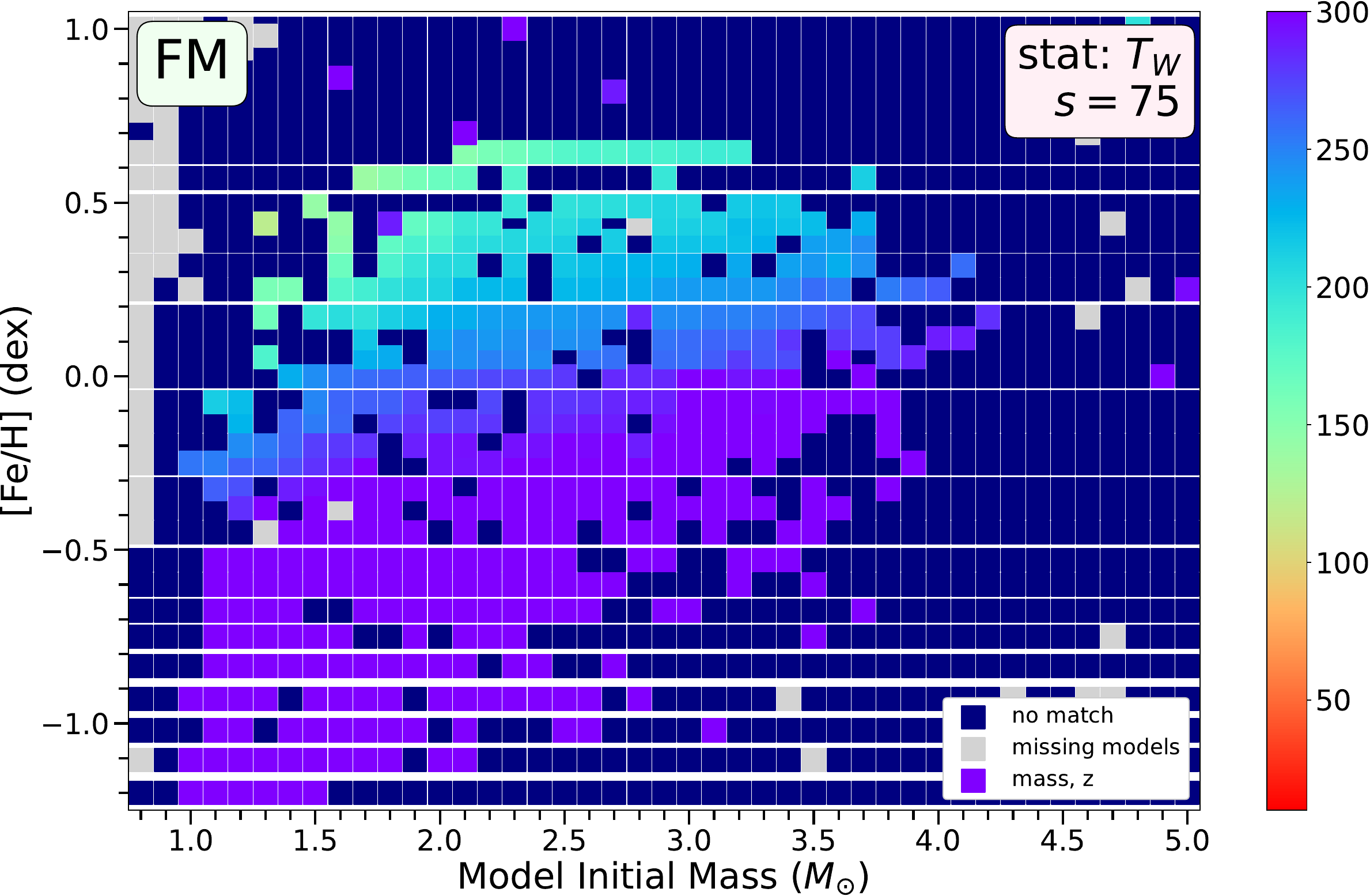}
\includegraphics[width=0.48\columnwidth]{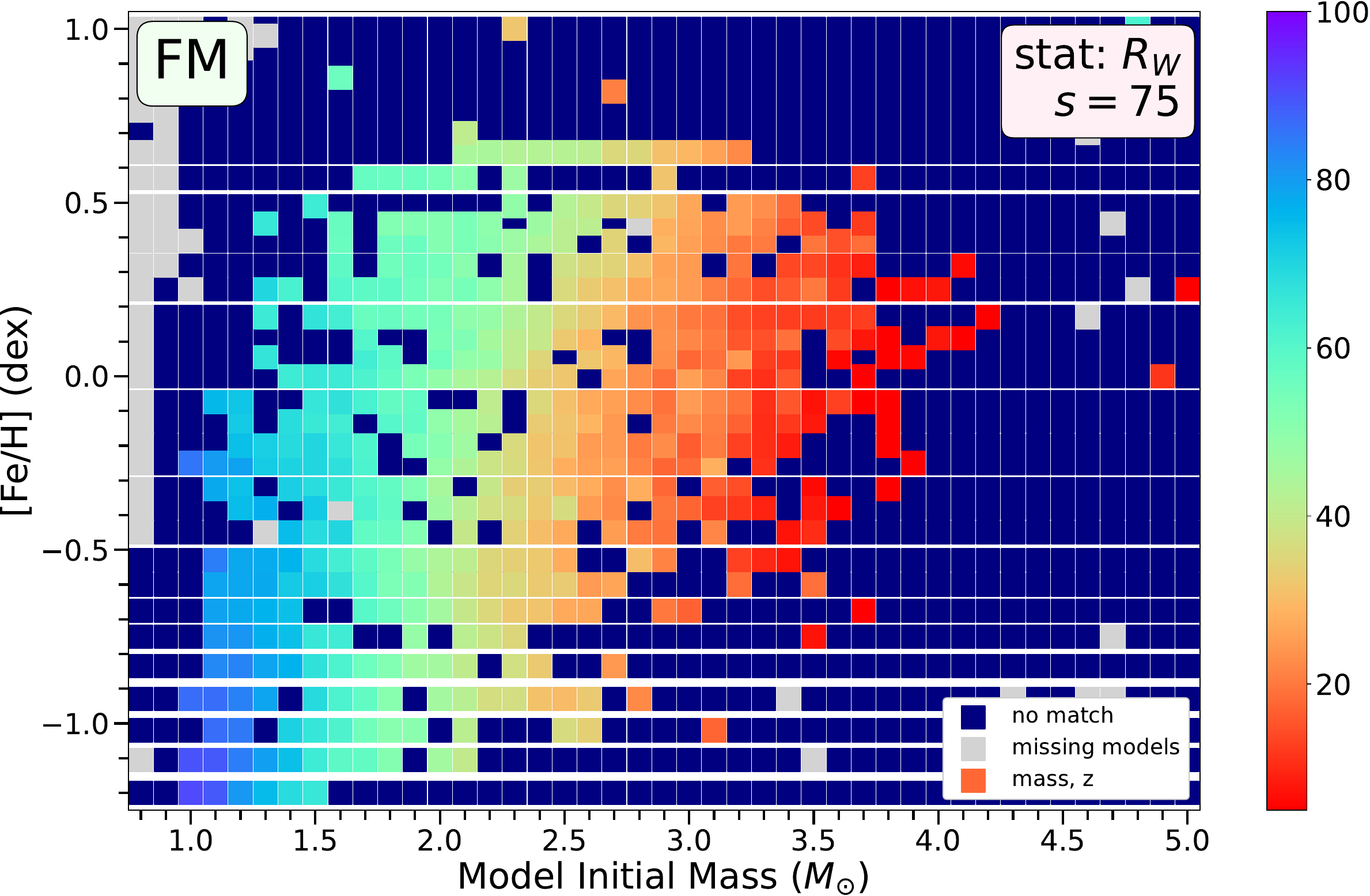}
\includegraphics[width=0.48\columnwidth]{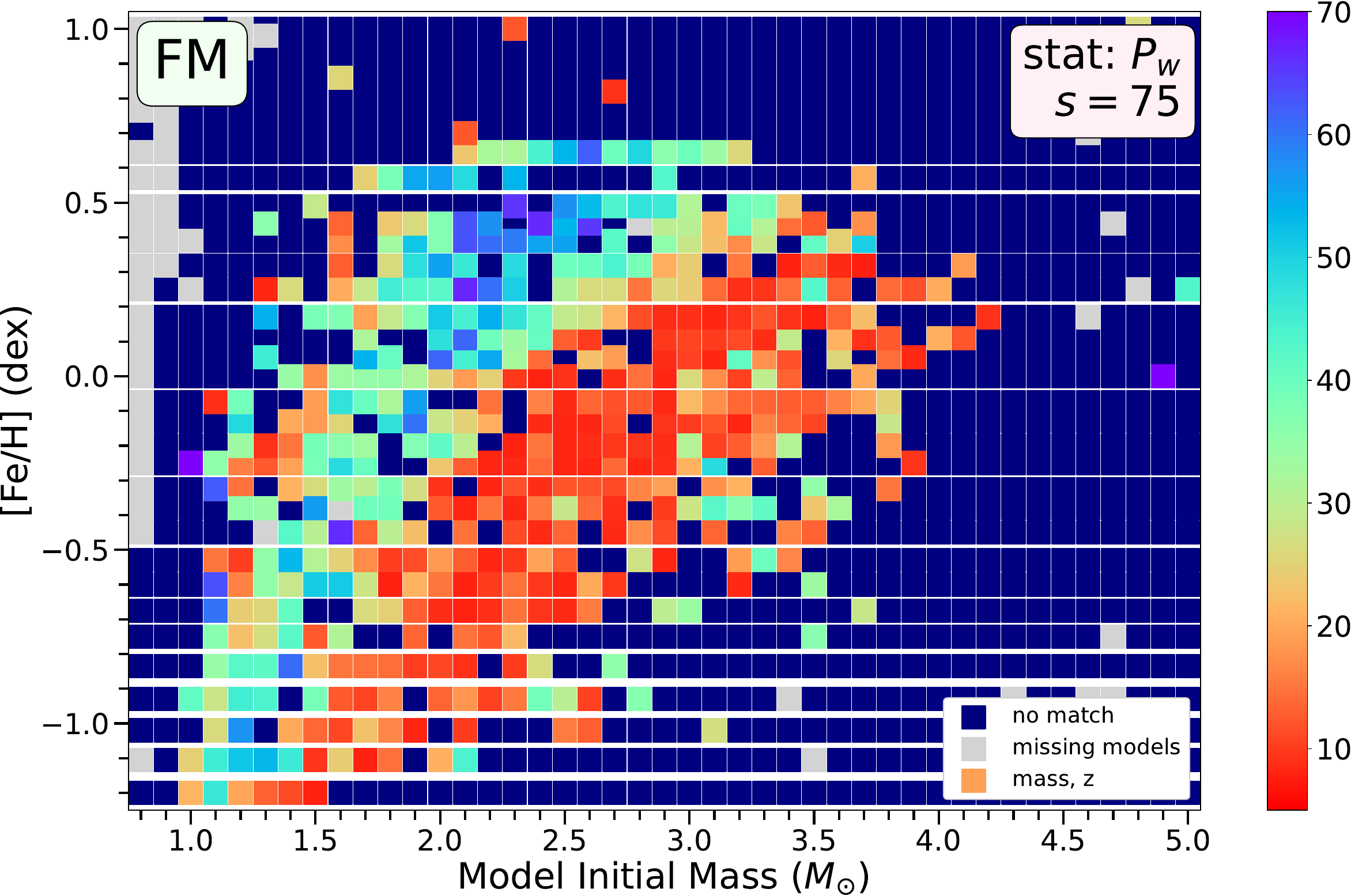}
\includegraphics[width=0.48\columnwidth]{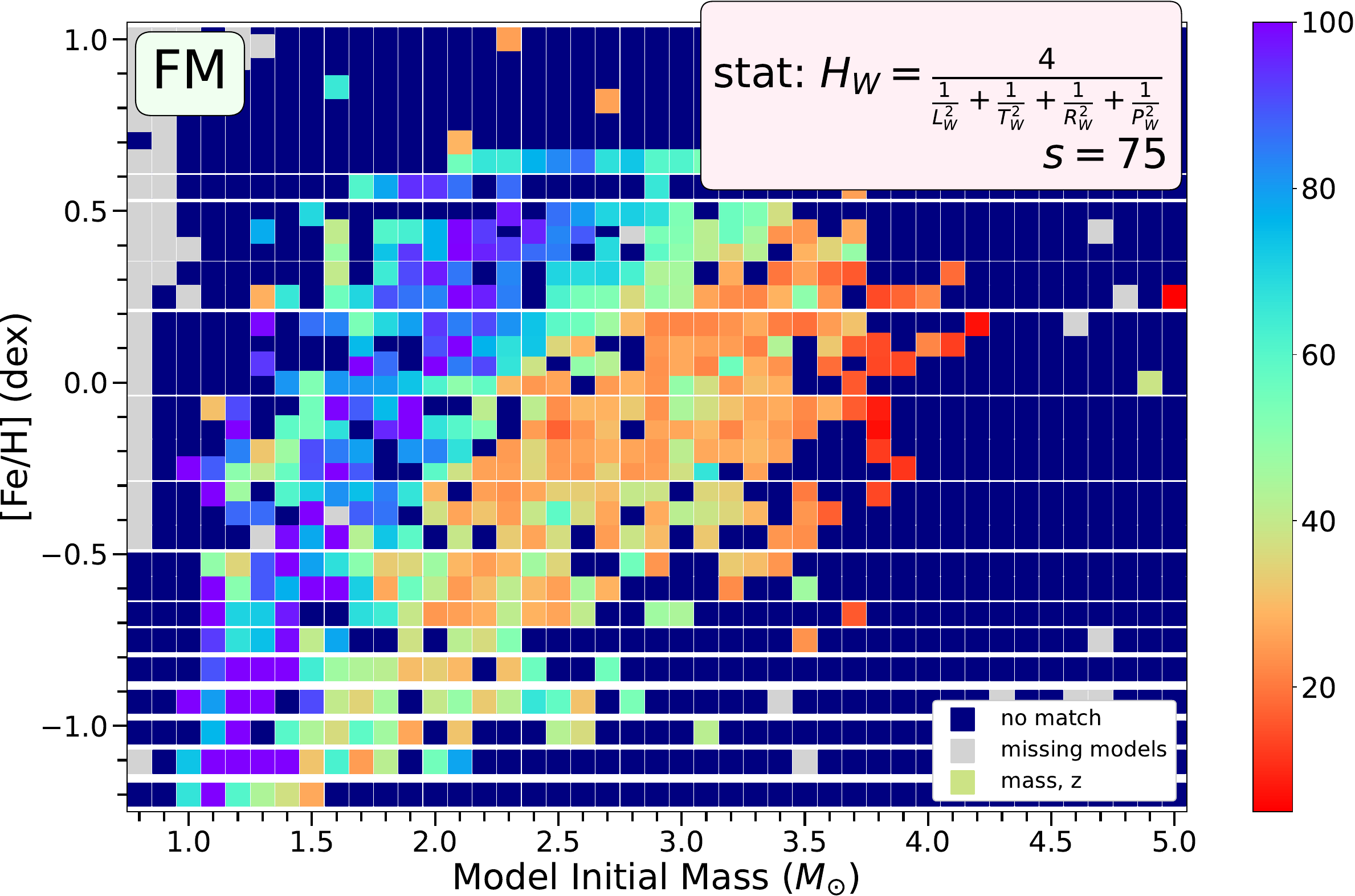}
\includegraphics[width=0.48\columnwidth]{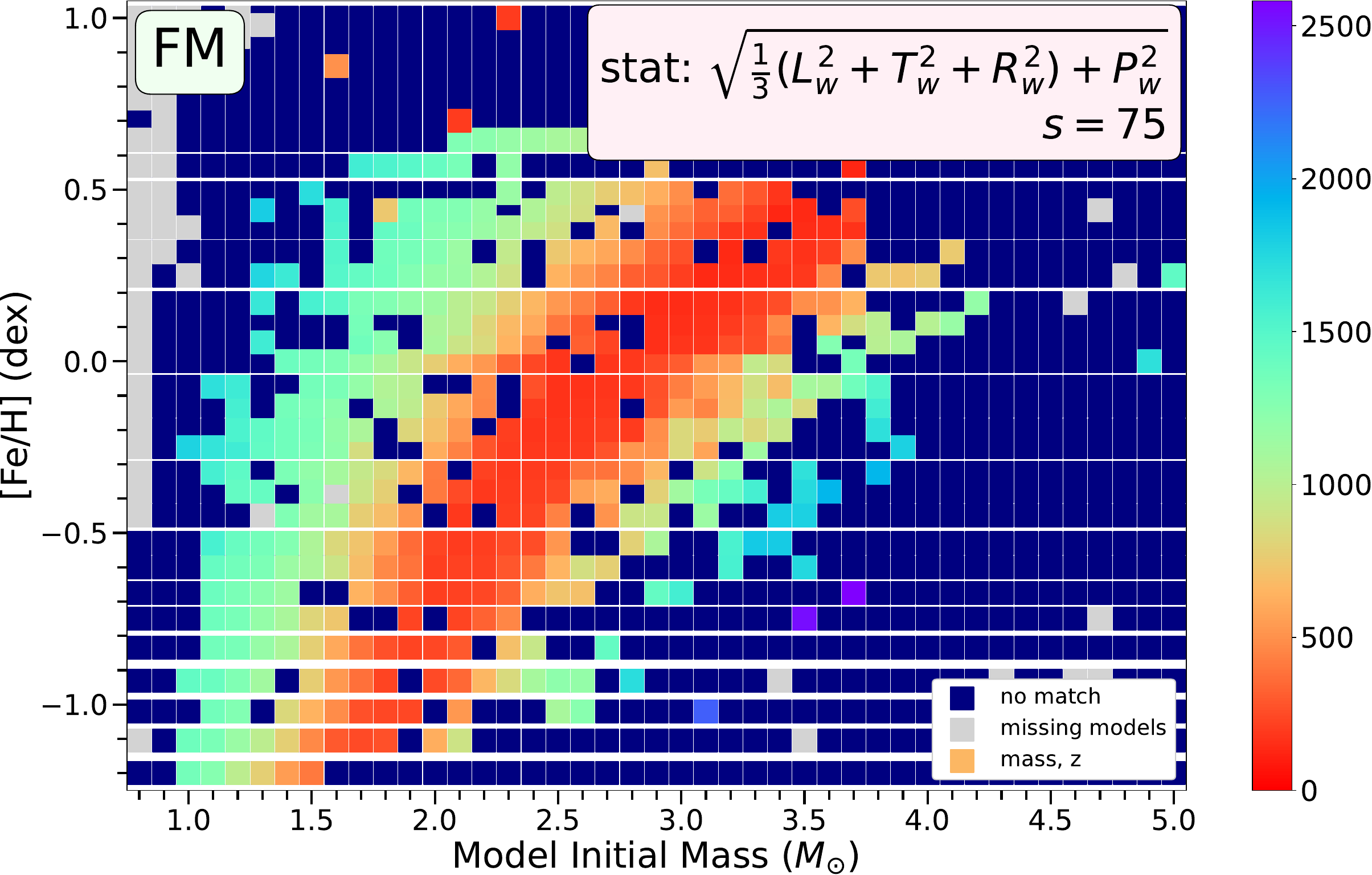}
\caption{Same as Figure \ref{fig:heatmap_FM}, but using the grid adopting the $Y_i \sim Z_i$ helium scaling assumption.}
\label{fig:heatmap_FM_Yi}
\end{figure*}

\begin{figure*}
\centering
\includegraphics[width=0.48\columnwidth]{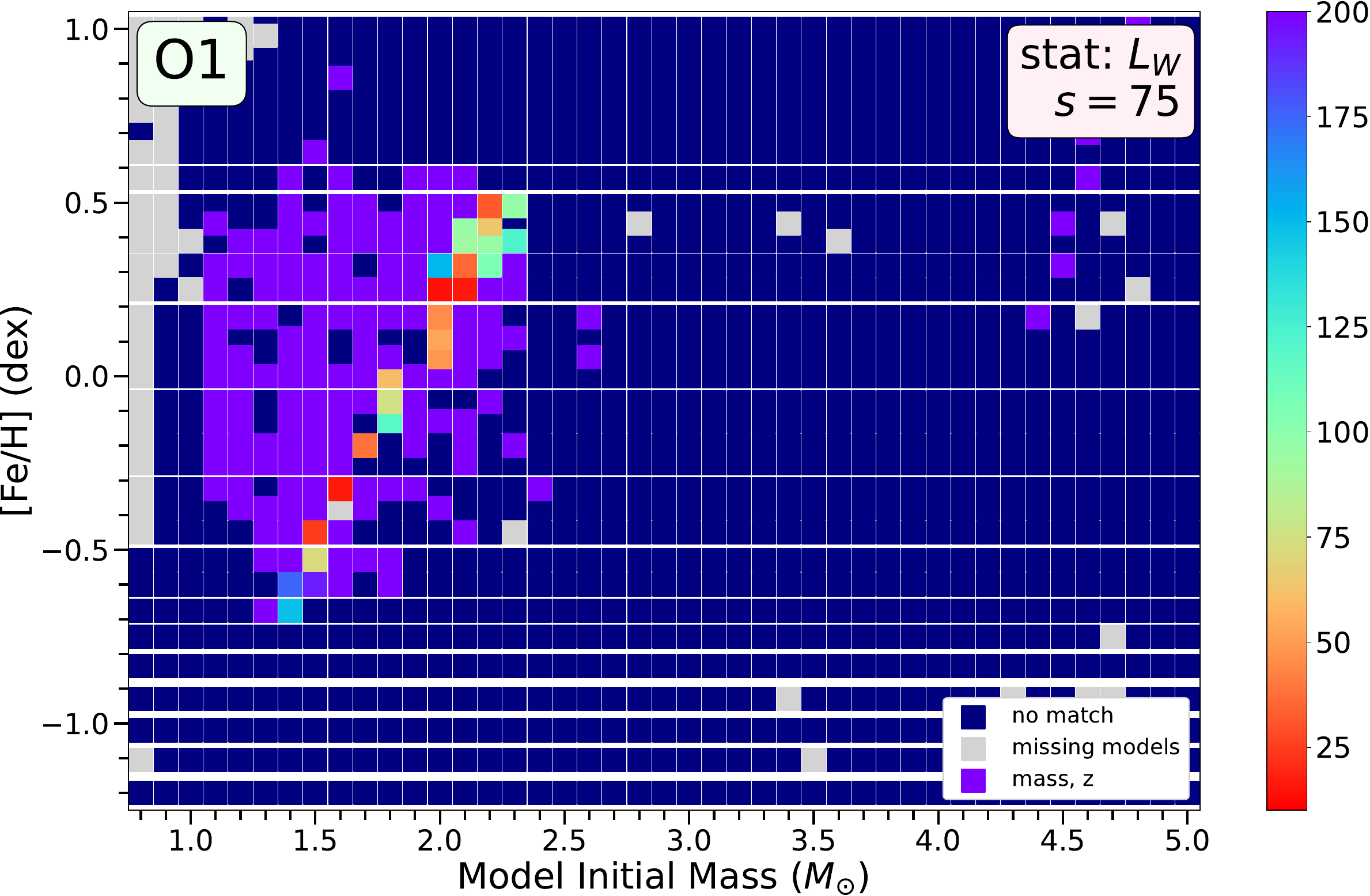}
\includegraphics[width=0.48\columnwidth]{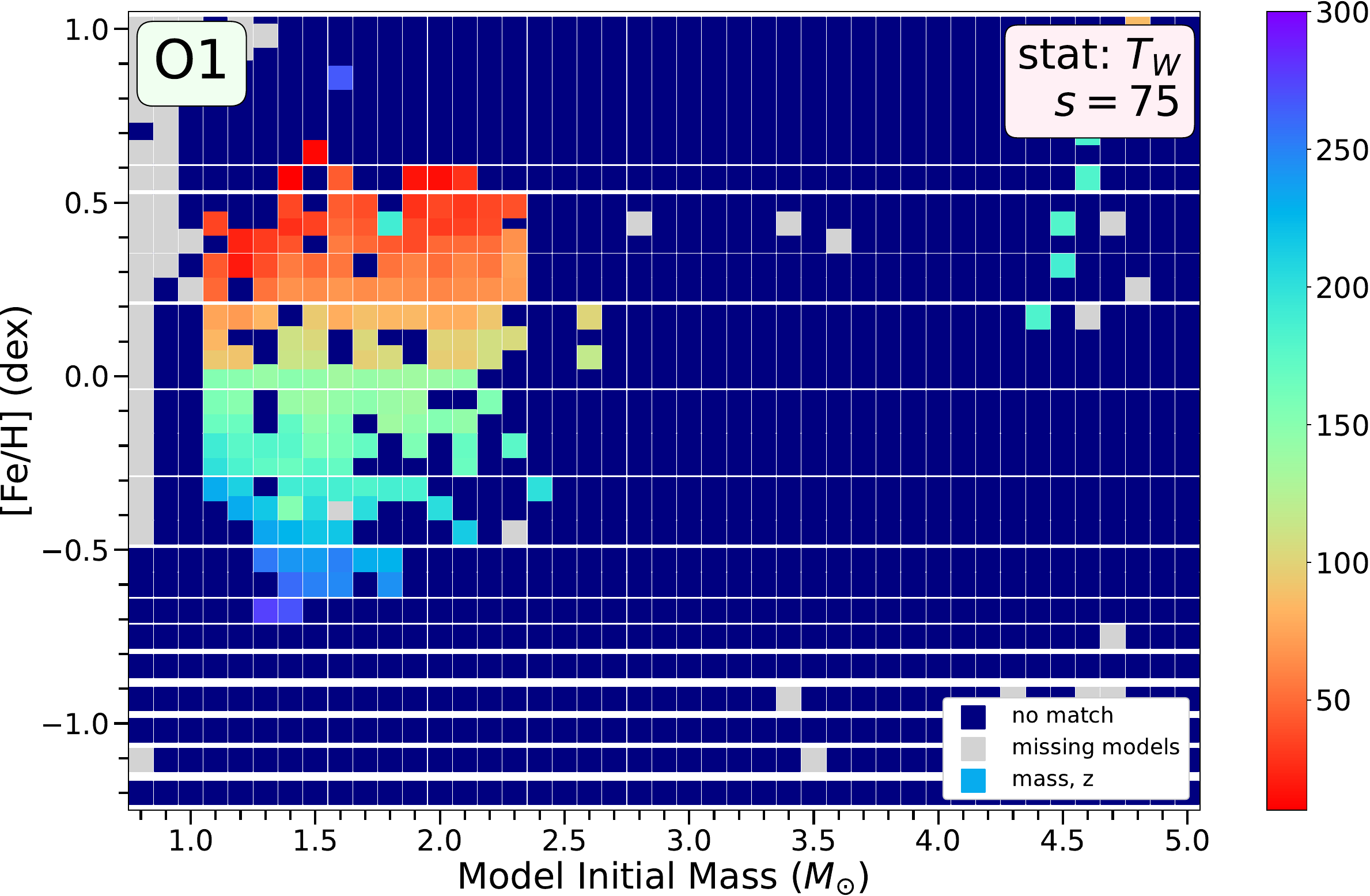}
\includegraphics[width=0.48\columnwidth]{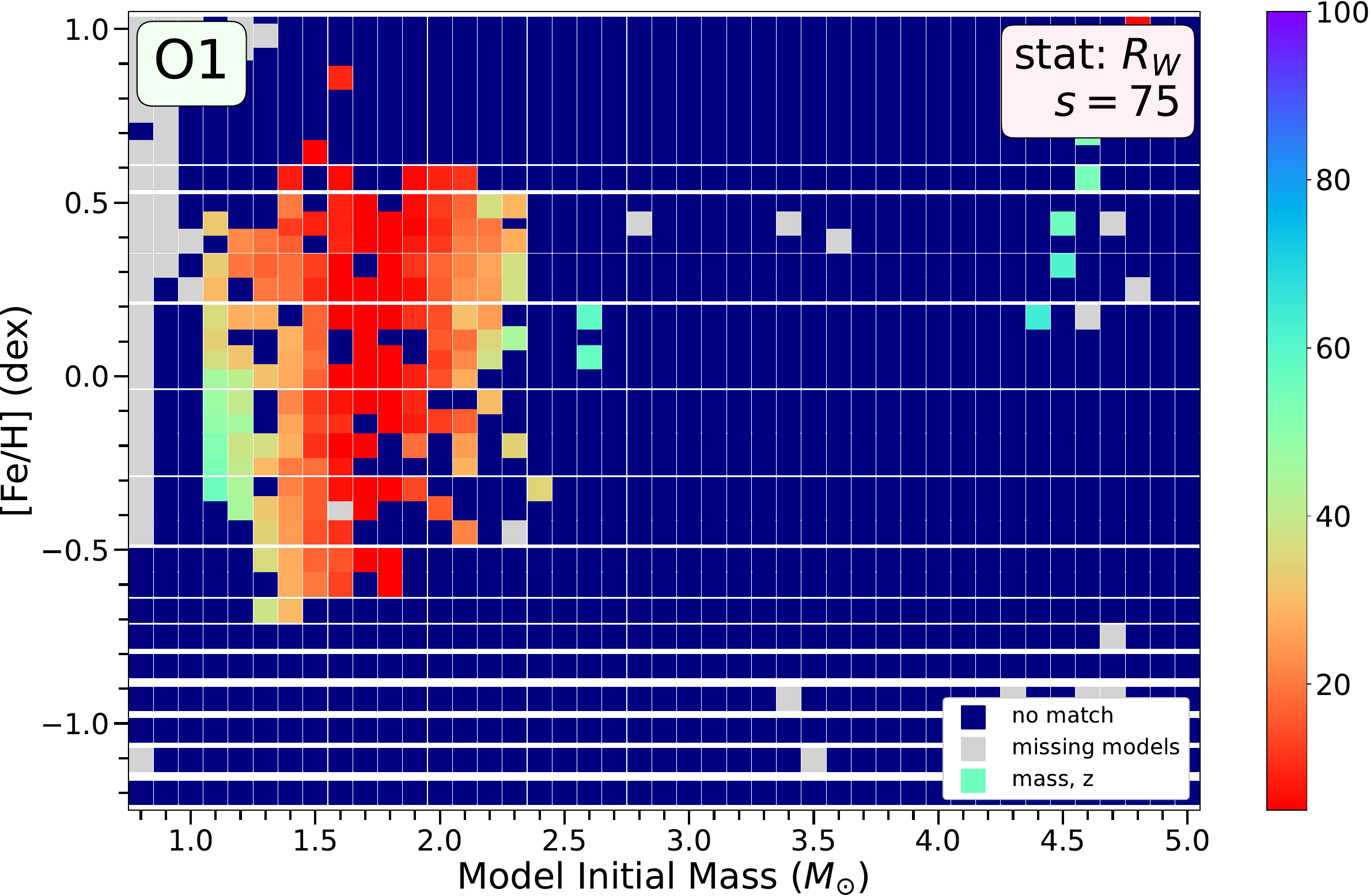}
\includegraphics[width=0.48\columnwidth]{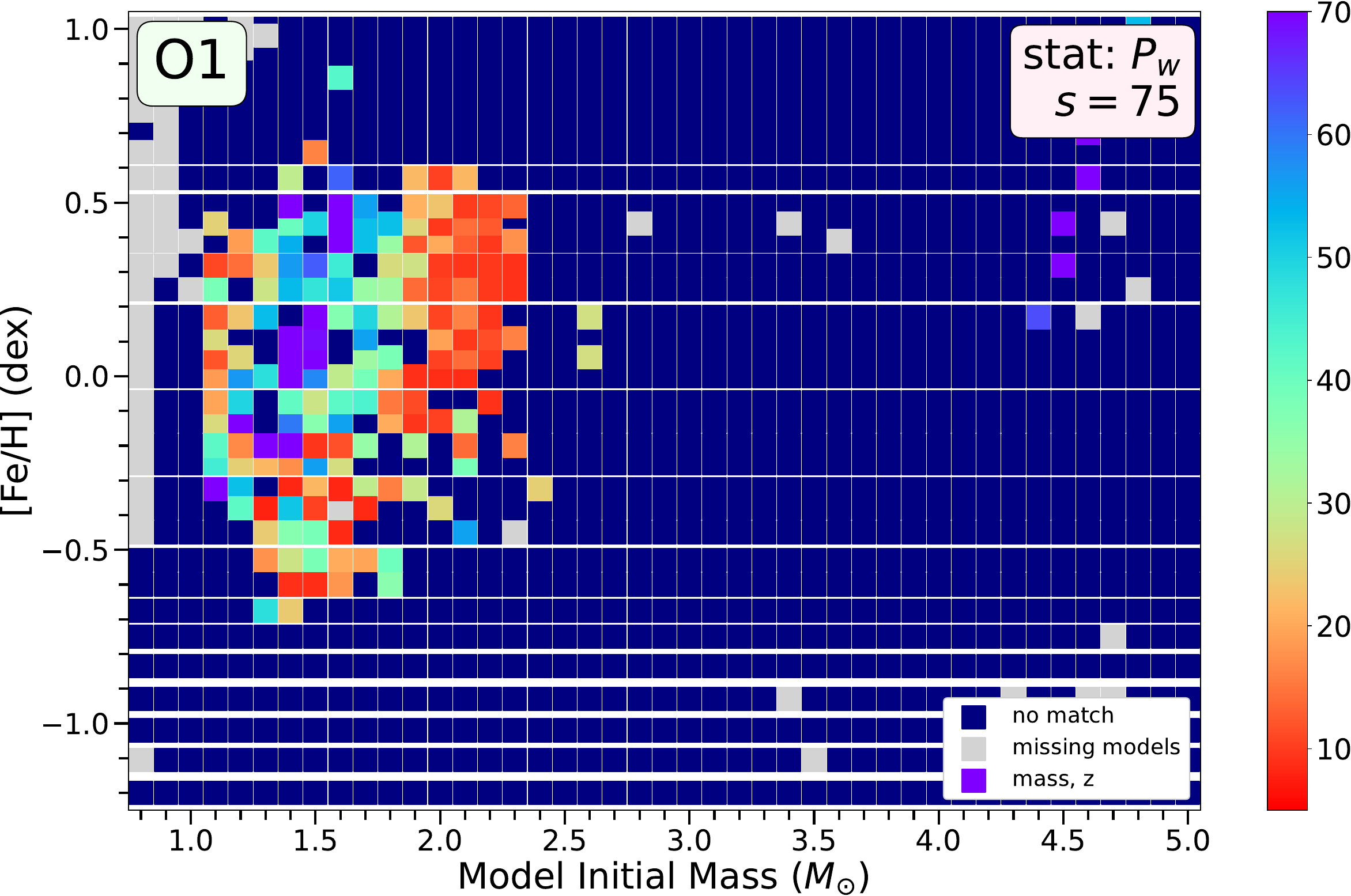}
\includegraphics[width=0.48\columnwidth]{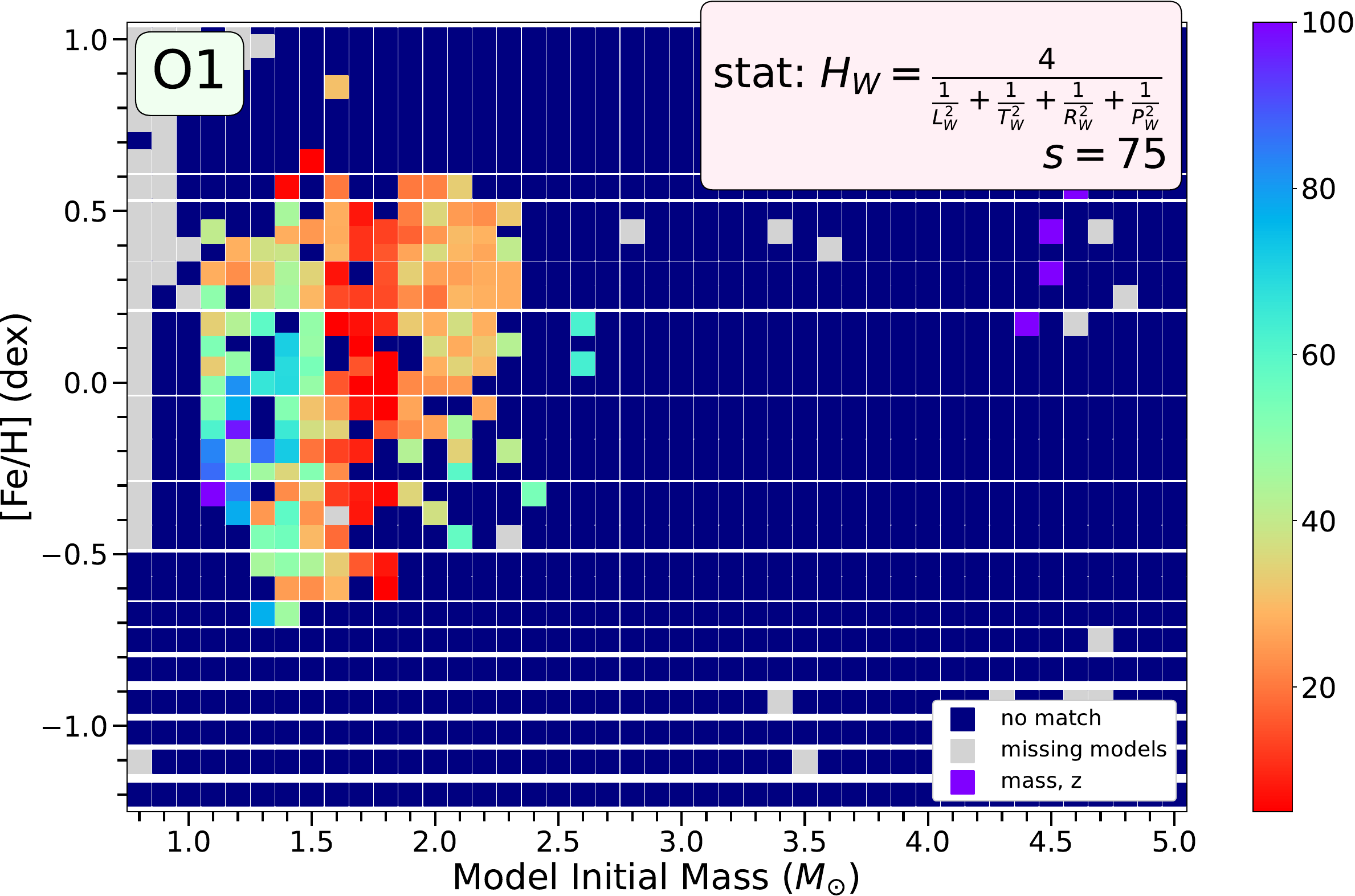}
\includegraphics[width=0.48\columnwidth]{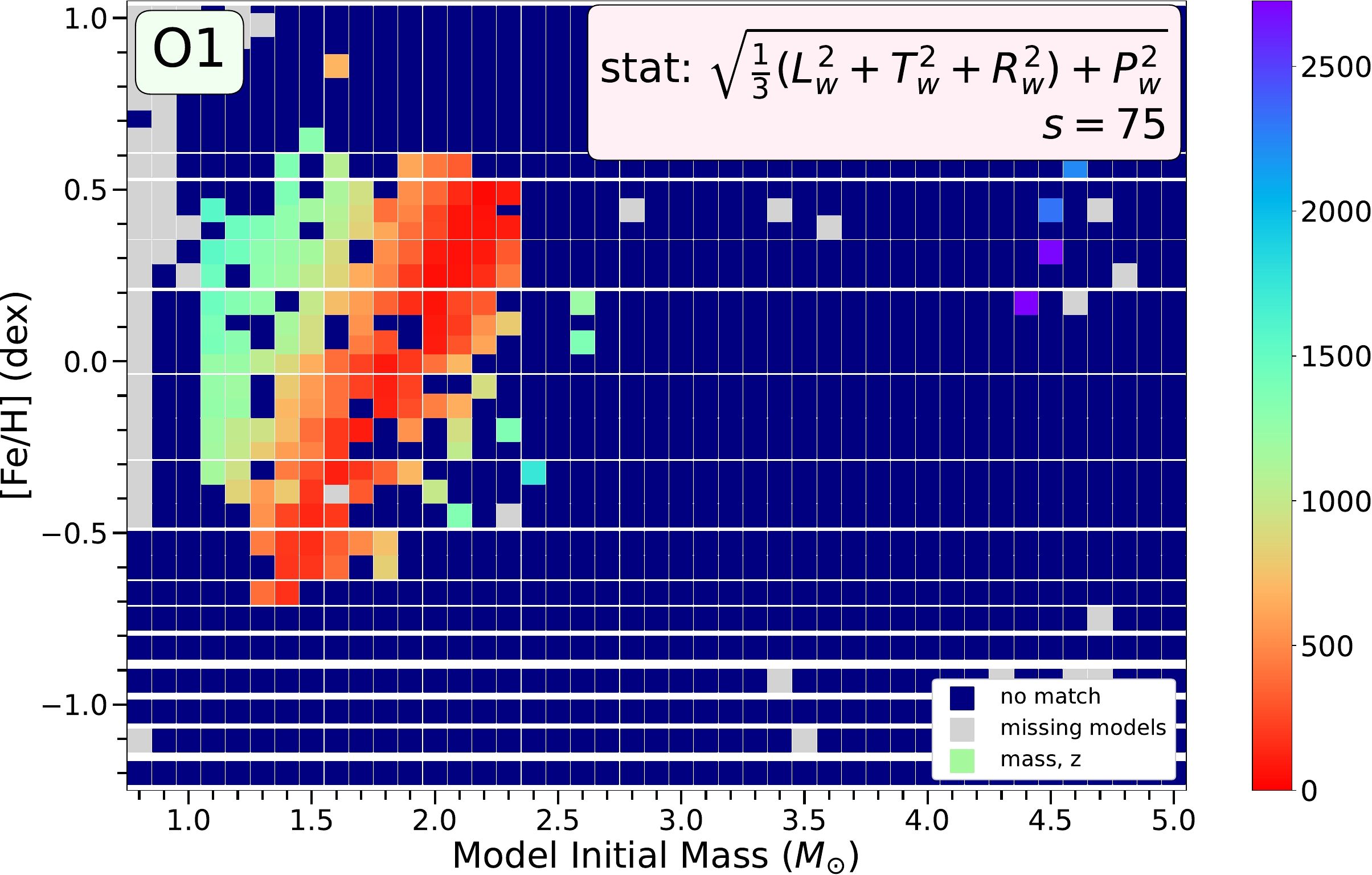}
\caption{Same as Figure \ref{fig:heatmap_O1}, but using the grid adopting the $Y_i \sim Z_i$ helium scaling assumption.}
\label{fig:heatmap_O1_Yi}
\end{figure*}

Results for other parameter choices in the fitting procedure are shown. Figures~\ref{fig:heatmap_FM_Yi} and \ref{fig:heatmap_O1_Yi} reproduce Figures~\ref{fig:heatmap_FM} and \ref{fig:heatmap_O1}, using the grid that adopts the $Y_i \sim Z_i$ relation.

\begin{figure*}
\centering
\includegraphics[width=0.48\columnwidth]{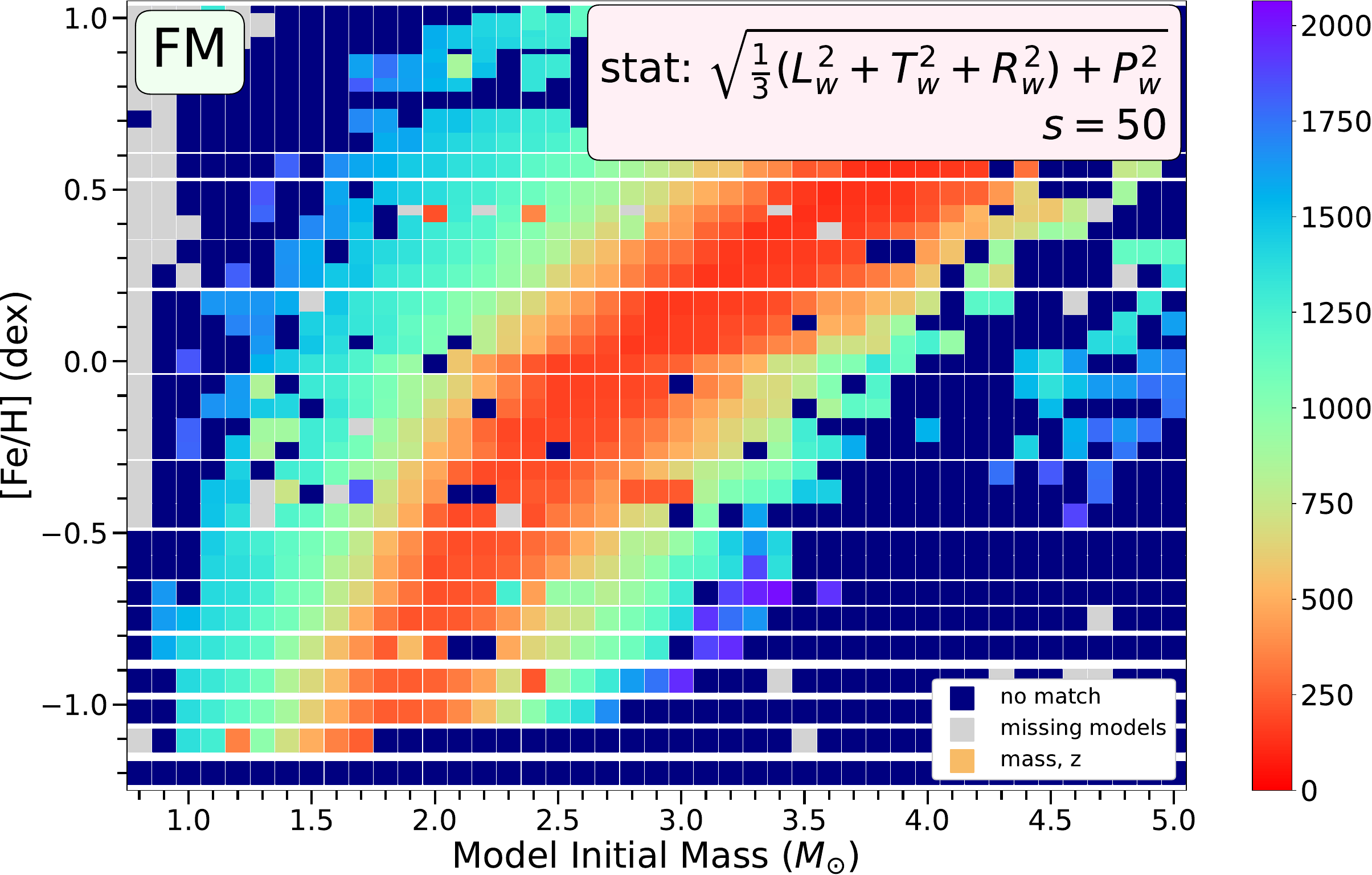}
\includegraphics[width=0.48\columnwidth]{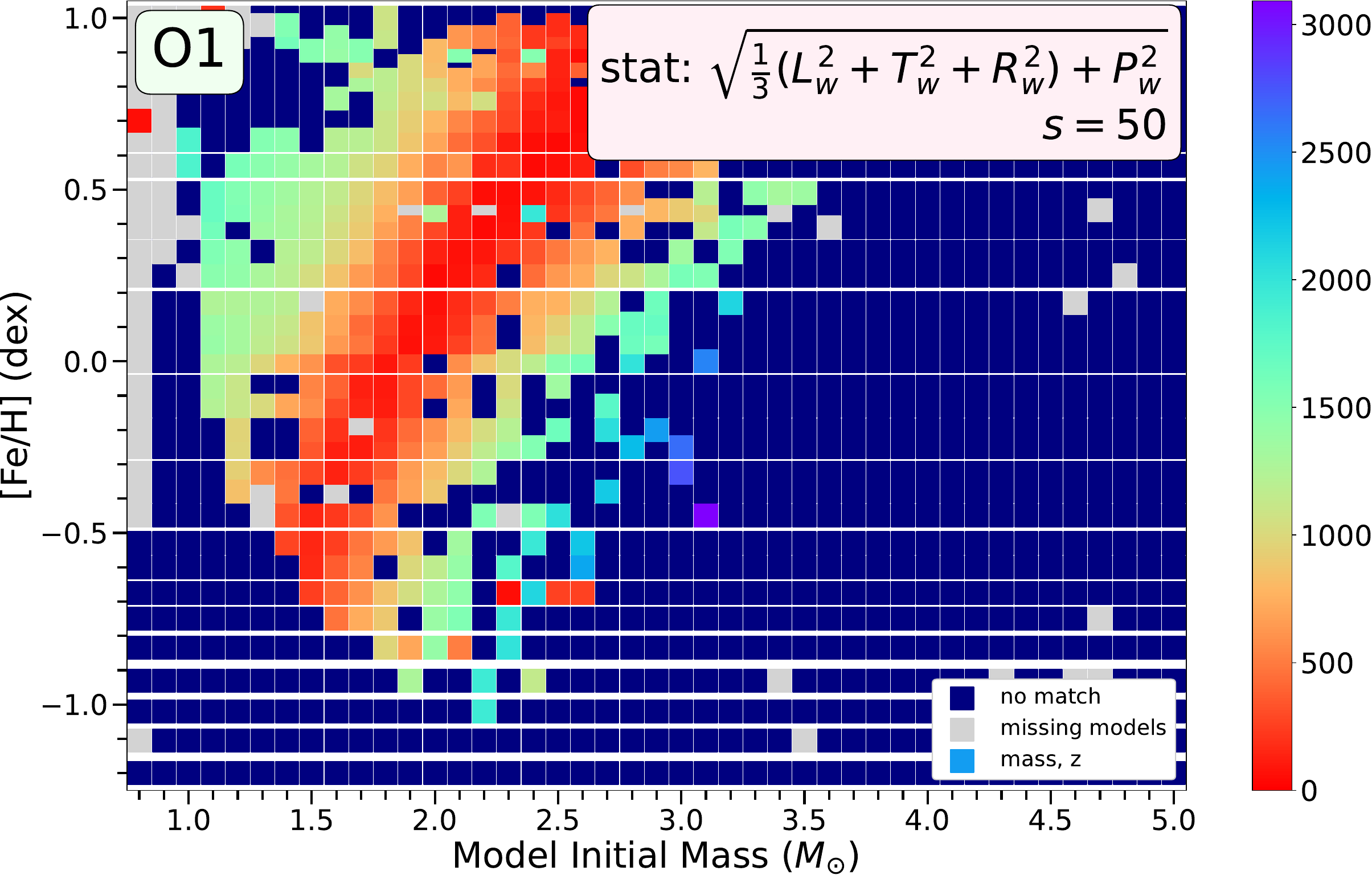}
\includegraphics[width=0.48\columnwidth]{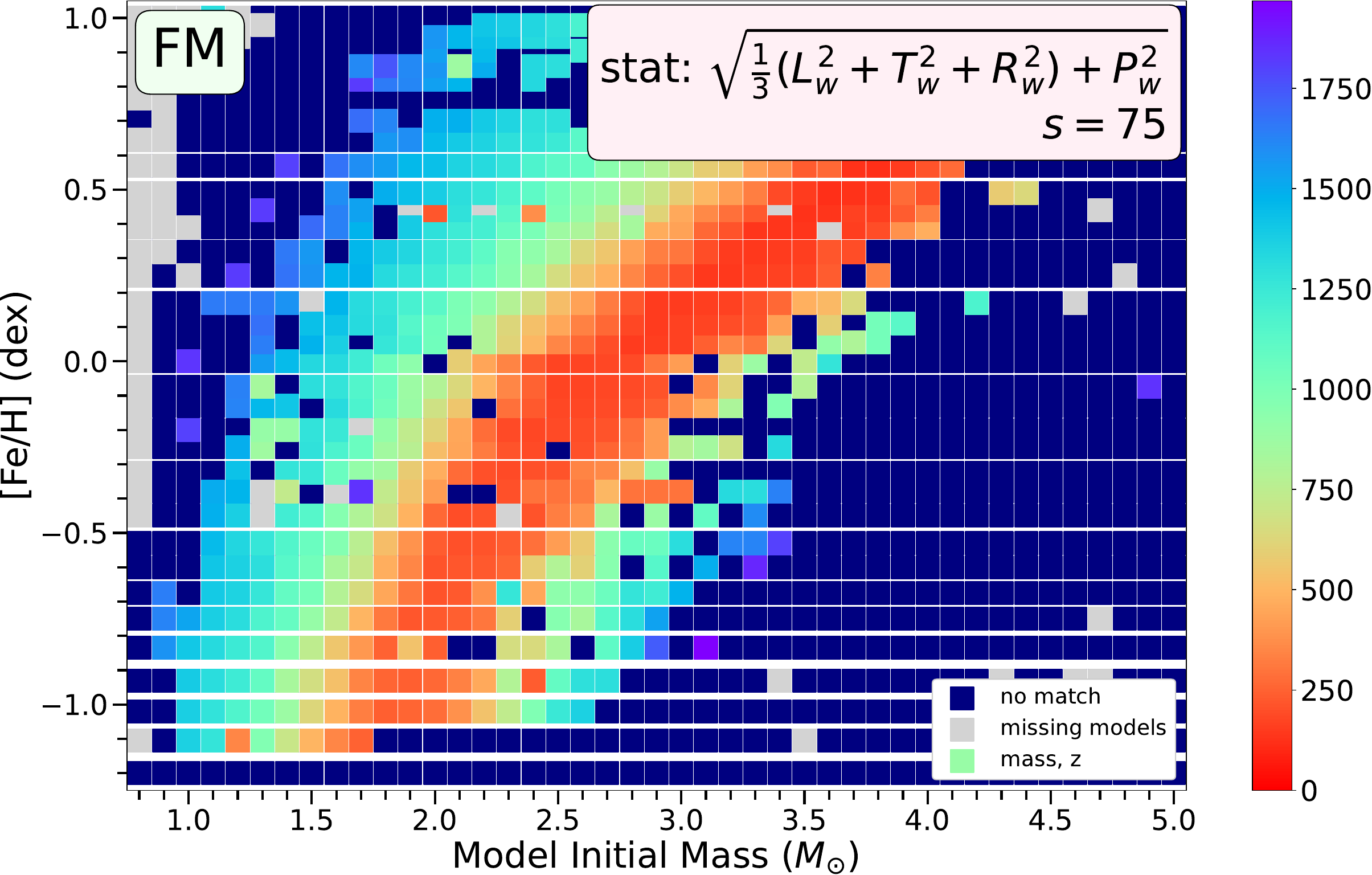}
\includegraphics[width=0.48\columnwidth]{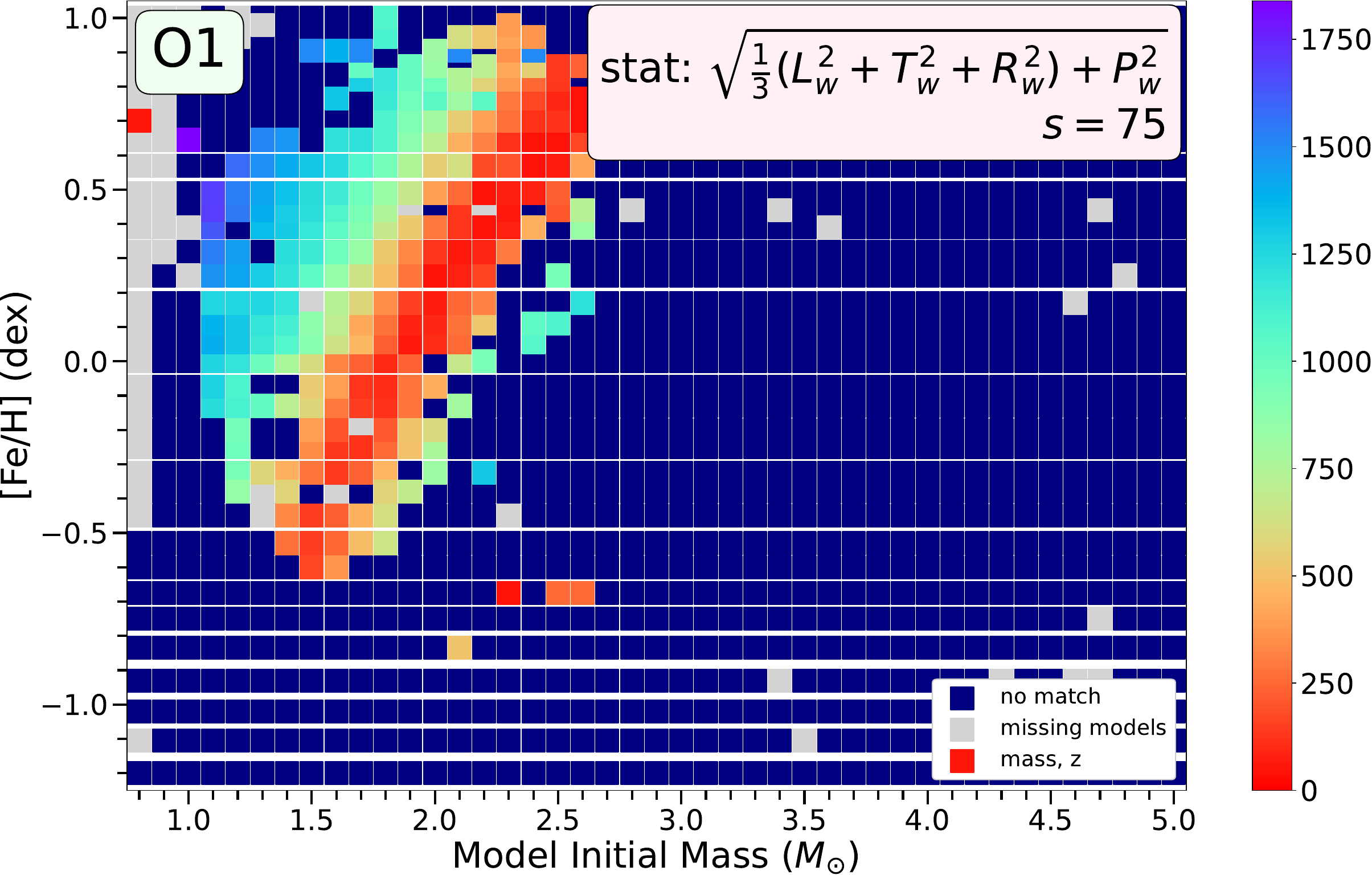}
\includegraphics[width=0.48\columnwidth]{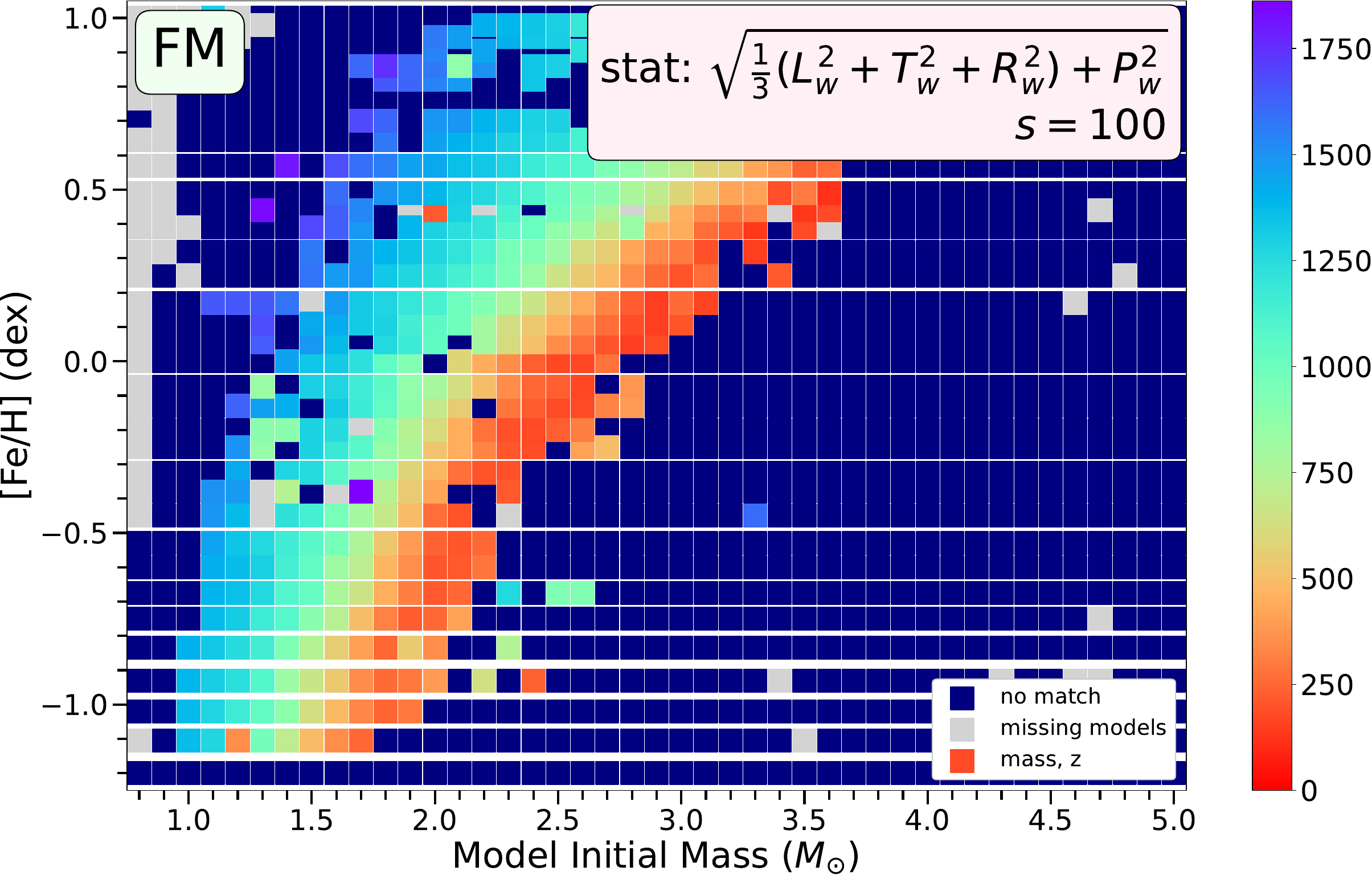}
\includegraphics[width=0.48\columnwidth]{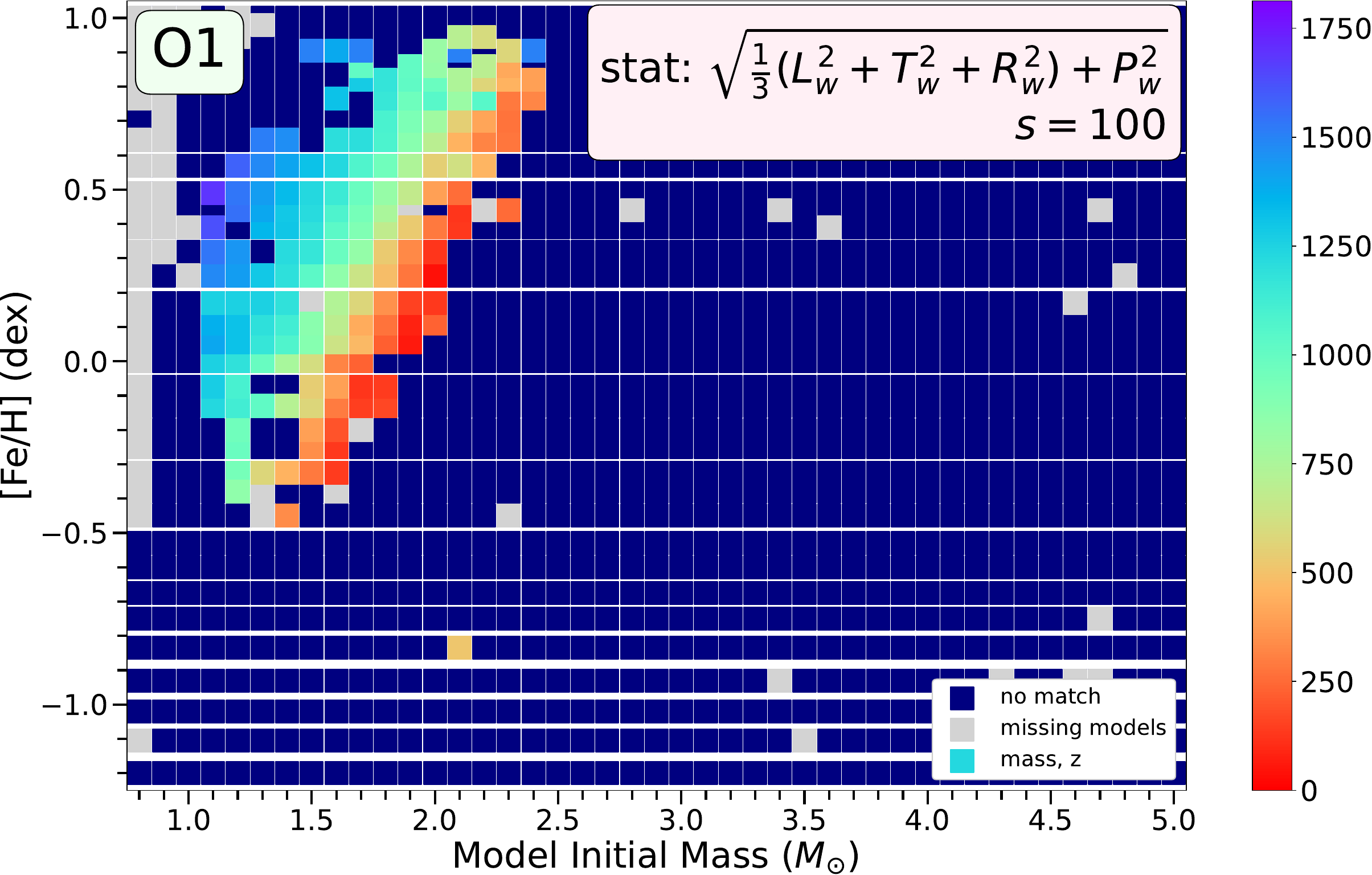}
\caption{The impact of changing the slope hardness parameter, $s$, is shown for the FM (left) and O1 (right) maps using the $P_\text{W}$ composite statistic and fixed-helium grid. While the difference between $s=75$ (adopted in the main analysis) and $s=50$ does not significantly affect the regions hosting the best solutions, enforcing $s=100$---the strictest slope requirement---truncates this region and removes many solutions. This is a good demonstration of why it is better not to assume we know the shape of the underlying period decline in the observations, as discussed in Section~\ref{subsec:meadering}.}
\label{fig:hardness}
\end{figure*}

We run this numerical experiment for three choices of slope hardness, $s$ such that $\dot{P} \approx \Delta P / \Delta t \ge s / \Delta t$ (Equation \ref{eq:hardness}), in all cases. 
The default value, shown throughout the rest of the paper, is $s=75$. A value of $s=50$ relaxes the slope constraint, making the standards for well-fitting pulses more permissive, and a value of $s=100$ tightens this constraint. Figure \ref{fig:hardness} demonstrates the effect of changing $s$ for the fixed-helium grid.

\begin{figure*}
\centering
\includegraphics[width=0.48\columnwidth]{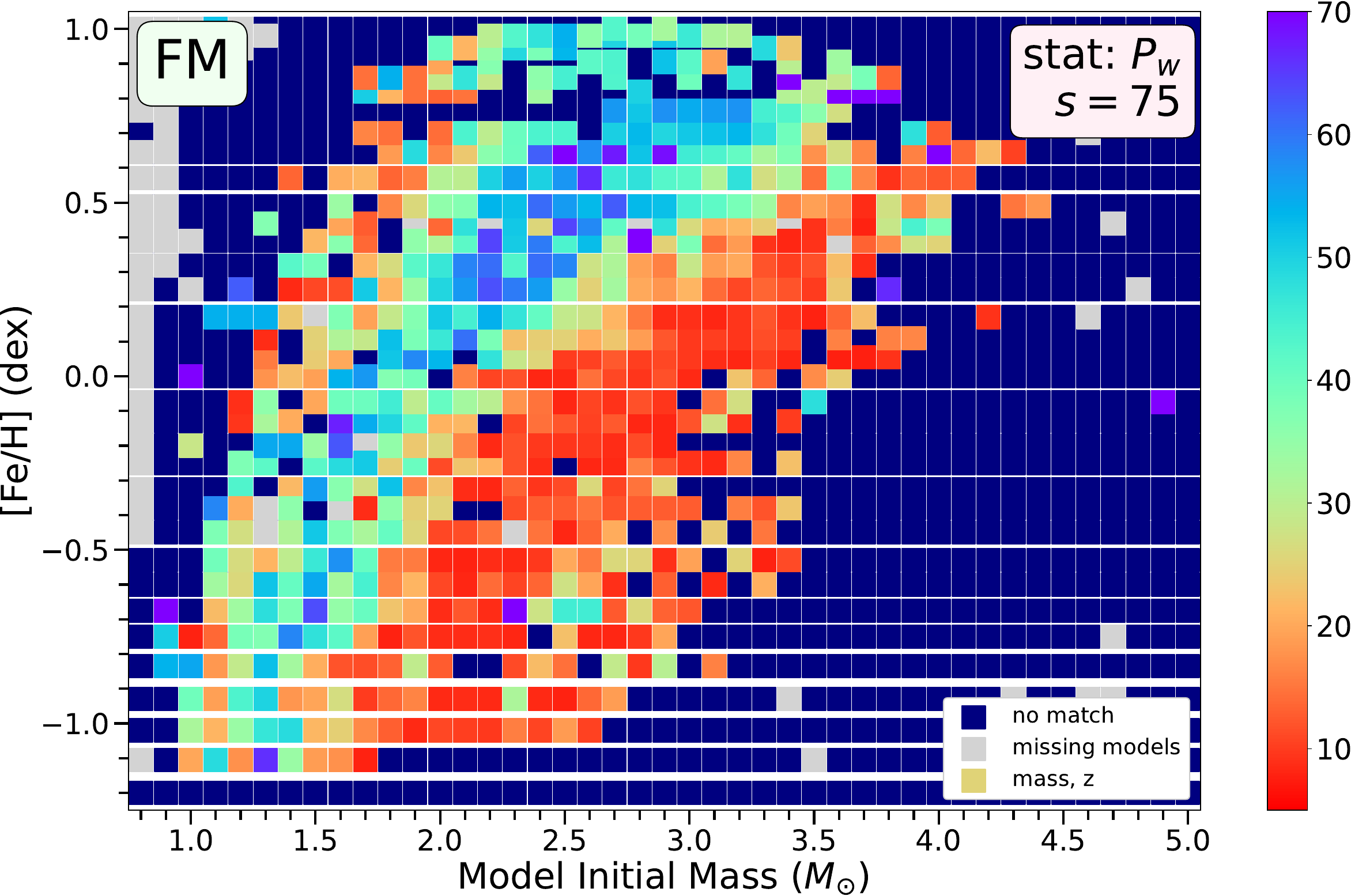}
\includegraphics[width=0.48\columnwidth]{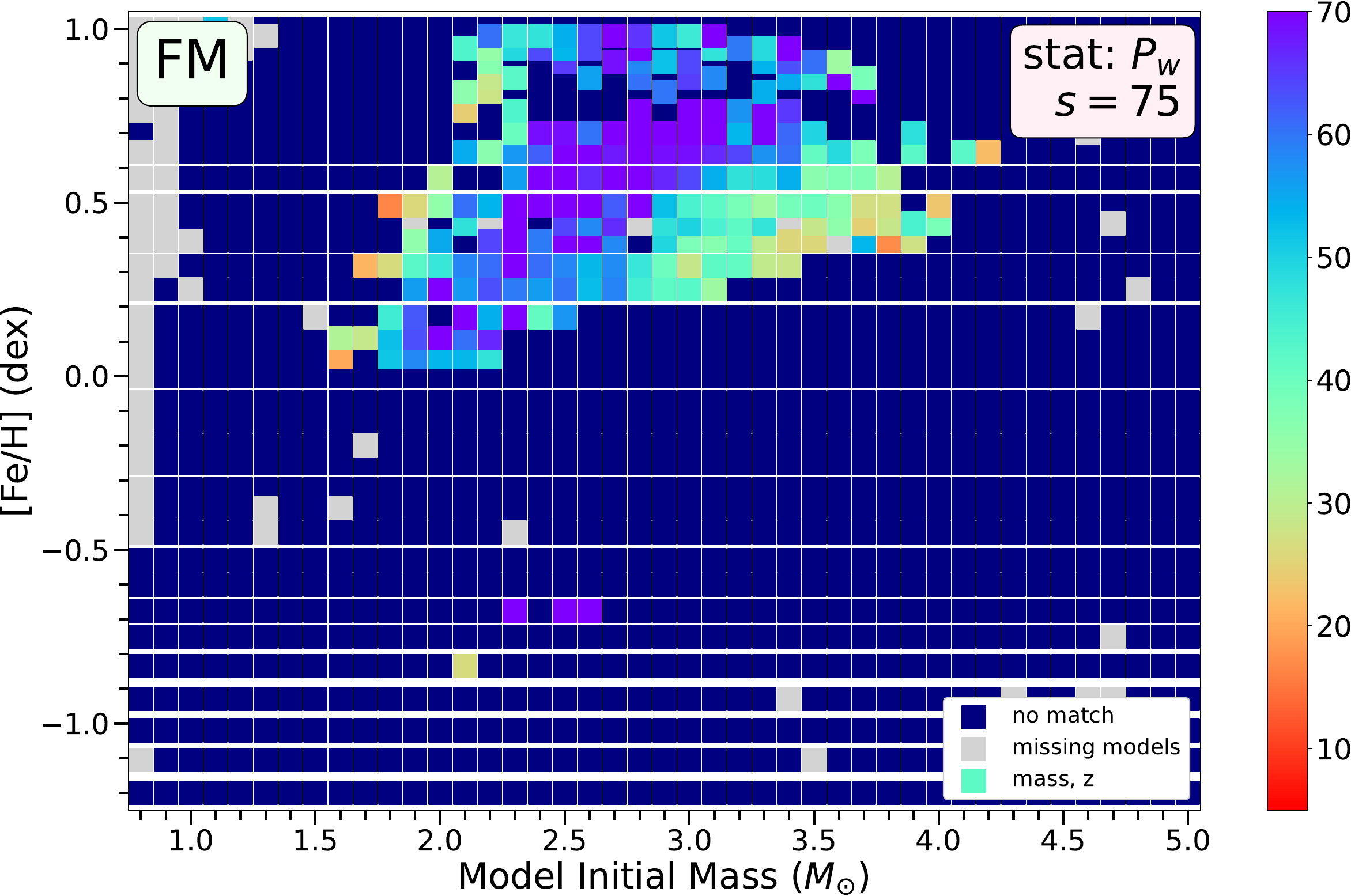}
\includegraphics[width=0.48\columnwidth]{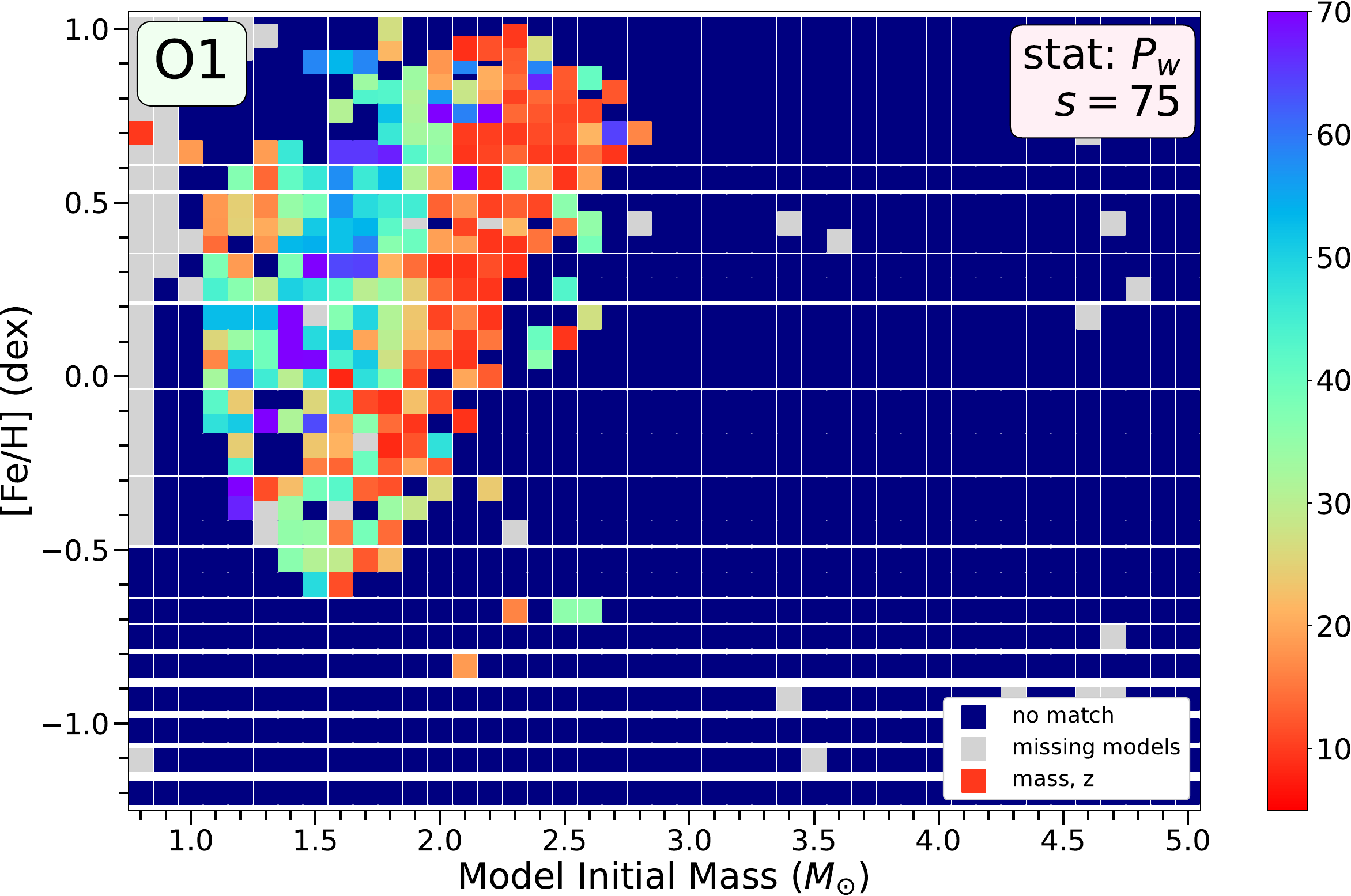}
\includegraphics[width=0.48\columnwidth]{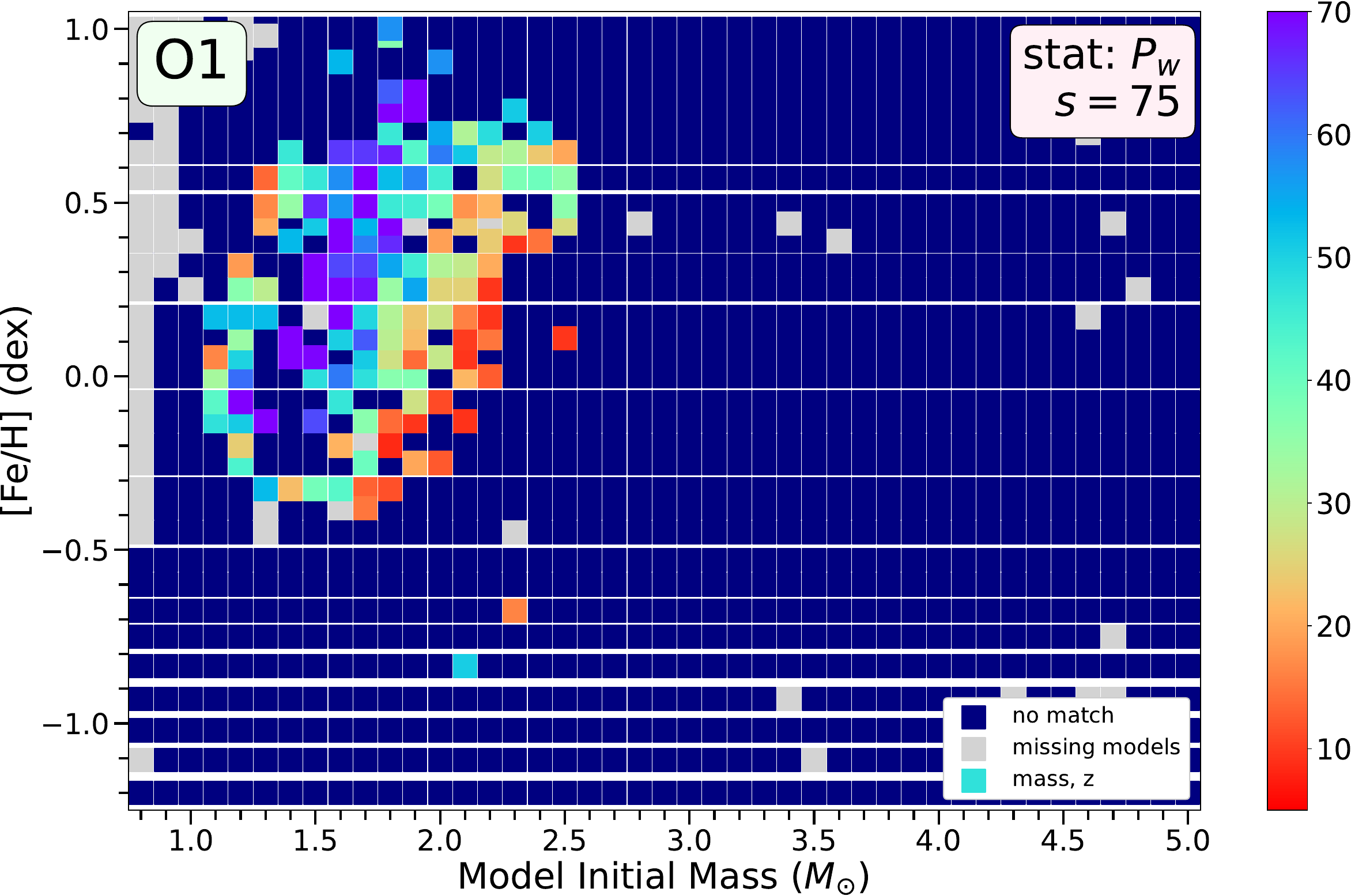}
\caption{Results for a more restrictive interpretation of the classical and seismic observational constraints are shown for the FM and O1 using the $P_\text{W}$ statistic (fixed helium grid). Panels on the left use the observational boundaries defined in Table \ref{table:obs_domain}, whereas panels on the right reduce the uncertainties by a factor of two in all parameters ($L$,$T_\text{eff}$,$R$, Period).
As we can see, such a strict interpretation of the observations removes most solutions, so it is not informative.}
\label{fig:strictness}
\end{figure*}

We briefly discuss in Section \ref{sec:fittingprocedure} that the outcomes are affected when we change the observational boundaries. This is demonstrated in Figure \ref{fig:strictness}, which shows what happens when we adopt tighter constraints on the classical and seismic observations. 

\FloatBarrier
\section{Data visualization}
\label{appendix:visualizer}

\begin{figure*}
\centering
\fboxsep0pt
\fbox{\includegraphics[width=0.98\textwidth]{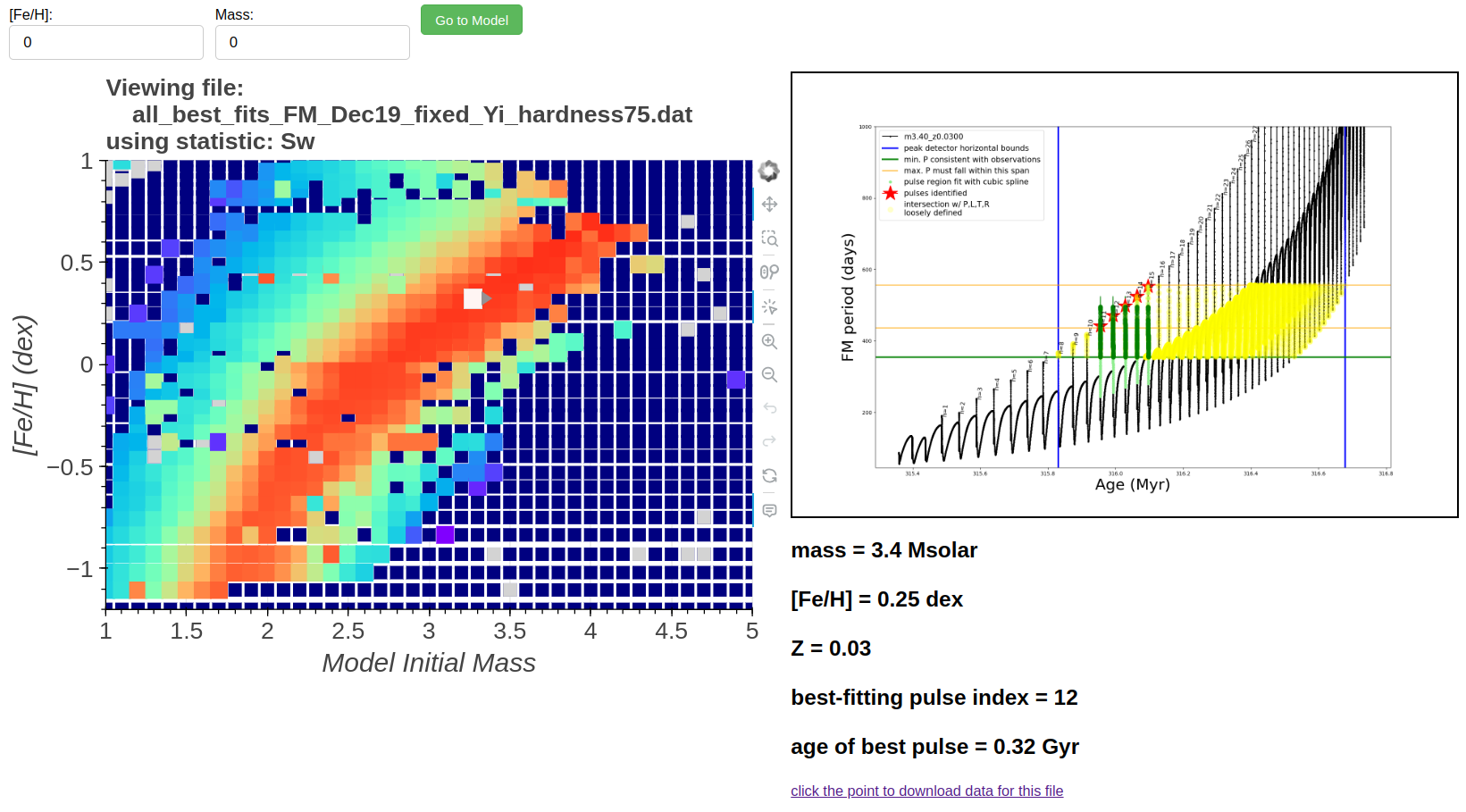}}
\fbox{\includegraphics[width=0.98\textwidth]{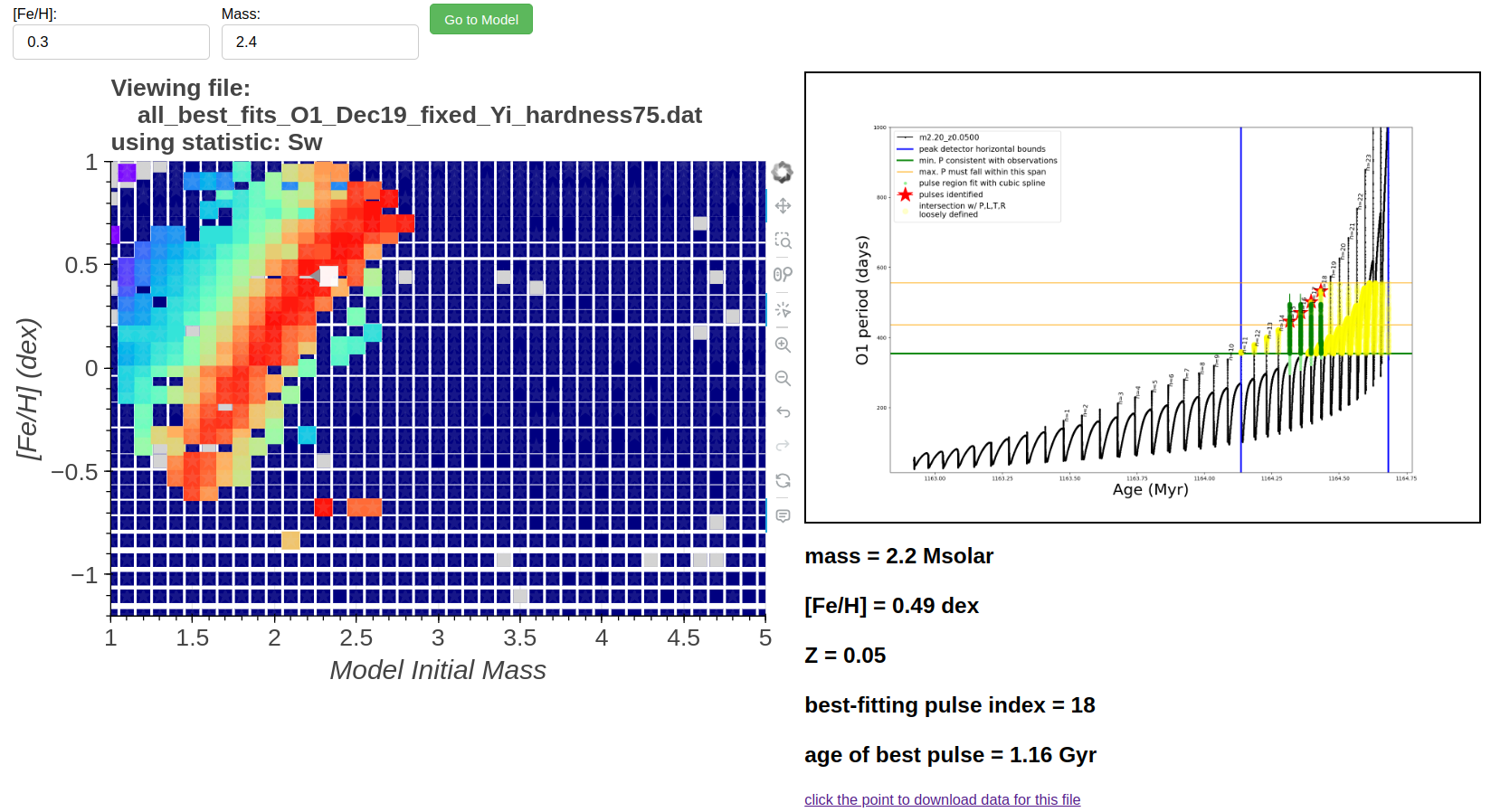}}
\caption{
Two examples from the data visualization tool packaged with this grid are shown. \textbf{UPPER:} The $S_\text{W}$ composite statistic is shown for the FM map with a model at $m= 3.4M_{\odot}, Z=0.0300$ highlighted. The TP-AGB spectrum associated to this model is shown to the right, with the observationally-compatible region highlighted in yellow and the pulses with power-down maxima in this regime identified by red stars. A similar figure is available for every model that ran successfully. 
\textbf{LOWER:} Same as above, but showing the $S_\text{W}$ statistic on the O1 map with an $m= 2.2M_{\odot}, Z=0.0500$ model highlighted.
}
\label{fig:data_visualizer}
\end{figure*}

Figure \ref{fig:data_visualizer} shows a screenshot of the data visualization application released with the model grid, available at the first author's GitHub:\footnote{Repository to be made public upon publication.} \url{https://github.com/mjoyceGR/AGB_grid_visualizer}. The pulse-fitting images are hosted at \url{https://www.meridithjoyce.com/images/AGB_grid/}, organized by date. Each grid contains realizations for the FM and O1 for three values of slope hardness. The statistical measures for each realization are kept in human-readable data files, also on GitHub.

Features of the visualization tool include:
\begin{itemize}
    \item interactive options including zoom, pan, reset, etc., on the tool bar to the right of the map
\item navigation boxes (upper left) into which model parameters can be entered, which then zoom in and align the map on that model
\item mouse-over images (pngs) for each successful model that appear automatically on the right when the cursor is hovered over that parameter combination
\item additional information about the model and best-fitting pulse, where applicable, appearing below the pulse spectrum image on the lower right
\item ability to choose from any combination of slope hardness, $s=\{50, 75, 100\}$, FM or O1 mode, helium varied or fixed, and all of the statistics presented in this paper (see Figures \ref{fig:heatmap_FM} and \ref{fig:heatmap_O1}) as well as from other properties, including age and best-fitting pulse index. Instructions for how to specify these settings are provided on GitHub. 
\end{itemize}

\end{document}